\documentclass[letter,11pt]{article}

\usepackage{amsmath}
\usepackage{amssymb}

\usepackage{cite}
\usepackage{float}  
\usepackage{hyperref} 

\usepackage{lineno}

\usepackage{microtype}
\DisableLigatures[f]{encoding = *, family = * }
\usepackage{authblk}

\usepackage{epsfig}
\usepackage{lscape,graphicx}
\usepackage{array}
\usepackage{caption}
\usepackage{algorithm}
\usepackage{algpseudocode}
\usepackage{centernot}
\usepackage{bbm}
\usepackage{chemarr}
\usepackage[margin=1in]{geometry}

      \newcommand {\mm}[1] {\ifmmode{#1}\else{\mbox{\(#1\)}}\fi}
      \newcommand{\real} {\mm{{\mathbb R}}}

      \newcolumntype{^}{>{\currentrowstyle}}

\newcommand{\utwi}[1]{\mbox{\boldmath $ #1$}}
\newcommand{\ba}{{\utwi{a}}}
\newcommand{\bb}{{\utwi{b}}}

\newcommand{\bee}{{\utwi{e}}}

\newcommand{\bp}{{\utwi{p}}}
\newcommand{\bq}{{\utwi{q}}}

\newcommand{\bs}{{\utwi{s}}}

\newcommand{\bv}{{\utwi{v}}}

\newcommand{\bx}{{\utwi{x}}}

\newcommand{\bA}{{\utwi{A}}}
\newcommand{\bB}{{\utwi{B}}}
\newcommand{\bC}{{\utwi{C}}}
\newcommand{\bD}{{\utwi{D}}}
\newcommand{\bE}{{\utwi{E}}}

\newcommand{\bG}{{\utwi{G}}}
\newcommand{\bH}{{\utwi{H}}}

\newcommand{\bL}{{\utwi{L}}}

\newcommand{\bO}{{\utwi{O}}}
\newcommand{\bP}{{\utwi{P}}}
\newcommand{\bQ}{{\utwi{Q}}}

\newcommand{\bS}{{\utwi{S}}}
\newcommand{\bT}{{\utwi{T}}}

\newcommand{\bV}{{\utwi{V}}}

\newcommand{\bpi}{{\utwi{\pi}}}

\DeclareMathOperator{\Err}{Err}

\DeclareMathAlphabet{\mathpzc}{OT1}{pzc}{m}{it}

\newtheorem{fact}{Fact}

\title{Accurate Chemical Master Equation Solution Using Multi-Finite Buffers}
\author[1,2]{Youfang Cao\thanks{ycao@lanl.gov}}
\author[1]{Anna Terebus}
\author[1,3]{Jie Liang\thanks{jliang@uic.edu}}
\affil[1]{Department of Bioengineering, University of Illinois at Chicago, Chicago IL}
\affil[2]{Current address: Theoretical Biology and Biophysics (T-6), Center for Nonlinear Studies (CNLS), Los Alamos National Laboratory, Los Alamos, NM}
\affil[3]{Corresponding author}

\begin{document}

\date{}
\maketitle




\begin{abstract}
The discrete chemical master equation (dCME) provides a fundamental framework
for studying stochasticity in mesoscopic networks. Because of the multi-scale
nature of many networks where reaction rates have large disparity, directly
solving dCMEs is intractable due to the exploding size of the state space. It
is important to truncate the state space effectively with quantified errors, so
accurate solutions can be computed. It is also important to know if all major
probabilistic peaks have been computed. Here we introduce the Accurate CME
(ACME) algorithm for obtaining direct solutions to dCMEs. With multi-finite
buffers for reducing the state space by $O(n!)$, exact steady-state and
time-evolving network probability landscapes can be computed. We further
describe a theoretical framework of aggregating microstates into a smaller
number of macrostates by decomposing a network into independent aggregated
birth and death processes, and give an \textit{a priori} method for rapidly determining
steady-state truncation errors. The maximal sizes of the finite buffers for a
given error tolerance can also be pre-computed without costly trial solutions
of dCMEs. We show exactly computed probability landscapes of three multi-scale
networks, namely, a 6-node toggle switch, 11-node phage-lambda epigenetic
circuit, and 16-node MAPK cascade network, the latter two with no known
solutions. We also show how probabilities of rare events can be computed from
first-passage times, another class of unsolved problems challenging for
simulation-based techniques due to large separations in time scales. Overall,
the ACME method enables accurate and efficient solutions of the dCME for a
large class of networks.
\end{abstract}

\section{Introduction}

Biochemical reaction networks are intrinsically stochastic~\cite{Ao2005,Stewart2012,Qian2012} 
and often multi-scale when there exists large disparity in reaction rates.
When genes, transcription factors, signaling
molecules, and regulatory proteins are in small quantities ($10 \sim
100$ nM), stochasticity plays
important roles~\cite{Arkin1998,Swain2002,Elowitz2002,Cao2010}. 
Deterministic models based on chemical mass action kinetics cannot
capture the stochastic nature of these
networks~\cite{McAdams1999,Wilkinson2009,Cao2010}.  Instead, the
discrete Chemical Master Equations (dCME) that describe the
probabilistic jumps between discrete states due to the firing of
reactions can fully describe these mesoscopic stochastic processes in
a well mixed
system~\cite{Gillespie_JPC77,Gillespie-PhysicaA-1992,vanKampen2007,Beard2008,Gillespie2009-jcp}.

However, studying the stochastic behavior of a multi-scale network is
challenging. The rate constants of different reactions often have large
separations in time scale by a few orders of magnitude. Copy numbers of
molecular species can also span across a number of orders of magnitude, further
exacerbating the problem of time separations between slow and fast reactions.
Even with a correctly constructed model of a stochastic network,
it is generally unknown if an accurate solution has been
found. One does not know if a computed probabilistic landscapes
is overall erroneous and how such errors can be quantified.  For
example, it is difficult to know if all major probabilistic peaks have
been identified, or important peaks in the usually high dimensional
space with significant probability mass are undetected.  
Furthermore, the best possible accuracy one can
hope to achieve with given finite computing resources is generally unknown. In
addition, one does not know what  is required so
accurate solutions with errors smaller than a predefined tolerance can
be obtained.

While the time-evolving probability landscape over discrete states
governed by the dCME  provides detailed information of the underlying 
dynamic stochastic processes, the dCME cannot be solved
analytically, except for a few very simple
cases~\cite{Darvey-1966-jcp,McQuarrie-1967-JAppProb,Laurenzi-2000-jcp,Vellela2007}. 
Approximations to the dCME such as the chemical
Fokker-Planck equation (FPE) and the chemical Langevin equation (CLE)
are widely used to study stochastic
reactions~\cite{VanKampen-1961,Gillespie_JCP2000,Gillespie2002,Haseltine2002,Gardiner2004-book,Ao2010,Shi2012}.
However, these approximations assume relatively large copy numbers of molecules, so the states
can be regarded as continuous, and higher order terms of the Kramers-Moyal expansion of the
dCME can be truncated~\cite{vanKampen2007}. 
These approximations do not provide a full account of the 
stochasticity of the system and are not valid when copy numbers of
molecular species are small~\cite{Gillespie_JCP2000}.  
Although errors of these approximations have been assessed for simple reactions~\cite{Grima-2011-jcp,Grima-2013-bcmgenomics} 
and a recent study showed that CLE failed to converge to the correct steady
state probability landscape (see the Appendix of ref~\cite{Cao2010}), the consequences of
such approximations for realistic problems involving many molecular species and
with complex reactions across multiple temporal scales are largely unknown.

The stochastic simulation algorithm (SSA) is widely used to study stochasticity
in biological networks. It generates reaction trajectories dictated by the
underlying dCME of the network~\cite{Gillespie_JPC77}. The stochastic properties of the network can
then be inferred through analysis of a large number of simulation trajectories.
However, as the SSA follows high-probability reaction paths, it is therefore
inefficient for sampling biologically critical rare events that often occur in
stiff multi-scale reaction networks, in which slow and fast reactions are
well-separated in time scale~\cite{Allen2005Sampling,Kuwahara2008,Daigle2011,Jiao2011,Cao2013JCP,Wang2014}. 
In addition, assessment of its convergence of
simulation trajectories is also difficult. Recent development in biased
sampling aims to address this problem~\cite{Allen2005Sampling,Kuwahara2008,Daigle2011,Cao2013JCP}.

An attractive approach to study stochastic networks is to directly
solve the dCME numerically. By computing the exact probability
landscape of a stochastic network, its properties, including those
involving rare events, can be studied accurately in details. The
finite state projection (FSP) method and the sliding window method are
among several methods that have been developed to solve the dCME
directly~\cite{Munsky2006,CaoBMCSB08,Cao2010,Wolf2010,Jahnke2011}.  

The finite state projection (FSP) method is based on a truncated
projection of the state space and uses numerical techniques to compute
direct solution to the dCME~\cite{Sidje1998,Munsky2006}.  Although the
error due to state space truncation can be captured by the absorption
state, to which all truncated states are projected~\cite{Munsky2006},
there is no systematic guidance as to which states and how many of
them should be incorporated so the error can be minimized to remain
within an acceptable tolerance~\cite{Munsky2006,Munsky2007}.
Furthermore, the introduction of the absorption state leads to
accumulation of errors as time proceeds, as this state would
eventually absorb all probability mass.  
Designed to study transient behavior of stochastic networks, the FSP method
therefore is challenged to compute the steady state probability landscape and
the first passage time distribution of rare events in a multi-scale network.

The sliding window method for solving the dCME is also based on
truncation of the state space. In this case, the state space is
adaptively restricted to those that are likely relevant within a small
time-window, with the assumption that most of the probability mass is
contained within a set of pre-selected states~\cite{Wolf2010}. 
However, to ensure that the truncation error is small, a large number
of states need to be included, as the size of the state space takes
the form of a $d$-dimensional hypercube, with the upper and lower
bounds of copy numbers of each of the $d$ molecular species pre-determined by a
Poisson model~\cite{Wolf2010}.  

The main difficulty of all these methods is to have an adequate and accurate account of
the discrete state space.  As the copy number of each of the $d$
molecular species takes an integer value, conventional hypercube-based
methods incorporate all vertices in a $d$-dimensional hypercubic
integer lattice, which has an overall size of $ O
(\prod_{i=1}^{d}{m_i})$, where $m_i$ is the maximally allowed copy
number of molecular species $i$.  State enumeration rapidly becomes
intractable, both in storage and in computing time.  For example,
assuming a system has $16$ molecular species, each with maximally $9$
copies of molecules, a state space of size $(9+1)^{16} = 10^{16} $
would be required. This makes the direct solution of the dCME
impossible for many realistic problems.

To address the issue of prohibitive size of the discrete state space,
the finite buffer discrete CME (fb-dCME) method was developed for
efficient state enumeration~\cite{CaoBMCSB08}.  This algorithm is
provably optimal in both memory usage and in time required for
enumeration when a single buffer queue is used.  Instead of including
every states in a hypercube, it examines only states that can be
reached from a given initial state.  It can be used to compute the
exact probability landscape of a closed network, or an open network
when the net gain in newly synthesized molecules does not exceed a
predefined finite capacity.  However, as the available memory is
limited, state truncation will eventually occur for open systems when
synthesis reactions outpaces degradation reactions, and for closed
system whose full enumeration requires memory that exceeds available
capacity.  In these cases, it is unclear whether the error associated
with a truncated state space is within a tolerance threshold.
Furthermore, similar to other methods aimed to solve the dCME
directly, it is unclear how to minimize the error of a truncated state
space, thus limiting the scope of applications of this method.

In this study, we introduce the \underline{A}ccurate
\underline{C}hemical \underline{M}aster \underline{E}quation method
(the ACME method) for solving the dCME.  
Our method is based on the decomposition of the multi-scale stochastic reaction
network into multiple independent components, each is governed by its own
birth-death process, and each has a unique pattern of generation and
degradation of molecules. In the ACME method, each independent component is
equipped with its own finite state sub-space controlled by a separate buffer
queue.  Similar to the original fb-dCME method, it is optimal in space and in
time required for state enumeration, but has the advantage of more effective
usage of the overall finite state space, and allows detailed analysis.  This
approach improves computing efficiency significantly and can generate state
spaces of much larger effective sizes.

We also provide a method for rapid estimation of the
errors in the computed steady state probability landscape upon
truncation of the state space when using a buffer bank with a finite
capacity.  An estimation of the required buffer sizes can also be
computed so the truncation error is within a pre-defined tolerance.
These estimations are derived conservatively, so that the actual
errors will not be larger than the estimated errors.  A strategy for
optimized buffer allocation is also given.  
Furthermore, the error bounds and required buffer sizes for each individual
independent component can all be rapidly computed {\it a priori} without costly
computation of trial solutions to the dCME.
These are based on results of theoretical analysis of
the upper bound of the truncation error of the probability landscape
at the steady state, which will be discussed in details.  The ACME
algorithm, along with the error estimation are implemented in the
ACME package.  Overall, the ACME method allows accurate solutions to
the dCME with small and controlled errors for a much larger class of
biological problems than previously feasible.

Our paper is organized as follows.  We first review basic concepts of
the discrete chemical master equation and issues associated with the
finite discrete state space. We then describe the concept of reaction
graph, its decomposition, and how independent birth-death components
can be identified.  We further introduce the ACME algorithm
in which multi-finite buffers are used
for state enumeration.  This is followed by a discussion of
results of theoretical analysis of errors in the steady state probability
landscape due to state truncation, and how probability of boundary
states can be used to construct upper bounds of the truncation errors.
We then give detailed examples of three biological networks, namely,
the toggle switch, the epigenetic circuit of lysis-lysogeny decision
of phage lambda, and a model of MAPK cascade.  We discuss the computed
time-evolving and the steady state probability landscapes, along with
the significant state space reduction achieved for these networks.
Results on the challenging problem of estimating rare event
probability through the computation of the first-passage times of
these networks are also reported.  We conclude with summaries and
discussions.

\section{Methods and Theory}

\subsection{Background}

\subsubsection{Reaction Network, State Space and Probability Landscape}


In a well-mixed biochemical system with constant volume and
temperature, we assume there are $n$ molecular species, denoted as ${\mathcal X}
= \{ X_1 , X_2, \cdots, X_n \}$, and $m$ reactions, denoted as
${\mathcal R} = \{ R_1, R_2, \cdots, R_m \}$. Each reaction $R_k$ has
an intrinsic reaction rate constant $r_k$.  The microstate of the
system at time $ t $ is given by the non-negative integer column
vector $\bx(t) \in \mathbb{Z}_{\geq 0}^n$ of copy numbers of each molecular
species: $\bx (t) = (x_1(t), x_2(t), \cdots, x_ n(t) )^T$, where
$x_i(t)$ is the copy number of molecular species $X_i$ at time $t$. An
arbitrary reaction $R_k$ with intrinsic rate $r_k$ takes the general
form of
$$
c_{1k}X_1 + c_{2k}X_2 + \cdots + c_{nk}X_n
\overset{r_k}{\rightarrow} c'_{1k}X_1 + c'_{2k}X_2 + \cdots +
c'_{nk}X_n,
$$
which brings the system from a microstate $\bx_j$ to
$\bx_i$. The difference between $\bx_i$ and $\bx_j$ is the
stoichiometry vector $\bs_k$ of reaction $R_k$: $ \bs_k = \bx_i -
\bx_j = (s_{1k}, s_{2k},\cdots, s_{nk})^T =( c'_{1k}-c_{1k},\,
c'_{2k}-c_{2k},\, \cdots,\, c'_{nk}-c_{nk})^T \in { \mathbb Z}^n.$ The
stoichiometry matrix $\bS$ of the network is defined as:
$\bS = (\bs_1, \bs_2, \cdots, \bs_m) \in {\mathbb Z}^{n \times m}$,
where each column correspond to one reaction. The rate $A_k(\bx_i,
\bx_j)$ of reaction $R_k$ that brings the microstate from $\bx_j$ to
$\bx_i$ is determined by $r_k$ and the
combination number of relevant reactants in the current microstate
$\bx_j$: $$A_k(\bx_i, \bx_j) = A_k (\bx_j) = r_k \prod_{l=1}^n
\binom{x_l}{c_{lk}},$$ assuming the convention $\binom{0}{0} = 1$.

All possible microstates that a system can visit from a given
initial condition form the state space 
${\Omega} = \{\bx(t) | \bx(0), \, t \in (0, \, \infty)\}.$ We
denote the probability of each microstate at time $t$ as $p(\bx(t))$,
and the probability distribution at time $t$ over the full state
space as ${\bp}(t) = \{(p(\bx(t)) | \bx(t) \in \Omega) \}.$ We
also call ${\bp}(t)$ the {\it probability landscape\/} of the
network~\cite{Cao2010}.

\subsubsection{Discrete Chemical Master Equation}

The discrete chemical master equation (dCME) can be written as a set
of linear ordinary differential equations describing the change in probability
of each discrete state over time:
\begin{equation}
\frac{d p(\bx, t)}{dt} = \sum_{\bx',\, \bx' \neq \bx  } 
[ A(\bx, \bx') p (\bx', t) - A(\bx', \bx) p (\bx, t)].
\label{eqn:dcme1}
\end{equation}
Note that $p(\bx, t)$ is
continuous in time, but is over states that are discrete.
In matrix form, the dCME can be written as: 
\begin{equation}
\frac{d \bp(t)}{dt} = \bA \bp(t), 
\label{eqn:dcme2}
\end{equation}
where $\bA  \in 
\real^{|\Omega| \times |\Omega|}$
is the transition rate matrix formed by the collection of
all $A(\bx_i, \bx_j)$: 
\begin{equation}
 A(\bx_i,\bx_j) = \left\{ 
  \begin{array}{l l}
    - \sum_{\substack{\bx' \in \Omega, \\ \bx' \neq \bx_j}} A_k(\bx', \bx_j) & \quad \text{if $\bx_i = \bx_j$}, \\
    A_k(\bx_i, \bx_j) & \quad \text{if $\bx_i \neq \bx_j$ and $\bx_j \stackrel{R_k}{\longrightarrow} \bx_i$}, \\
		0 & \quad \text{otherwise}. \\
  \end{array} \right.
\label{eqn:matA}
\end{equation}

\subsection{Finite Buffer for State Space Enumeration}

Enumeration of the state space is a prerequisite for directly solving 
the dCME.  The method of finite-buffer dCME (fb-dCME) provides an
efficient algorithm for state enumeration~\cite{CaoBMCSB08,Cao2010}.
By treating states as nodes and reactions as edges, the problem of
state enumeration is transformed into that of a graph traversal
problem~\cite{Cormen2001}. The fb-dCME algorithm uses the depth-first
search (DFS) to enumerate states that can be reached from an
initial state~\cite{CaoBMCSB08}.  For closed
networks with no synthesis reactions, the finite state space can be
fully enumerated, assuming the capacity of available computer memory is adequate.

For open networks with synthesis and degradation reactions found in a 
biological system, the size of the state space is also finite, as the
total mass of molecules in a reaction system is conserved and the
duration of reactions is bounded by the life-time of a cell.
Therefore, the net number of synthesized molecules 
that need to be modeled is finite.  However, errors due to state space
truncation will occur when the compute capacity is insufficient to 
fully account for the finite state space, as synthesis reaction can
no longer proceed after memory exhaustion.  Similarly, truncation
error will occur when the size of the full state space of a closed
network cannot be contained in the available memory. 

The fb-dCME algorithm uses a buffer of a predefined capacity as a counter 
to keep track of the total number of molecules in the reaction system. 
Once the buffer capacity is determined, the maximum number of
molecules in the system is given, which is the number of molecules that 
can be synthesized in the model.  The buffer capacity is
dictated by the available computer memory. 
When a synthesis reaction occurs, one buffer token is spent.
When a degradation reaction occurs, one buffer token is deposited
back.  Multiple buffer tokens are taken or deposited when synthesis
and degradation involve higher-order reactions such as homo- or
hetero-oligomers, with the number of tokens equivalent to that of the
monomers. The fb-dCME algorithm has been successfully applied in studying the 
stability and efficiency problem of phage lambda lysogeny-lysis 
epigenetic switch~\cite{Cao2010}, as well as in direct computation of
probabilities of critical rare events in the birth and death process, 
the Schl\"{o}gl model, and the enzymatic futile cycle~\cite{Cao2013JCP}.


\subsection{Multi-Finite Buffers for State Space Enumeration}

Reaction rates in a network can vary greatly: many steps of fast
reactions can occur within a given time period, while only a few steps
of slow reactions can occur in the same time period.  The efficiency
of state enumeration can be greatly improved if memory allocation is
optimized based on different behavior of these reactions.

\paragraph{Independent Birth-Death (iBD) Processes} 
It is useful to examine the reaction network in terms of birth and
death processes, as birth (synthesis) and death (degradation)
are the only reactions that can change the total mass of an open
network by adding or removing molecules. These processes correspond to
spending or depositing buffer tokens, respectively.  Below we first
introduce the concept of \textit{reaction graph} and its partition into
disjoint components.  We then examine those components equipped with 
their own birth-death processes.

\paragraph{Reaction Graph and Independent Reaction Components}
We first construct an undirected graph $\bG_R$, with reactions form the
set of vertices $\bV$.  A pair of reactions $R_i$ and $R_j$ are then
connected by an edge $e_{ij}$ if they share either reactant(s) or
product(s).  
To correctly discover related reactions through the 
stoichiometry matrix, all molecular species in the network are represented 
using the combination of their most elementary form. 
For example, if a molecular species $C$ is a complex formed by 
$A$ bounded with $B$, we use the original form $A+B$ to represent $C$. 
Collectively, these reaction pairs sharing reactants or
products form the edge set of the graph: $\bE = \{ e_{ij} \} $.
The reaction graph $\bG_R$ can be decomposed into $u$ number of disjoint
{\it independent reaction components} $\{\bH_i\}$: $\bG_R =
\bigcup^u_{i=1} \bH_i$, with $\bE(\bH_i) \cap \bE(\bH_j) = \emptyset$ for $i \ne
j$.

We are interested in those independent reaction components $\bH_j$s that
contain at least one synthesis reaction.  These are
called {\it independent Birth-Death\/} (iBD) components
$\{ \bH^{iBD}_j\}$. 
The number $w$ of iBD components necessarily does not exceed the
number $u$ of connected components in $\bG_R$: $w \leq u$.

A number of methods can be used to decompose $\bG_R$ into
independent reaction components.  For example, the standard
disjoint-set data structure and the {\sc Union-Find} algorithm can be
used for this purpose~\cite{Cormen2001}.  Another method is to represent $\bG_R$ by an
$m \times m$ adjacency matrix $\bC$ or a Laplacian matrix $\bL$.
According to spectral graph theory, the connectedness of $\bG_R$ is encoded in the
eigenvalue spectrum of its Laplacian $\bL$~\cite{Chung1997}: the
number of connected components of $\bG_R$ is the multiplicity $u$ of
the $0$ eigenvalue of $\bL$, and the corresponding $u$ orthogonal
eigenvectors $(\bv_1, \cdots, \bv_u )$ gives  memberships for
reaction to be in each connected independent component.  Specifically, the
non-zero elements of the vector $\bv_i$ correspond to
the member reactions of an independent reaction component
$\bH_i$ of $\bG_R$.
Algorithm~\ref{alg:IBD} 
can be used to decompose $\bG_R$.  Additional information
on calculating $\bG_R$ can be found in the Appendix.

\begin{algorithm}
\caption{Determination of Independent Birth-Death Processes (iBDs) (${\mathcal X}, {\mathcal R}$)}
\label{alg:IBD}
\begin{algorithmic}
\State Network model: $\bO \leftarrow \{{\mathcal X}, {\mathcal R}\}$;
\State Initialization of number of iBDs $w = 0$;
\State Obtain the stoichiometry matrix $\bS$ of network $\bO$;
\State Construct adjacency matrix $\bC$ of reaction-centered graph $\bG_R$ following Eqn.~(\ref{eqn:adjC}) in Appendix ;
\State Construct degree matrix $\bD$ of $\bG_R$ following Eqn.~(\ref{eqn:degD}) in Appendix;
\State Construct the Laplacian matrix $\bL$ following Eqn.~(\ref{eqn:lapL}) in Appendix;
\State Calculate the eigenvalue spectrum of $\bL$ and obtain the multiplicity $u$ of eigenvalue $0$;
\State Calculate all $u$ orthogonal eigenvectors $\bv_i, \, i = 1, \cdots, u$ of the eigenvalue $0$;
\For {$i=1$ to $u$}
   \State Construct connected reaction sets $\bH_i = \left\{R_j | \text{ if } v_{i,\,j} \ne 0 \right\}$;
\EndFor
\For {$i=1$ to $u$}
   \If {there exists a synthesis or degradation reaction in $\bH_i$}
			\State $w \leftarrow w + 1$
			\State $\bH^{iBD}_w = \bH_i$
	 \EndIf
\EndFor
\State Output number of iBDs and buffers $w$, and iBDs: $\bH^{iBD}_i, \, i = 1, \cdots, w$. 
\end{algorithmic}
\end{algorithm}

\paragraph{Relationship between States and iBDs}
The iBDs are components of partitioned reactions according to how 
they share reactants/products, or equivalently, how they 
contribute to the change of the total mass of the network. The iBDs can be 
viewed as aggregated reactions and are dictated only by 
the topology of the network that connects reactions through shared 
reactants/products.  Once the stoichiometry matrix of a 
reaction network is defined, its iBDs are also determined. 

In contrast, a state is a physical realization of the network at a
particular time instance. It describes the number of molecules in the
system, regardless of which iBD(s) each may participate.  For a
mesoscopic system, the state of the system changes with time.  It is
possible a state can participate in transitions in multiple iBDs.
There are many ways states can be aggregated,  the aggregations we
study in later sections are by the total net number of synthesized
molecules in an individual iBD.

\paragraph{ACME Multi-Buffer Algorithm for State Enumeration}
To enumerate the state space more effectively, we introduce the
multi-buffer state enumeration algorithm for solving the discrete
chemical master equation (mb-dCME). We assign a separate buffer queue
$B_i$ of size $b_i \in \mathbb{Z}_{\geq 0}$ to each of the $i$-th iBD component.
Collectively, they form a buffer bank $\mathcal{B} = (B_1, \cdots,
B_w)$.  The current sizes of the buffer queues, or the numbers of the
remaining buffer tokens,  form a vector $\bb = (b_1, b_2, \cdots, b_w) \in \mathbb{Z}_{\geq 0}^w$.
The $i$-th synthesis reaction cannot proceed if the $i$-th buffer
queue is exhausted, \textit{i.e.}, $b_i = 0$, resulting in state truncation.

When all iBDs have infinite buffer capacities, we have the infinite buffer bank 
$\mathcal{I} = (\infty, \infty, \cdots, \infty)$.  
The infinite state space $\Omega^{(\mathcal{I})}$ 
associated with buffer bank $\mathcal{I}$ gives the full 
state space, which will give the exact solution of the dCME: 
$\Omega^{(\mathcal{I})} \equiv \Omega = \{\bx(t) | \bx(0), t \in (0, \infty)\}$. 
We further use 
$\mathcal{I}_j = (\infty, \cdots, \infty, B_j, \infty, \cdots, \infty)$ to 
denote a buffer bank when only the $j$-th iBD is finite with capacity $B_j$. 
We can define a partial order $\mathcal{B}' \leq \mathcal{B}''$ for buffer banks, 
if $B'_j \leq B''_j$ for all $j = 1, \cdots, w$. 
We then have $\mathcal{B} \leq \mathcal{I}_j \leq \mathcal{I}$. 
We also have $\Omega^{(\mathcal{B})} \subseteq \Omega^{(\mathcal{I}_j)} \subseteq \Omega^{(\mathcal{I})}$.

With the total amount of available computer memory fixed, each enumerated state
$\bx \in \mathbb{Z}_{\geq 0}^n$ is associated with a vector of buffer sizes
$\bb (\bx) = (b_1(\bx), b_2(\bx), \cdots, b_w(\bx))$, which records
the remaining number of unspent tokens in each buffer queue.  We can
augment the state vector $\bx$ by concatenating $\bb(\bx)$ after $\bx$
to obtain the expanded state vector $\hat{\bx} = (\bx, \bb) \in
\mathbb{Z}_{\geq 0}^{n+w}$. With the buffer queues in $\mathcal{B}$ defined, we
list the mb-dCME algorithm in Algorithm~\ref{alg:fbsmb}.   The
associated transition rate matrix $\bA$ can also be calculated using
Algorithm~\ref{alg:fbsmb}.

Instead of truncating the state space by specifying a maximum allowed
copy number $B$ for each individual molecular species as in the
conventional hypercube approach, the multi-buffer method specifies a
maximum allowed copy number $B$ for each buffer.  Assume the $j$-th
buffer contains $n_j$ distinct molecular species, the number of all
possible states for the $j$-th buffer is then that of the number of
integer lattice nodes in  an
$n_j$-dimensional orthogonal corner simplex, with equal length $B$ for
all edges starting from the origin.  The total number of integer
lattice nodes in this $n_j$-dimensional simplex gives the precise
number of states of the $j$-th buffer, which is
the multiset number $\binom{B + n_j}{n_j}$.  The size of the
state space is therefore much smaller than the size of the state space
$B^{n_j}$ that would be generated by the hypercube method, with a
dramatic reduction factor of roughly $n_j !$ factorial.  Note that
under the constraint of mass conservation, each molecular species in
this buffer can still have a maximum of $B$ copies of molecules.  
With a conservative assumption that different buffers are independent,
the size of the overall truncated state space is then $O(\prod_j
\binom{B + n_j}{n_j})$.  This is much smaller than the $n$-dimensional
hypercube, which has an overall size of $O( \prod_j B^{n_j}) =
O(B^n)$, with $n$ total number of molecular species in the network.
Overall, the state spaces generated using the multi-buffer algorithm
are dramatically smaller than those generated using the conventional
hypercube method without loss of resolutions.

\begin{algorithm}
\caption{Multi-Finite Buffer Optimal State Space Enumeration and Transition Rate Matrix Generation (${\mathcal X}, {\mathcal R}$, $\{ \bH^{iBD}_i \}$,  buffer capacities: $\bb = (b_1, b_2, \cdots, b_w)$)}
\label{alg:fbsmb}
\begin{algorithmic}
\State Network model: $\bO \leftarrow \{{\mathcal X}, {\mathcal R}\}$;
\State Initialization of $w$ Independent Birth-Death processes: $\bH^{iBD}_1, \bH^{iBD}_2, \cdots, \bH^{iBD}_w$;
\State Buffer capacities: $\bb = (b_1, b_2, \cdots, b_w)$;
\State Initial state:  $\bx^{t=0} \leftarrow \{x_1^{0}, x_2^{0}, \ldots, x_n^{0}\}$;
\State Initialize the state space and the set of transitions: 
    $\Omega \leftarrow \emptyset$; 
    $\bT \leftarrow \emptyset$;
\State $\Omega \leftarrow \Omega \cup (\bx^{t=0}, \bb)$; $\quad$ Stack $ST \leftarrow \emptyset; \quad$ {\tt Push}($ST,\, \bx^{t=0}$);
\While{$ST \ne \emptyset$}
\State $StateGenerated \leftarrow ${\tt FALSE};  $\bx_i \leftarrow$ {\tt Pop} $(ST)$;
\For{$k=1$ to $m$}  \Comment{ {\sf There are $m$ reactions.}}
		\For{$j=1$ to $w$}  \Comment{{\sf Look up which iBD reaction $R_k$ belongs to.}}
		   \If {$R_k \in \bH^{iBD}_j$}
				\State Break;
			\EndIf
		\EndFor
\If{Reaction $R_k$ can occur in state $\bx_i$}
    \If{ $R_k$ is a synthesis reaction generating $g_k$ new copies of $X_i$}
       \If{$b_j \ge g_k$}  \Comment{{\sf Check if buffer tokens are sufficient for synthesis reaction.}}
          \State Generate state $\bx_j$ that is reached via reaction $R_k$ from $\bx_i$;
					\State $b_j \leftarrow b_j - g_k$;  $StateGenerated \leftarrow ${\tt TRUE};
       \EndIf
    \Else
       \If{ $R_k$ is a degradation and breaks down $d_k$ copies of $X_i$}
				\State $b_j \leftarrow b_j + d_k$;
       \EndIf
				\State Generate state $\bx_j$ that is reached via reaction $R_k$ from $\bx_i$;
				\State $StateGenerated \leftarrow ${\tt TRUE};
    \EndIf
    \If{($StateGenerated =$ {\tt TRUE})}
		   \State Combined state $\hat{\bx}_j = (\bx_j, \bb)$;
			 \If{($\hat{\bx}_j \notin \Omega$)}
         \State $\Omega \leftarrow \Omega \cup \hat{\bx}_j$; $\quad$ {\tt Push}($ST,\, \bx_j$); $\quad$ $\bT \leftarrow \bT \cup t_{\bx_i,\,\bx_j}$;
\Comment{{\sf $t_{\bx_i,\,\bx_j}$ records this transition.}}
         \State $A(\bx_i,\,\bx_j) \leftarrow$ {\tt ReactionRate}($\bx_i, \, \bx_j, \, R_k$)
			 \EndIf
    \EndIf
\EndIf
\EndFor
\EndWhile
\State Output $\Omega,\,\bT$ and $\bA = \{A(\bx_i, \, \bx_j) \}$.
\end{algorithmic}
\end{algorithm}

\subsection{Controlling Truncation Errors}

When one or more buffer queues are exhausted, no new states can be enumerated 
and synthesis reaction(s) cannot proceed, resulting in errors due to state truncation. 
Below we describe a theoretical framework for analyzing effects of truncating state space.  
We give an error estimate such that the truncation error is bounded from above, 
namely, the actual error will be smaller than the estimated error bound. 
Furthermore, we give an estimate on the minimal size of buffer required so the 
truncation error is within a specified tolerance. It is important to note that 
this error estimate is obtained \textit{a priori} without computing costly 
trial solutions. Detailed proofs for all statements of facts can be found in 
Ref~\cite{Cao2016a}.

\subsubsection{Overall description}

We briefly outline our approach to construct error bounds.  
  We first define truncation error
  $\Err^{(\mathcal{B})}$ when a finite state space
  $\Omega^{(\mathcal{B})}$ instead of a full infinite state space
  $\Omega^{(\mathcal{I})}$ 
is used to solve the dCME.  We then introduce the concept of boundary
states $\partial \Omega^{(\mathcal{B})}$ of the state space 
$\Omega^{(\mathcal{B})}$ and boundary states $\partial
\Omega^{(B_j)}$ of the individual $j$-th iBD, as well as the
corresponding steady state probabilities 
$\pi^{(\mathcal{B})}_{\partial, \, \mathcal{B}}$ and 
$\pi^{(\mathcal{B})}_{\partial, \, B_j}$. 
We show that the steady state probability
$\pi^{(\mathcal{B})}_{\partial, \, \mathcal{B}}$ 
provides an upper-bound for the truncation error
$\Err^{(\mathcal{B})}$.  This is established by first examining the
truncation error $\Err^{(\mathcal{B}_j)}$ when only one iBD is truncated. 
The techniques used 
include: (1) permuting the transition rate matrix $\bA$ and
lumping microstates into groups with the same number of net synthesized molecules
or buffer usage of the iBD; (2) constructing a quotient matrix $\bB$
on the lumped groups from the permuted matrix $\bA$ and its associated steady state probability distribution. 
We then show that the truncation error $\Err^{(\mathcal{B}_j)}$ can be asymptotically bounded 
by $\pi^{(\mathcal{B})}_{\partial, \, B_j}$ computed from 
the quotient matrix $\bB$.  We further analyze the 
asymptotic behavior of the boundary probability
$\pi^{(\mathcal{B})}_{\partial, \, \mathcal{B}_j}$, 
and show that this probability increases when additional iBDs are truncated. 
The upper and lower bounds for truncation error 
are then obtained based on known facts
of stochastic ordering.  We then generalize our results on error
bounds to truncation errors when two, three, and all buffer queues are
of finite capacity.

It is useful to also examine an intuitive picture of the
probability landscape governed by a dCME.  Starting from an initial
condition, the probability mass flows following 
a diffusion process dictated by the dynamics of the reaction network.  
At any given time $t$, the front of the probability flow
traces out a boundary $\partial_t$, which expands to a new boundary
$\partial_{t+\Delta t}$ at a subsequent time.  Given long enough
time, the probability distribution will reach a steady state. Since the
probability flows across the boundaries, we can compare the
difference in the probability mass between the boundary surfaces of 
$\partial_t$ and $\partial_{t+\Delta t}$ to infer how much total
probability mass has fluxed out of the finite volume of the state
space through its boundary.  Our asymptotic analysis is aided by
decomposing the overall probability flow into several different
fluxes, each governed by a different independent Birth-Death (iBD)
component.

\subsubsection{Truncation Error Decreases with Increasing Buffer Capacity}

Denote the true probability landscape governed by a dCME over $\Omega^{(\mathcal{I})}$
without truncation as $\bp^{(\mathcal{I})}(t)$.  When the state space is
truncated to $\Omega^{(\mathcal{B})} \subset
\Omega^{(\mathcal{I})} $ using a buffer bank $\mathcal{B}$, the
deviation of the summed probability mass of $\bp^{(\mathcal{I})}(t)$ over
$\Omega^{(\mathcal{B})}$ from $1$ gives the truncation error:
\begin{equation}
\Err^{(\mathcal{B})}(t)
= 1 - \sum_{\bx \in \Omega^{(\mathcal{B})}} p^{(\mathcal{I})}(\bx,\, t)
= 
\sum_{\bx \in \Omega^{(\mathcal{I})}, \, \bx \notin \Omega^{(\mathcal{B})}} p^{(\mathcal{I})}(\bx,\, t). 
\label{eqn:trueerr}
\end{equation}
As the overall buffer size of $\mathcal{B}$ increases, $\Err^{(\mathcal{B})}(t)$ decreases.  
Using $\Err^{(\mathcal{B})}$ to denote the steady state error, we have: 
$$\Err^{(\mathcal{B})} \equiv \Err^{(\mathcal{B})}(t=\infty) = 
1 - \sum_{\bx \in \Omega^{(\mathcal{B})}} \pi^{(\mathcal{I})}(\bx,\, t).$$ 
In addition, we consider error resulting from truncating only the
$j$-th buffer queue to the state space $\Omega^{(\mathcal{I}_j)} \subset \Omega^{(\mathcal{I})}$ 
using buffer bank $\mathcal{I}_j = (\infty, \cdots, \infty, B_j, \infty, \cdots, \infty)$. 
Similarly, we have 
$$
\Err^{(\mathcal{I}_j)} = 1 - \sum_{\bx \in \Omega^{(\mathcal{I}_j)}} \pi^{(\mathcal{I})}(\bx,\, t).
$$
\begin{fact}
\label{lm:1}
For any two truncated state spaces $\Omega^{(\mathcal{B}')}$ and $\Omega^{(\mathcal{B}'')}$, 
we have $\Err^{(\mathcal{B}')} (t) \geq \Err^{(\mathcal{B}'')} (t)$
if $\mathcal{B}' \leq \mathcal{B}''$ component-wise.
\end{fact}
Note that, if $\mathcal{B}' \leq \mathcal{B}'' \leq \mathcal{I}$, 
then $\Err^{(\mathcal{B}')} \geq \Err^{(\mathcal{B}'')} \geq \Err^{(\mathcal{I})} \equiv 0$.

\subsubsection{Probabilities of Boundary States of Finite State Space and Increments of Truncation Error}

It is difficult to compute the exact truncation error
$\Err^{(\mathcal{B})}(t)$, as it requires $\bp^{(\mathcal{I})}(t)$ to be known. 
However, only the computed probability landscape $\bp^{(\mathcal{B})}(t)$ 
using a finite state space $\Omega^{(\mathcal{B})}$ is known. 

We now consider the steady state probabilities 
$\bpi^{(\mathcal{I})} \equiv \bp^{(\mathcal{I})}(\infty)$, 
$\bpi^{(\mathcal{I}_j)} \equiv \bp^{(\mathcal{I}_j)}(\infty)$, and 
$\bpi^{(\mathcal{B})} \equiv \bp^{(\mathcal{B})}(\infty)$. 
We further consider the boundary states $\partial
\Omega^{(\mathcal{B})}$ of $\Omega^{(\mathcal{B})}$, and
show that $\pi^{(\mathcal{B})} (\partial \Omega^{(\mathcal{B})})$
can be used as a surrogate for 
estimating the steady state error $\Err^{(\mathcal{B})}$ and
for assessing the convergence behavior of $\Err^{(\mathcal{B})}$.

\paragraph{Boundary of state space $\Omega^{(\mathcal{B})}$ and 
boundary states of the $j$-th iBD}
The boundary states $\partial \Omega^{(\mathcal{B})}$ of
$\Omega^{(\mathcal{B})}$ are those states with at least one depleted
buffer queue:
\begin{equation}
\partial \Omega^{(\mathcal{B})}
= \{ \bx | b_i \mbox{ of } \bb(\bx)= 0,\, i \in (1, \cdots, w) \}.
\label{eqn:Ball}
\end{equation}
{\it i.e.}, there are exactly $B_i$ net synthesized molecules for at
least one of the iBDs.
The time-evolving and steady state probability mass of
$p^{(\mathcal{I})}(t)$ over $\partial \Omega^{(\mathcal{B})}$ is
denoted as $p^{(\mathcal{I})}_{\partial, \mathcal{B}}(t)$ and 
$\pi^{(\mathcal{I})}_{\partial, \mathcal{B}}$, respectively. 

In addition to boundary states of the full buffer bank, we also consider 
boundary states of individual buffer queues. 
We consider a subset of the boundary states $\partial
\Omega^{(\mathcal{B})}_{(B_j)} \in \partial \Omega^{(\mathcal{B})}$ that are 
associated with the $j$-th iBD component: 
\begin{equation}
\partial \Omega^{(\mathcal{B})}_{(B_j)} \equiv \{ \bx | b_j \mbox{ of } \bb(\bx)= 0\}. 
\label{eqn:BIBD}
\end{equation}

\paragraph{Probabilities of boundary states of $\partial
\Omega^{(\mathcal{B})}$ and $\partial \Omega^{(\mathcal{B})}_{(B_j)}$}
We name the summation of the true steady state probability $\pi^{(\mathcal{I})} (\bx)$ 
over all boundary states $\partial \Omega^{(\mathcal{B})}$
the {\it true total boundary probability} 
$\pi^{(\mathcal{I})}_{\partial, \, \mathcal{B}} $: 
$$\pi^{(\mathcal{I})}_{\partial, \, \mathcal{B}} 
\equiv \sum_{\bx \in \partial \Omega^{(\mathcal{B})}}
\pi^{(\mathcal{I})}(\bx).$$ 
The summation of the computed probability $\pi^{(\mathcal{B})} (\bx)$
using the truncated state space $\Omega^{(\mathcal{B})}$ over the same boundary
states $\partial \Omega^{(\mathcal{B})}$ 
is the {\it computed total boundary probability} 
$\pi^{(\mathcal{B})}_{\partial,\, \mathcal{B}}$:
$$\pi^{(\mathcal{B})}_{\partial,\, \mathcal{B}} \equiv
\sum_{\bx \in \partial \Omega^{(\mathcal{B})}} \pi^{(\mathcal{B})}(\bx).
$$

Similarly, we call the summation of the true probability
$\pi^{(\mathcal{I})} (\bx)$
associated with the boundary states of the $j$-th iBD 
the {\it true boundary probability of $j$-th iBD} 
$\pi^{(\mathcal{I})}_{\partial, \, B_j}$: 
$$\pi^{(\mathcal{I})}_{\partial, \, B_j} 
\equiv \sum_{\bx \in \partial \Omega^{(\mathcal{B})}_{(B_j)}}
\pi^{(\mathcal{I})}(\bx).$$ 
The summation of the computed probability $\pi^{(\mathcal{B})} (\bx)$ 
using the truncated state space $\Omega^{(\mathcal{B})}$ over the same boundary
states associated with the $j$-th iBD in $\partial \Omega^{(\mathcal{B})}_{(B_j)}$
is the {\it computed boundary probability of the $j$-th iBD} 
$\pi^{(\mathcal{B})}_{\partial,\, B_j}$:
$$\pi^{(\mathcal{B})}_{\partial,\, B_j} \equiv
\sum_{\bx \in \partial \Omega^{(\mathcal{B})}_{(B_j)}} \pi^{(\mathcal{B})}(\bx).
$$ 

It is also useful to examine the total boundary probability 
$\pi^{(\mathcal{I}_j)}_{\partial,\, B_j}$ of the $j$-th iBD 
on the state space $\Omega^{(\mathcal{I}_j)}$: 
$$
\pi^{(\mathcal{I}_j)}_{\partial, B_j} \equiv
\sum_{\bx \in \partial \Omega^{(\mathcal{I}_j)}_{B_j}}
\pi^{(\mathcal{I}_j)}(\bx). 
$$
Note that when $B_j$ goes to infinity, the probability 
$\pi^{(\mathcal{I}_j)}_{\partial, B_j}$ approaches 
$\pi^{(\mathcal{I})}_{\partial, B_j}$.

\begin{figure}[ht]
\centering{\includegraphics[scale=0.3]{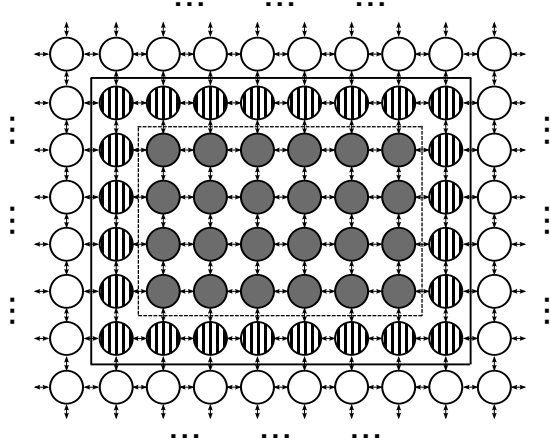}}
\caption{ Probability of boundary states and truncation errors.  The
  gray states in the center box of dashed lines form the state space
  $\Omega^{(\mathcal{B}-\mathbbm{1})}$, with the rest as states 
  truncated from $\Omega^{(\mathcal{B}-\mathbbm{1})}$. The stripe-filled
  states are the newly added states when the buffer capacity is
  increased from $\mathcal{B}-\mathbbm{1}$ to $\mathcal{B}$. These new
	states and those gray states, both enclosed in the box in solid lines, 
  form the state space $\Omega^{(\mathcal{B})}$. The summed true 
	probability mass over the white states outside the solid-lined box 
  is the error $\Err^{(\mathcal{B})}$ of the truncated state
  space $\Omega^{(\mathcal{B})}$.  The summed true probability mass over 
  all states outside of the dashed line
  box is the error $\Err^{(\mathcal{B}-\mathbbm{1})}$.  The summed
  true probability over stripe-filled states $\pi^{(\mathcal{I})}_{\partial, \, \mathcal{B}}$ 
	is the incremental error $\Delta
  \Err^{(\mathcal{B})}(t)$ when the buffer capacity of all buffer
  queues is increased by $1$ from $\mathcal{B}-\mathbbm{1}$.  We have:
  $\pi^{(\mathcal{I})}_{\partial, \, \mathcal{B}} = 
	\Delta \Err^{(\mathcal{B})} (t) = |\Err^{(\mathcal{B})}(t) -
  \Err^{(\mathcal{B}-\mathbbm{1})}(t)|$.  }
\label{fig:errdiff}
\end{figure}

\paragraph{Incremental truncation errors}
The state space $\Omega^{(\mathcal{B})}$ is obtained from enumeration by adding
$1$ to the capacity of every buffer queue used to obtain the 
state space $\Omega^{(\mathcal{B}-\mathbbm{1})}$. 
Let $\mathbbm{1} = (1, 1, \cdots, 1) \in \mathbb{Z}^w$. 
The boundary of $\Omega^{(\mathcal{B})}$ can then be written as: 
$\partial \Omega^{(\mathcal{B})} = \Omega^{(\mathcal{B})} - \Omega^{(\mathcal{B}-\mathbbm{1})}$. 
It is obvious that the true total boundary probability 
$\pi^{(\mathcal{I})}_{\partial, \, \mathcal{B}}$ is the
increment of the truncation error between
$\Omega^{(\mathcal{B}-\mathbbm{1})}$ and
$\Omega^{(\mathcal{B})}$:
\begin{equation}
\pi^{(\mathcal{I})}_{\partial, \, \mathcal{B}} 
       = \Delta \Err^{(\mathcal{B})} = \Err^{(\mathcal{B}-\mathbbm{1})}
                                           - \Err^{(\mathcal{B})}.
\label{eqn:bdryall}
\end{equation}
Fig.~\ref{fig:errdiff} gives an illustration.

Let $\bee_j = (0, \cdots, 0, \, 1, \, 0, \cdots, 0) \in \mathbb{Z}_{\geq 0}^w$ be 
an elementary vector with only the $j$-th element as $1$ and all others $0$. 
The boundary states of the $j$-th iBD is given by: 
$\partial \Omega^{(\mathcal{B})}_{(B_j)} = \Omega^{(\mathcal{B})} - \Omega^{(\mathcal{B}-\bee_j)}$. 
Analogous to Eqn.~(\ref{eqn:bdryall}), the boundary 
probability $\pi^{(\mathcal{I})}_{\partial,\, B_j}$ is therefore the increment of 
the truncation error between $\Omega^{(\mathcal{B}-\bee_j)}$ and
$\Omega^{(\mathcal{B})}$:
\begin{equation}
\pi^{(\mathcal{I})}_{\partial,\,B_j} = \Delta \Err^{(B_j)} 
    = \Err^{(\mathcal{B}  -\bee_j)} - \Err^{(\mathcal{B})},
\label{eqn:bdryj}
\end{equation}
as the only difference between $\Omega^{(\mathcal{B}-\bee_j)}$ and $\Omega^{(\mathcal{B})}$ 
are those states containing exactly $B_j$ net synthesized molecules in the $j$-th iBD, 
namely, the states with the $j$-th buffer queue depleted.

\paragraph{Total true error is no greater than summed errors over all iBDs}
Overall, we have $\partial \Omega^{(\mathcal{B})}
 = \bigcup_{j=1}^w \partial \Omega^{(\mathcal{B})}_{(B_j)}$.
As some boundary states may have multiple depleted
buffer queues, it is possible 
$
\partial \Omega^{(\mathcal{B})}_{(B_i)} \cap \partial \Omega^{(\mathcal{B})}_{(B_j)} \neq \emptyset, \, \cdots, \,
\bigcap_{i=1}^w \partial \Omega^{(\mathcal{B})}_{(B_i)}  \neq \emptyset.
$
Therefore, the actual total boundary probability 
$\pi^{(\mathcal{I})}_{\partial,\, \mathcal{B}}$ 
is smaller than or equal to the
summation of individual $\pi^{(\mathcal{I})}_{\partial,\,B_j}$: 
\begin{equation}
\pi^{(\mathcal{I})}_{\partial,\, \mathcal{B}}
\leq
\sum_{j=1}^w
\pi^{(\mathcal{I})}(\partial \Omega^{(\mathcal{B})}_{(B_j)}) \equiv
\sum_{j=1}^w
\pi^{(\mathcal{I})}_{\partial,\, B_j}.
\label{eqn:toterr}
\end{equation}

As the state space $\Omega^{(\mathcal{B})} = \bigcap_{i=1}^w \Omega^{(\mathcal{I}_j)}$ 
and the buffer capacity of the $j$-th iBD in $\Omega^{(\mathcal{I}_j)}$ 
is the same as that in $\Omega^{(\mathcal{B})}$, 
we have that the total true error of the state space $\Omega^{(\mathcal{B})}$ 
is bounded by the summation of true errors from individually truncated state spaces
$\Omega^{(\mathcal{I}_j)}$:
\begin{equation}
\Err^{(\mathcal{B})} \leq \sum_{j=1}^w \Err^{(\mathcal{I}_j)}. 
\label{eqn:bd1}
\end{equation}

\begin{figure}[ht]
\centering{\includegraphics[scale=0.4]{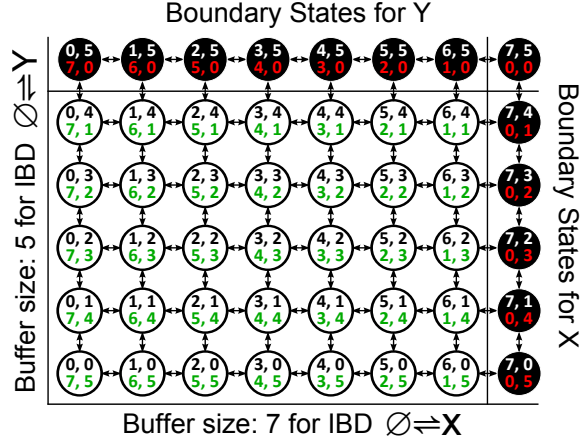}}
\caption{ An illustration of the enumerated state space and the 
  boundary states of a simple network with two reactions $\emptyset
  \rightleftharpoons X$ and $\emptyset \rightleftharpoons Y$.  There
  are two iBDs in this network, with two buffer queues $B_1$ and $B_2$
  of size 5 and 7 assigned to the first and second iBD, respectively.  
	Each circle represents an
  enumerated state. Filled circles are boundary states, in which at
  least one of the two buffer queues is depleted.  There are four
  integers inside each circle.  The two at the top are copy numbers
  $x$ and $y$ of molecular species $X$ and $Y$, namely, $\bx = (x,
  y)$.  The two at the bottom are the remaining numbers $b_1$ and
  $b_2$ of tokens in the buffer queues $B_1$ and $B_2$, namely, $\bb =
  (b_1, b_2)$.  }
\label{fig:fbsss}
\end{figure}

\paragraph{An example}
Fig.~\ref{fig:fbsss} shows an example of the enumerated state space
using Algorithm~\ref{alg:fbsmb} for a simple network with reversible
reactions $\emptyset \rightleftharpoons X$ and 
$\emptyset \rightleftharpoons Y$.  
The network is partitioned into two iBD components, one for $\emptyset
\rightleftharpoons X$ and another for $\emptyset \rightleftharpoons
Y$.  A buffer bank $\mathcal{B} = (B_1, B_2)$ with two buffer queues
is assigned to the network, with the size vector $(B_1,\, B_2) =(7,\,
5)$.  A synthesis reaction is halted once its buffer queue is
depleted, resulting in truncation error.  Boundary states, in which at
least one of the two buffer queues is depleted, are shown as filled black
circles, with states of the buffer queues shown in red numbers. The
union of all black filled circles in Fig.~\ref{fig:fbsss} form the
boundary $\partial \Omega^{(\mathcal{B})}$ of the state space.
The boundary states associated with the buffer queue corresponding to
the iBD of reaction 
$\emptyset \rightleftharpoons X$  
are:
\begin{eqnarray*}
\partial \Omega^{(\mathcal{B})}_{(B_1)} =& \{ (x=7, y=5), (x=7, y=4), (x=7, y=3), \\ 
&(x=7, y=2), (x=7, y=1), (x=7, y=0) \},
\end{eqnarray*}
in which the buffer queue $B_1$ is depleted.
The boundary states associated with the iBD of reaction 
$\emptyset \rightleftharpoons Y$  
are:
\begin{eqnarray*}
\partial \Omega^{(\mathcal{B})}_{(B_2)} =& \{ (x=7, y=5), (x=6, y=5), (x=5, y=5), (x=4, y=5), \\
&(x=3, y=5), (x=2, y=5), (x=1, y=5), (x=0, y=5) \},
\end{eqnarray*}
in which the buffer queue $B_2$ is depleted (Fig.~\ref{fig:fbsss}).  We have
$\partial \Omega^{(\mathcal{B})} = \partial \Omega^{(\mathcal{B})}_{(B_1)}
\cup \partial \Omega^{(\mathcal{B})}_{(B_2)}$.  We also observe that $\partial
\Omega^{(\mathcal{B})}_{(B_1)} \cap \partial \Omega^{(\mathcal{B})}_{(B_2)} = (x=7, y=5)$ is
none-empty.  Those states that are not on the boundary are shown as unfilled circles.

\subsubsection{Bounding Errors Due to A Truncated Buffer Queue}
\label{sec:convergence}

We show how to construct an error bound after truncating an individual buffer queue. 
We first examine the steady state boundary probability 
$\pi^{(\mathcal{I}_j)}_{\partial, B_j}$,
For ease of discussion, we use $N$ instead of $B_j$ to denote the buffer
capacity of the $j$-th iBD, and use $\pi_N^{(\mathcal{I}_j)} \equiv \pi^{(\mathcal{I}_j)}_{\partial, \, B_j}$ to denote the boundary probability of $\Omega^{(\mathcal{I}_j)}$. 
The true error $\Err^{(\mathcal{I}_j)}$ associated with buffer bank 
$\mathcal{I}_j = (\infty, \cdots, \infty, B_j = N, \infty, \cdots, \infty)$ 
for the steady state is unknown, as it 
requires knowledge of $\pi^{(\mathcal{I})}(\bx)$ for all $\bx \in
\Omega^{(\mathcal{I})}$.  Here, we show that $\Err^{(\mathcal{I}_j)}$ 
converges to the true boundary probability $\pi^{(\mathcal{I}_j)}_N$ 
asymptotically as the size of the buffer queue $N$ increases. 
Specifically, if the size of the buffer queue is sufficiently large, 
$\Err^{(\mathcal{I}_j)}$ is bounded by 
$\pi^{(\mathcal{I}_j)}_N$ up to a constant factor.  As $N$ further increases,
$\Err^{(\mathcal{I}_j)}$ converges to $\pi_N^{(\mathcal{I}_j)}$.

\paragraph{Aggregating states by buffer queue usage}
To show how boundary probability $\pi_N^{(\mathcal{I}_j)}$ can be used to construct truncation
error bound, we first aggregate states in the original state space $\Omega^{(\mathcal{I}_j)}$ 
into $N+1$ non-intersecting subsets according to the net number of tokens in
use from buffer $B_j$:  $\Omega^{(\mathcal{I}_j)}
\equiv \{ \mathcal{G}_0, \mathcal{G}_1, \cdots, \mathcal{G}_N \}$.
Here states in each aggregated subset
$\mathcal{G}_s \subseteq \Omega^{(\mathcal{I}_j)}$, 
$s = 1, \cdots, N$,
all have the same $s$ number of buffer tokens
spent from buffer queue $B_j$, or equivalently, $(N-s)$ tokens unused in buffer
$B_j$. Note that each $\mathcal{G}_s$ can be of infinite size if the capacity of any other
buffer queues are infinite. Conceptually disregard the practical issue of time
complexity for now, the states in the state space $\Omega$ can be sorted according to
the buffer token from buffer queue $B_j$ in use.  This can be done using any
sorting algorithm, such as the bucket sort algorithm with $N+1$ buckets, with
each bucket $\mathcal{G}_s$ contain only states with exactly $s$ buffer tokens spent.

With this partition, we can construct a transition rate matrix $\tilde{\bA}$ from the
sorted state space $\Omega^{(\mathcal{I}_j)}$. The new transition rate matrix $\tilde{\bA}$ is a permutation
of the original dCME matrix $\bA$ Eqn.~(\ref{eqn:dcme2}):
\begin{equation}
\tilde{\bA} = \left( {\begin{array}{*{20}c}
   {\bA_{0,0}} & {\bA_{0,1}} & \cdots & {\bA_{0,N}}  \\
   {\bA_{1,0}} & {\bA_{1,1}} & \cdots & {\bA_{1,N}}  \\
   \cdots & \cdots & \cdots & \cdots \\
   {\bA_{N,0}} & {\bA_{N,1}} & \cdots & {\bA_{N,N}}  \\
\end{array}} \right),
\label{eqn:Aaggreg1}
\end{equation}
where each block sub-matrix $\bA_{i,\,j}$ includes all transitions
from states in group $\mathcal{G}_j$ to states in group $\mathcal{G}_i$,
and can be defined as: $\bA_{i,j} = \{\ba_{m,n}\}_{||\mathcal{G}_i||\times ||\mathcal{G}_j||}$,
and each entry $\ba_{m,n}$ in $\bA_{i,j}$ is the transition rate from a 
state $\bx_n \in \mathcal{G}_j$ to a state $\bx_m \in \mathcal{G}_i$. 

Although in principle one can obtain the sorted state space partition 
$\Omega^{(\mathcal{I}_j)} \equiv \{\mathcal{G}_0, \mathcal{G}_1, \cdots, \mathcal{G}_N \}$ 
and the permuted transition rate matrix $\tilde{\bA}$, there is no need
to do so in practice. The construction of $\Omega^{(\mathcal{I}_j)}$ and $\tilde{\bA}$ only serves the
purpose for proving lemmas and theorems. Specifically, we only need to know
that conceptually the original state space can be sorted and partitioned, and a
permuted transition rate matrix $\tilde{\bA}$ can be constructed from the sorted state
space according to the aggregation.

\begin{figure}[ht]
\centering{\includegraphics[scale=0.4]{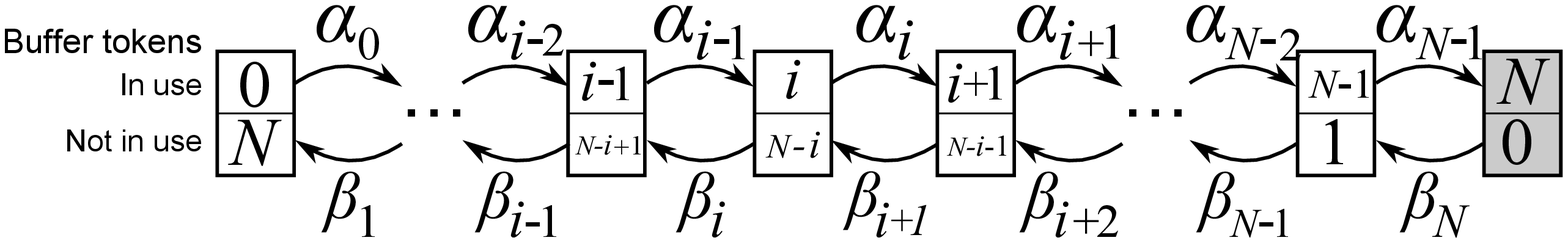}}
\caption{ The birth-death system associated with the aggregated rate
  matrix $\bB$. Each box represents an aggregated state consisting of
  all microstates with the same number of buffer tokens in use. The
  top half of each box lists the number of buffer tokens in use, and
  the bottom half lists the number of remaining free buffer tokens in
  the buffer queue. The gray box contains the boundary states. The
  total number of spent and free tokens sums to the buffer capacity
  $N$. These aggregated states are connected by aggregated birth and
  death reactions, with apparent synthesis rates $\alpha_i$ and
  degradation rates $\beta_{i+1}$ (see Fact~\ref{lm:22}).  }
\label{fig:bfbd}
\end{figure}

Assume the partition and the steady state probability distribution over the state space 
$\Omega^{(\mathcal{I}_j)}$ are known, we can construct an aggregated synthesis
rate $\alpha^{(N)}_i$ for the group $\mathcal{G}_i$ and an aggregated
degradation rate $\beta^{(N)}_{i+1}$ for the group $\mathcal{G}_{i+1}$
at the steady state as two constants (Fig~\ref{fig:bfbd}):
\begin{equation}
\alpha^{(N)}_i \equiv \left( \mathbbm{1}^T \bA_{i+1,i} \right) \cdot
\frac{\bpi^{(\mathcal{I}_j)}(\mathcal{G}_{i})}{\mathbbm{1}^T \bpi^{(\mathcal{I}_j)}(\mathcal{G}_{i})} \quad \text{and} \quad
\beta^{(N)}_{i+1} \equiv \left( \mathbbm{1}^T \bA_{i,i+1} \right)
\cdot \frac{\bpi^{(\mathcal{I}_j)}(\mathcal{G}_{i+1})}{\mathbbm{1}^T \bpi^{(\mathcal{I}_j)}(\mathcal{G}_{i+1})},
\label{eqn:abdef}
\end{equation}
where vector $\bpi^{(\mathcal{I}_j)}(\mathcal{G}_{i})$ and $\bpi^{(\mathcal{I}_j)}(\mathcal{G}_{i+1})$
are steady state probability vectors over the permuted microstates in the lumped group
$\mathcal{G}_{i}$ and $\mathcal{G}_{i+1}$, respectively. 
Row vectors $\mathbbm{1}^T \bA_{i+1,i}$ and $\mathbbm{1}^T \bA_{i,i+1}$ 
are summed columns of block sub-matrices $\bA_{i+1,i}$ and $\bA_{i,i+1}$, respectively. 

Similarly, if the buffer queue $B_j$ has infinite capacity, we have
\begin{equation}
\alpha^{(\infty)}_i \equiv \left( \mathbbm{1}^T \bA_{i+1,i} \right) \cdot
\frac{\bpi^{(\mathcal{I})}(\mathcal{G}_{i})}{\mathbbm{1}^T \bpi^{(\mathcal{I})}(\mathcal{G}_{i})} \quad \text{and} \quad
\beta^{(\infty)}_{i+1} \equiv \left( \mathbbm{1}^T \bA_{i,i+1} \right)
\cdot \frac{\bpi^{(\mathcal{I})}(\mathcal{G}_{i+1})}{\mathbbm{1}^T \bpi^{(\mathcal{I})}(\mathcal{G}_{i+1})}.
\label{eqn:abdefinf}
\end{equation}

We can then construct an aggregated transition rate matrix $\bB$ from 
the permuted matrix $\tilde{\bA}$ based on Fact~\ref{lm:22}:
\begin{fact}
\label{lm:22}
Consider a homogeneous continuous-time Markov process with the
infinitesimal generator rate matrix $\bA$ on the infinite state space
$\Omega^{(\mathcal{I}_j)}$ equipped with buffer queues $\mathcal{I}_j = (\infty, \cdots, B_j, \cdots, \infty)$ 
with a finite buffer capacity $B_j = N$ for the $j$-th iBD, and infinite 
capacities for all other iBDs.  Denote its steady state probability
distribution as $\bpi^{(\mathcal{I}_j)} \equiv \bpi (\Omega^{(\mathcal{I}_j)})$.
An aggregated continuous-time Markov process with a finite size rate matrix
$\bB_{(N+1)\times(N+1)}$ can be constructed on the partition
$\tilde{\Omega}^{(\mathcal{I}_j)}_{B_j} = \{ \mathcal{G}_0,\,
\mathcal{G}_1,\, \cdots,\, \mathcal{G}_N \}$ with respect to the buffer
queue $B_j$.  Denote $\tilde{\pi}^{(N)}_s \equiv \tilde{\pi}(\mathcal{G}_s) = 
\sum_{\bx \in \mathcal{G}_s} \pi^{(\mathcal{I}_j)}(\bx)$. 
The steady state probability vector 
$\tilde{\bpi}(\tilde{\Omega}^{(\mathcal{I}_j)}_{B_j}) = 
(\tilde{\pi}^{(N)}_0, \cdots, \tilde{\pi}^{(N)}_N) = 
( \tilde{\pi}(\mathcal{G}_0), \cdots, \tilde{\pi}(\mathcal{G}_N) )$ 
of the aggregated Markov process gives the
same steady state probability distribution for the partitioned groups
$\{\mathcal{G}_s\}$ as that given by the original matrix $\bA$,
for all $s = 0, 1, \cdots, N$.
Furthermore, the $(N+1) \times (N+1)$ transition rate matrix $\bB$ can
be constructed as:
\begin{equation}
\bB^{(N)} = \left( {\begin{array}{c}
   \boldsymbol\alpha^{(N)}, \boldsymbol\gamma^{(N)}, \boldsymbol\beta^{(N)}
\end{array}} \right),
\label{eqn:bdmatapp}
\end{equation}
with the lower off-diagonal vector $$\boldsymbol\alpha^{(N)} = (\alpha^{(N)}_i), \, i = 0,\cdots,N-1.$$
the upper off-diagonal vector $$\boldsymbol\beta^{(N)} = (\beta^{(N)}_{i}), \, i = 1,\cdots,N.$$
and the diagonal vector $$\boldsymbol\gamma^{(N)} = (\gamma^{(N)}_i) = (-\alpha^{(N)}_i - \beta^{(N)}_{i}), \, i = 0,\cdots,N.$$
It is equivalent to transforming the transition rate matrix
$\tilde{\bA}$ in Eqn.~(\ref{eqn:Aaggreg1}) to $\bB$ by substituting each
block sub-matrix $\bA_{i+1,\, i}$ of synthesis reactions with the
corresponding aggregated synthesis rate $\alpha^{(N)}_i$, and each
block $\bA_{i,\, i+1}$ of degradation reactions with the aggregated
degradation rate $\beta^{(N)}_{i+1}$ in Eqn.~(\ref{eqn:abdef}),
respectively.
\end{fact}

Detailed proof for Fact~\ref{lm:22} can be found in Lemma 1 in Ref~\cite{Cao2016a}.

\paragraph{Computing steady state boundary probabilities}
Following Refs~\cite{Vellela2007,Taylor1998} on birth-death 
processes (Fig.~\ref{fig:bfbd}), the analytic solution for the steady state 
$\tilde{\pi}_i^{(N)}$ and $\tilde{\pi}_0^{(N)}$ can be written as:
\begin{equation}
\tilde{\pi}_i^{(N)} = 
  \prod\limits_{k = 0}^{i-1}
    \frac{\alpha^{(N)}_{k}}{\beta^{(N)}_{k+1}} \tilde{\pi}_0^{(N)} 
\label{eqn:pini}
\end{equation}
and
\begin{equation}
\tilde{\pi}_0^{(N)} = 
\frac{1}
{1 + {\sum\limits_{j = 1}^{N} {\prod\limits_{k = 0}^{j-1}
     {\frac{\alpha^{(N)}_{k}}{\beta^{(N)}_{k+1}}}}
    }
}. 
\label{eqn:pin0}
\end{equation}
The boundary probability $\tilde{\pi}_{N}^{(N)}$ is then:
\begin{equation}
\tilde{\pi}_N^{(N)} \equiv \pi^{(\mathcal{I}_j)}_{\partial, \, B_j} = \frac{\prod\limits_{k = 0}^{N-1}
    \frac{\alpha^{(N)}_{k}}{\beta^{(N)}_{k+1}}}
{1 + {\sum\limits_{j = 1}^{N} {\prod\limits_{k = 0}^{j-1}
     {\frac{\alpha^{(N)}_{k}}{\beta^{(N)}_{k+1}}}} 
    }
}. 
\label{eqn:pinn}
\end{equation}
If we have infinite buffer capacity for the $j$-th iBD, we will have the true
probability mass over the same fixed set of states in $\mathcal{G}_N$ as
\begin{equation}
\tilde{\pi}_N^{(\infty)} \equiv \tilde{\pi}_N^{(\mathcal{I})} \equiv \tilde{\pi}_{\partial, \, B_j}^{(\mathcal{I})} = \frac{\prod\limits_{k = 0}^{N-1}
    \frac{\alpha^{(\infty)}_{k}}{\beta^{(\infty)}_{k+1}}}
{1 + {\sum\limits_{j = 1}^{\infty} {\prod\limits_{k = 0}^{j-1}
     {\frac{\alpha^{(\infty)}_{k}}{\beta^{(\infty)}_{k+1}}}}
    }
}. 
\label{eqn:piinfn}
\end{equation}


\paragraph{Boundary probability as error bound of state truncation}

According to Fact~\ref{lm:1}, the error $\Err^{(\mathcal{I}_j)}$ converges
to $0$ as the buffer capacity $B_j = N$ increases to infinity.  
For a truncated state space, the series of the true boundary probabilities
$\{{\tilde{\pi}^{(\mathcal{I})}_{N}} | N = 1, 2, \cdots, \}$ (Eqn.~(\ref{eqn:piinfn})) also converges to
$0$, as the sequence of its partial sums converges to $1$. That is,
the $N$-th member $\ {\tilde{\pi}^{(\mathcal{I})}_{N}}$ of this series converges to
$0$ while the residual sum of this series 
$\Err^{(\mathcal{I}_j)} \equiv \sum\limits_{i = N + 1}^\infty {\tilde{\pi} _i^{(\infty )}}$ 
also converges to $0$.

We now examine the convergence behavior of the truncation error $\Err^{(\mathcal{I}_j)}_{(N)}$ 
and the true boundary probability $\tilde{\pi}^{(\infty)}_N$. 
\begin{fact}
For a truncated state space associated with a buffer
bank $\mathcal{I}_j$, if the buffer capacity $N$ for 
queue $B_j$ increases to infinity, the truncation error of $B_j$
obeys the following inequality:
\begin{equation}
\Err^{(\mathcal{I}_j)}_{(N)}
\leq \frac{\alpha^{(\infty)}_{N} / \beta^{(\infty)}_{N+1}}{1-\alpha^{(\infty)}_{N} / \beta^{(\infty)}_{N+1}} \cdot \tilde{\pi}_{\partial, \, B_j}^{(\mathcal{I})}. 
\label{eqn:thm}
\end{equation}
\label{thm:1}
\end{fact}

Detailed proof for Fact~\ref{thm:1} can be found in Theorem 1 in Ref~\cite{Cao2016a}.

That is, the true error $\Err^{(\mathcal{I}_j)}_{(N)}$ is bounded by a simple function 
of $\alpha^{(\infty)}_{N}$ and $\beta^{(\infty)}_{N+1}$ multiplied by the boundary probability 
$\tilde{\pi}_{\partial, \, B_j}^{(\mathcal{I})}$. 
We can use this
inequality to construct an upper-bound for $\Err^{(\mathcal{I}_j)}_{(N)}$.  
We take advantage of the following fact: 
\begin{fact}
For any biological system in which the total amount of mass is finite, 
\textit{e.g.}, cells with finite mass and growth
the aggregated synthesis rate $\alpha^{(\infty)}_{N}$ becomes smaller 
than the aggregated degradation rate $\beta^{(\infty)}_{N+1}$ 
when the buffer capacity $N$ is sufficiently large:
$$
\lim_{N \rightarrow \infty} \frac{\alpha^{(\infty)}_{N}}{\beta^{(\infty)}_{N+1}} < 1.
$$ 
\label{lm:3}
\end{fact}

Detailed proof for Fact~\ref{lm:3} can be found in Lemma 2 in Ref~\cite{Cao2016a}.

Let $C \equiv \frac{\alpha^{(\infty)}_{N} / \beta^{(\infty)}_{N+1}}{1-\alpha^{(\infty)}_{N} / \beta^{(\infty)}_{N+1}}$. 
If $\alpha^{(\infty)}_{N}/\beta^{(\infty)}_{N+1} < 0.5$, we have $C < 1$, and
the true error $\Err^{(\mathcal{I}_j)}_{(N)}$ is always less than the true boundary
probability $\tilde{\pi}_{\partial, \, B_j}^{(\mathcal{I})}$.  If
$\alpha^{(\infty)}_{N}/\beta^{(\infty)}_{N+1} = 0.5$, then $C = 1$, and the true error
converges asymptotically to the true boundary probability $\tilde{\pi}_{\partial, \, B_j}^{(\mathcal{I})}$.  
If $ 0.5 < \frac{\alpha^{(\infty)}_{N}}{\beta^{(\infty)}_{N+1}} < 1.0 $, then $C > 1$, and the error
is larger than $\tilde{\pi}_{\partial, \, B_j}^{(\mathcal{I})}$ but is 
bounded by $\tilde{\pi}_{\partial, \, B_j}^{(\mathcal{I})}$
up to the constant factor $C \equiv \frac{\alpha^{(\infty)}_{N} / \beta^{(\infty)}_{N+1}}{1-\alpha^{(\infty)}_{N} / \beta^{(\infty)}_{N+1}}$.  Therefore, we can conclude that the true boundary
probability $\tilde{\pi}_{\partial, \, B_j}^{(\mathcal{I})}$ provides 
an error bound to the state space truncation. 

Note that in real biological reaction networks, the inequality 
$\alpha^{(\infty)}_{N}/\beta^{(\infty)}_{N+1} < 0.5$ usually holds 
when buffer capacity $N$ is sufficiently large. 
This is because synthesis reactions usually have constant rates, while 
rates of degradation reactions depend on the copy number of net molecules 
in the network. As a result, the ratio between aggregated synthesis and degradation rates 
decreases monotonically when the total number of molecules 
in the system increases.

\subsubsection{True Boundary Probability and Computed Boundary Probability on Truncated Space}

However, it is not possible to calculate the true boundary probability 
$\tilde{\pi}_{\partial, \, B_j}^{(\mathcal{I})}$ 
on the infinite state space.  
We have the following fact: 
\begin{fact}
\label{lm:2}
The total probability $\tilde{\pi}_N^{(\mathcal{I}_j)}$ of the 
boundary states $\partial \Omega^{(\mathcal{I}_j)}_{B_j}$ of 
the $j$-th iBD with buffer capacity $B_j \equiv N$ 
obtained from the truncated state space $\Omega^{(\mathcal{I}_j)}$
is greater than or equal to the true probability 
$\tilde{\pi}_N^{(\mathcal{I})}$ over the same boundary states, 
i.e., $\tilde{\pi}_N^{(\mathcal{I})} \leq \tilde{\pi}_N^{(\mathcal{I}_j)}$. 
\end{fact}

Detailed proof for Fact~\ref{lm:2} can be found in Theorem 2 in Ref~\cite{Cao2016a}.

We can therefore conclude that the true boundary probability is no greater than 
the truncated boundary probability given in Eqn.~({\ref{eqn:pinn}) in the general case when
$\alpha^{(N)}_i \ne 0$ and $\beta^{(N)}_{i+1} \ne 0$. 
We further consider two additional cases.  When reactions 
associated with the $j$-th iBD has zero synthesis and
nonzero degradation constants, namely, $\alpha^{(N)}_i = 0$ and
$\beta^{(N)}_{i+1} \ne 0$, the aggregated system with respect to
$j$-th iBD is a death process and there is no synthesis
reactions. The associated iBD is closed and a finite buffer works once
all states of the closed iBD are enumerated.
When reactions associated with the $j$-th iBD has nonzero synthesis but zero
degradation constants, we have $\alpha^{(N)}_i \ne 0$ but
$\beta^{(N)}_{i+1} = 0$. The aggregated system with respect to
the $j$-th iBD is a birth process without degradation reactions.
In this case, the error for the time evolving probability can be
estimated using a Poisson distribution with parameter $\alpha^{(N)}_i \cdot
t$, where $\alpha^{(N)}_i$ is the maximum aggregated rate, and $t$ is
the elapsed time used for computing the time evolution of the probability landscape~\cite{Fox1988,Wolf2010}.  
We dispense with details here.

\subsubsection{Bounding Errors When Truncating Multiple Buffer Queues}

We now consider truncating one additional buffer queue at the $i$-th iBD. 
We denote the buffer bank as $\mathcal{I}_{i,j} = (\infty, \cdots, B_i, \cdots, B_j, \cdots, \infty)$,
with $B_i$ and $B_j$ as the buffer capacities of the $i$-th and $j$-th iBDs, 
respectively.  The rest of the buffer queues all have infinite capacities. We denote the 
corresponding state space as $\Omega^{\mathcal{I}_{i,j}}$, the transition 
rate matrix as $\bA^{\mathcal{I}_{i,j}}$, and the steady state probability 
distribution as $\bpi^{\mathcal{I}_{i,j}}$. We have the fact that 
the probability of each state in the state space $\Omega^{\mathcal{I}_{i,j}}$ 
is no less than the corresponding probability on $\Omega^{\mathcal{I}_{j}}$, \textit{i.e.}, 
$\pi^{\mathcal{I}_{i,j}}(\bx) \geq \pi^{\mathcal{I}_{j}}(\bx)$ for all $\bx \in \Omega^{\mathcal{I}_{i,j}}$. 

\begin{fact}
\label{lm:Iij}
At steady state, $\bpi^{\mathcal{I}_{i,j}} \geq \bpi^{\mathcal{I}_{j}}$ and 
$\bpi^{\mathcal{I}_{i,j}} \rightarrow \bpi^{\mathcal{I}_{j}}$ component-wise 
over state space $\Omega^{\mathcal{I}_{i,j}}$ when buffer capacity $B_i \rightarrow \infty$. 
\end{fact}

Detailed proof for Fact~\ref{lm:Iij} can be found in Theorem 3 in Ref~\cite{Cao2016a}.

That is, the computed boundary probability of the $j$-th iBD after introducing an 
additional truncation at the $i$-th iBD will be no smaller than when the buffer 
capacity is sufficiently large. Therefore, the boundary probability 
from double truncated state space $\Omega^{\mathcal{I}_{i,j}}$ can be 
conservatively and safely used to bound the 
truncation error. We can further show by induction that boundary probability 
computed from state space truncated at multiple iBDs $\Omega^{(\mathcal{B})}$ 
will not be smaller, and therefore can be used to bound the true boundary probabilities.

\paragraph{Error Bound Inequality}
According to Eqn.~(\ref{eqn:bd1}), and Facts 1--6,
we have the following inequality to bound the true error of state space truncation 
using the finite buffer bank $\mathcal{B} = (B_1, \cdots, B_w)$: 
\begin{equation}
\Err^{(\mathcal{B})} 
\leq 
\sum_{j=1}^w \Err^{(\mathcal{I}_j)}
\leq
\sum_{j=1}^w C_j \, \tilde{\pi}^{(\mathcal{I})}_{\partial, B_j}
\leq
\sum_{j=1}^w C_j \, \tilde{\pi}^{(\mathcal{I}_j)}_{\partial, B_j}
\leq
\sum_{j=1}^w C_j \, \tilde{\pi}^{(\mathcal{B})}_{\partial, B_j}, 
\label{eqn:errbd}
\end{equation}
where $C_j \equiv \frac{\alpha^{(\infty)}_{B_j-1} / \beta^{(\infty)}_{B_j}}{1-\alpha^{(\infty)}_{B_j-1} / \beta^{(\infty)}_{B_j}}$, $j = 1, \cdots, w$, are finite constants for each individual buffer queue.

\subsubsection{Upper and Lower Bounds for Steady State Boundary Probability}
However, the boundary probability $\tilde{\pi}^{(\mathcal{B})}_{\partial, B_j}$ 
cannot be calculated \textit{a priori} without solving the dCME. 
To efficiently estimate if the size of the truncated state space 
is adequate to compute the steady state probability landscape 
with errors smaller than a predefined tolerance, we now introduce an 
easy-to-compute method to obtain the upper- 
and lower-bounds of the boundary probabilities 
$\tilde{\pi}^{(N)}_N$ 
\textit{ a priori} without solving the dCME.

Denote the maximum and minimum aggregated synthesis rates from the block 
sub-matrix $\bA_{i+1,\,i}$ as $\overline{\alpha}^{(N)}_i$ and 
$\underline{\alpha}^{(N)}_i$, respectively.  They can be computed as 
the maximum and minimum element of the row vector obtained from the column sums: 
\begin{equation}
\overline{\alpha}^{(N)}_i = \max\{\mathbbm{1}^T \bA_{i+1,i}\} \quad \mbox{ and } \quad
\underline{\alpha}^{(N)}_i = \min\{\mathbbm{1}^T \bA_{i+1,i}\},
\label{eqn:alu}
\end{equation}
respectively. The maximum and minimum aggregated degradation rates can be 
computed similarly from the block sub-matrix $\bA_{i,\,i+1}$ as: 
\begin{equation}
\overline{\beta}^{(N)}_{i+1} = \max\{\mathbbm{1}^T \bA_{i,i+1}\} \quad \mbox{ and } \quad
\underline{\beta}^{(N)}_{i+1} = \min\{\mathbbm{1}^T \bA_{i,i+1}\},
\label{eqn:blu}
\end{equation}
respectively.  Note that $\overline{\alpha}^{(N)}_i$,
$\underline{\alpha}^{(N)}_i$, $\overline{\beta}^{(N)}_{i+1}$, and
$\underline{\beta}^{(N)}_{i+1}$ can be easily calculated {\it a priori}\/
without the need 
for explicit state enumeration and generation of the partitioned transition rate matrix $\tilde{\bA}$.  
The block sub-matrix $\bA_{i+1,i}$ only contains synthesis reactions, 
$\bA_{i,i+1}$ only contains degradation reactions. The maximum total 
copy numbers of reactants are fixed at each aggregated state group 
when the maximum buffer capacity is specified, therefore 
$\overline{\alpha}^{(N)}_i$,
$\underline{\alpha}^{(N)}_i$, $\overline{\beta}^{(N)}_{i+1}$, and
$\underline{\beta}^{(N)}_{i+1}$
can be easily calculated by examining the maximum and minimum 
synthesis and degradation reaction rates. 
As the original $\alpha^{(N)}_i$ and $\beta^{(N)}_{i+1}$ given in Eqn.~(\ref{eqn:abdef})
are weighted sums of vector $\mathbbm{1}^T \bA_{i+1,i}$ and
$\mathbbm{1}^T \bA_{i,i+1}$ with regard to the steady state
probability distribution $\tilde{\bpi}^{(N)}(\mathcal{G}_i)$, respectively, we
have 
$$
\underline{\alpha}^{(N)}_i \leq \alpha^{(N)}_i \leq \overline{\alpha}^{(N)}_i \quad \mbox{ and } \quad
\underline{\beta}^{(N)}_{i+1} \leq \beta^{(N)}_{i+1} \leq \overline{\beta}^{(N)}_{i+1}. 
$$

We use results from the theory of stochastic ordering 
for comparing Markov processes to bound  $\tilde{\pi}^{(N)}_N$. Stochastic ordering 
``$\leq_{st} $''
between two infinitesimal generator 
matrices $\bP_{n \times n}$ and $\bQ_{n \times n}$ of Markov processes is defined as~\cite{Truffet1997,Irle2003} 
$$
\bP \leq_{st} \bQ \quad \text{if and only if } \sum_{k=j}^n P_{i,k} \leq \sum_{k=j}^n Q_{i,k} \text{ for all } i, j. 
$$ 
Stochastic ordering between two vectors are similarly defined as: 
$$
\bp \leq_{st} \bq, \quad \text{if and only if } \sum_{k=j}^n p_{k} \leq \sum_{k=j}^n q_{k} \text{ for all } j. 
$$ 
To derive an upper bound for $\tilde{\pi}^{(N)}_N$ in
Eqn.~(\ref{eqn:pinn}), we construct a new matrix $\overline{\bB}$ by
replacing $\alpha^{(N)}_{k}$ with the corresponding
$\overline{\alpha}^{(N)}_k$ and $\beta^{(N)}_{k+1}$ with the
corresponding $\underline{\beta}^{(N)}_{k+1}$ in the matrix $\bB$.
Similarly, to derive an lower bound for $\tilde{\pi}^{(N)}_N$, we construct the
matrix $\underline{\bB}$ by replacing $\alpha^{(N)}_{k}$ with the
corresponding $\underline{\alpha}^{(N)}_k$ and replace
$\beta^{(N)}_{k+1}$ with $\overline{\beta}^{(N)}_{k+1}$ in $\bB$.  We
then have the following stochastic ordering:
$$
\underline{\bB} \leq_{st} \bB \leq_{st} \overline{\bB}. 
$$ 
All three matrices $\underline{\bB}$, $\bB$, and $\overline{\bB}$
are ``$\leq_{st}-\rm{monotone}$'' according to the definitions in
Truffet~\cite{Truffet1997}.  The steady state probability
distributions of matrices $\underline{\bB}$, $\bB$, and
$\overline{\bB}$ are denoted as $\bpi_{\underline{\bB}}$, 
$\bpi_{\bB}$, and $\bpi_{\overline{\bB}}$, respectively. 
They maintain the same stochastic ordering (Theorem 4.1 of
Truffet~\cite{Truffet1997}):
$$
\bpi_{\underline{\bB}} \leq_{st} \bpi_{\bB} \leq_{st} \bpi_{\overline{\bB}}. 
$$ 
Therefore, we have the inequality for the $j$-th buffer queue with capacity $N$:
$$
\underline{\tilde{\pi}}^{(N)}_N
\leq 
\tilde{\pi}^{(N)}_N
\leq 
\overline{\tilde{\pi}}^{(N)}_N. 
$$ Here the upper bound $\overline{\tilde{\pi}}^{(N)}_N$ is the boundary
probability computed from  $\tilde{\bpi}_{\overline{\bB}}$, the lower bound
$\underline{\tilde{\pi}}^{(N)}_N$ is the boundary probability computed from
$\tilde{\bpi}_{\underline{\bB}}$, and $\tilde{\pi}^{(N)}_N$ is the boundary
probability from $\tilde{\bpi}_{\bB}$. From Eqn.~(\ref{eqn:pinn}), the upper
bound $\overline{\tilde{\pi}}^{(N)}_N$ can be calculated \textit{a
  priori}\/ from reaction rates:
\begin{equation}
\overline{\tilde{\pi}}^{(N)}_N = 
\frac{\prod\limits_{k = 0}^{N-1}
    \frac{\overline{\alpha}^{(N)}_{k}}{\underline{\beta}^{(N)}_{k+1}}} 
{1 + {\sum\limits_{j = 1}^{N} {\prod\limits_{k = 0}^{j-1}
     {\frac{\overline{\alpha}^{(N)}_{k}}{\underline{\beta}^{(N)}_{k+1}}}} 
    }
}, 
\label{eqn:upb}
\end{equation}
and the lower bound $\underline{\tilde{\pi}}^{(N)}_N$ can be calculated as:
\begin{equation}
\underline{\tilde{\pi}}^{(N)}_N = 
\frac{\prod\limits_{k = 0}^{N-1}
    \frac{\underline{\alpha}^{(N)}_{k}}{\overline{\beta}^{(N)}_{k+1}}}
{1 + {\sum\limits_{j = 1}^{N} {\prod\limits_{k = 0}^{j-1}
     {\frac{\underline{\alpha}^{(N)}_{k}}{\overline{\beta}^{(N)}_{k+1}}}} 
    }
}. 
\label{eqn:lowerb}
\end{equation} 
These are general upper and lower bounds of truncation error valid
for any iBD in a reaction network.
The upper and lower bounds for the total error of a reaction network
with multiple iBDs can be obtained straightforwardly by taking
summations of bounds for each individual iBDs:
\begin{equation}
\sum_{i=1}^w \underline{\tilde{\pi}}^{(\mathcal{B})}_{B_i}
\leq 
\sum_{i=1}^w \tilde{\pi}^{(\mathcal{B})}_{B_i}
\leq 
\sum_{i=1}^w \overline{\tilde{\pi}}^{(\mathcal{B})}_{B_i}.
\label{eqn:ubd}
\end{equation}

In summary, we have shown from Eqn.~(\ref{eqn:bd1}), Facts 1--6, Eqn.~(\ref{eqn:errbd}), and
Eqn.~(\ref{eqn:ubd}) the truncation error of the steady state 
probability landscape from each individual iBD $\Err^{(B_j)}$ 
using finite buffer bank $\mathcal{B} = (B_1, \cdots, B_w)$ can be 
bounded using the following inequality: 
\begin{align}
\begin{split}
\Err^{(\mathcal{I}_j)} 
\leq
C_j \, \tilde{\pi}^{(\mathcal{I})}_{\partial, B_j} 
\leq
C_j \, \tilde{\pi}^{(\mathcal{I}_j)}_{\partial, B_j} 
\leq
C_j \, \tilde{\pi}^{(\mathcal{B})}_{\partial, B_j}  
\leq
C_j \, \overline{\tilde{\pi}}^{(\mathcal{B})}_{B_j} 
\leq
\overline{C}_j \, \overline{\tilde{\pi}}^{(\mathcal{B})}_{B_j} 
=
\frac{\frac{\overline{\alpha}^{(B_j)}_{B_j-1}}{\underline{\beta}^{(B_j)}_{B_j}}}{1-\frac{\overline{\alpha}^{(B_j)}_{B_j-1}}{\underline{\beta}^{(B_j)}_{B_j}}} \cdot 
\frac{\prod\limits_{k = 0}^{B_j-1}
    \frac{\overline{\alpha}^{(B_j)}_{k}}{\underline{\beta}^{(B_j)}_{k+1}}} 
{1 + {\sum\limits_{j = 1}^{B_j} {\prod\limits_{k = 0}^{j-1}
     {\frac{\overline{\alpha}^{(B_j)}_{k}}{\underline{\beta}^{(B_j)}_{k+1}}}} 
    }
},
\label{eqn:errbdj}
\end{split}
\end{align}
and the overall truncation error $\Err^{(\mathcal{B})}$
using the finite buffer bank $\mathcal{B} = (B_1, \cdots, B_w)$ 
can therefore be bounded by the following inequality:
\begin{align}
\begin{split}
\Err^{(\mathcal{B})} 
\leq 
\sum_{j=1}^w \Err^{(\mathcal{I}_j)} 
\leq
\sum_{j=1}^w \overline{C}_j \, \overline{\tilde{\pi}}^{(\mathcal{B})}_{B_j} 
=
\sum_{j=1}^w \frac{\frac{\overline{\alpha}^{(B_j)}_{B_j-1}}{\underline{\beta}^{(B_j)}_{B_j}}}{1-\frac{\overline{\alpha}^{(B_j)}_{B_j-1}}{\underline{\beta}^{(B_j)}_{B_j}}} \cdot 
\frac{\prod\limits_{k = 0}^{B_j-1}
    \frac{\overline{\alpha}^{(B_j)}_{k}}{\underline{\beta}^{(B_j)}_{k+1}}} 
{1 + {\sum\limits_{j = 1}^{B_j} {\prod\limits_{k = 0}^{j-1}
     {\frac{\overline{\alpha}^{(B_j)}_{k}}{\underline{\beta}^{(B_j)}_{k+1}}}} 
    }
},
\label{eqn:errbdall}
\end{split}
\end{align}
where $C_j \equiv \frac{\alpha^{(\infty)}_{B_j-1} / \beta^{(\infty)}_{B_j}}{1-\alpha^{(\infty)}_{B_j-1}/\beta^{(\infty)}_{B_j}}$ and $\overline{C}_j \equiv \frac{\overline{\alpha}^{(B_j)}_{B_j-1} / \underline{\beta}^{(B_j)}_{B_j}}{1-\overline{\alpha}^{(B_j)}_{B_j-1}/\underline{\beta}^{(B_j)}_{B_j}}$, 
$j = 1, \cdots, w$, are finite constants for each individual buffer queue, 
and we have $C_j \leq \overline{C}_j$ as $\frac{\alpha^{(\infty)}_{B_j-1}}{\beta^{(\infty)}_{B_j}} \leq \frac{\overline{\alpha}^{(B_j)}_{B_j-1}}{\underline{\beta}^{(B_j)}_{B_j}}$.

\subsection{Optimizing Buffer Allocation}

\subsubsection{Determining Minimal Buffer Sizes Satisfying Pre-defined Error Tolerance} 

To determine the minimal buffer sizes for the $w$ iBDs so a
pre-defined error tolerance $\epsilon$ is satisfied, we first
calculate \textit{a priori}\/ the upper bound from the boundary
probability $\overline{\tilde{\pi}}^{(N)}_N$ of each iBD using
Eqn.~(\ref{eqn:upb}) for different buffer sizes.  The minimal $N$ for
each iBD with $\overline{\tilde{\pi}}^{(N)}_N < \epsilon/w$ is then chosen as
the size of that buffer queue.  Other weighted scheme is also possible. 
We then proceed to enumerate the state space using a buffer bank 
whose sizes have been thus determined {\it a priori} to 
numerically solve the dCME .  

It is possible that this
\textit{a priori}\/ upper bound is overly conservative, and buffer
sizes can be further decreased based on numerical results.
Specifically, if the boundary probability computed from numerical solution for an iBD
with an \textit {a priori}\/ determined buffer size is much smaller
than the pre-defined error tolerance $ \epsilon$, it is possible to
further decrease the buffer size of that iBD to gain in memory space
and improve computing efficiency.

\subsubsection{Optimized Memory Allocation Based on Error Bounds} 

Our method can also be used to optimize the allocation of memory space to 
improve the accuracy or computing efficiency of the solution to the dCME.  
When the total size of the state space is fixed, we can allocate buffer 
capacities for buffer queues differently, so that the total error of the dCME 
solution is minimized.  A simple strategy is to distribute the errors equally 
to all buffer queues or according to some weight scheme, for example, based 
on the error bounds of individual iBDs, or the complexity of computing the 
rates of individual iBDs, or the effects on numerical efficiency.  We then 
determine the buffer size of each iBD.  The relative ratio of buffer sizes of 
different iBDs can be used to allocate memory. When the state space to be 
enumerated is too large to fit into the computer memory, we can further 
decrease buffer capacities for all iBDs simultaneously according to the 
allocation ratio. Such optimization can be done \textit{a priori}\/ without trial 
computations.

\subsection{Numerical Solutions of dCME}

\paragraph{Time-evolving probability landscape}
The time evolving probability landscape derived from a dCME of
Eqn.~(\ref{eqn:dcme2}) can be expressed in the form of a matrix
exponential: $\bp(t) = e^{\bA t} \bp(0),$ where $\bp(0)$ is the
initial probability landscape and $\bA$ is the transition rate matrix
over the enumerated state space $\Omega^{(\mathcal{B})}$.  Once
$\bp(0)$ is given, $\bp(t)$ can be calculated using numerical methods
such as the Krylov subspace projection method, {\it e.g.}, as implemented in
the {\sc Expokit} package of Sidjie {\it et al}~\cite{Sidje1998}.  Other 
numerical techniques can also be applied~\cite{Kazeev2014}.  All
results of the time-evolving probability landscape in this study are
computed using the {\sc Expokit} package.

\paragraph{Steady state probability landscape} 
The steady state probability landscape $\bpi$ is of great
general interests.  It is governed by the equation $\bA \bpi 
= 0$, and corresponds to the right eigenvector of the $0$ eigenvalue.
With the states enumerated by the mb-dCME method, 
$\bpi$ can be computed using numerical techniques such as iterative
solvers~\cite{Lehoucq_Arpack,CaoBMCSB08,Stewart1994,Saad1986gmres}.
In this study, we use the  Gauss-Seidel solver to compute all
steady state probability landscape.  To our knowledge, the ACME 
method and its predecessor are the only known methods 
for computing the steady state probability landscape for an arbitrary
biological reaction network.

\paragraph{First passage time distribution} 

The first passage time from a specific initial state to a given end
state is of great importance in studying rare events.
 The probability that a network transits from the starting
state $\bx_s$ to the end state $\bx_e$ within time $t$ is
the {\it first passage time probability $p(t, \bx_e|\bx_s)$}.

The distribution of $p(t, \bx_e|\bx_s)$ at all possible time intervals
and the corresponding cumulative probability distribution $F(t,
\bx_e|\bx_s)$, namely, the probability distribution that the system
transits from $\bx_s$ to $\bx_e$ within time $t$, can be computed
using the ACME method.  To obtain $F(t, \bx_e|\bx_s)$, we use the
absorbing matrix $\bA_{abs}$ instead of the original rate matrix $\bA$
by simply replacing the end state $\bx_e$ with an absorbing
state~\cite{Gross1984}.  In addition, we assign the initial state
$\bx_i$ with a probability of $1$.  The time evolving probability
landscape of this absorbing system then can be computed as described earlier. 
The cumulative first passage probabilities $F(t, \bx_e|\bx_s)$ at time
$t$ is the marginal probability of the end state
$\bx_e$ at time $t$~\cite{Gross1984}.  The probability of a rare event
can be easily found from the cumulative distribution of first passage
time between the appropriate states.

\section{Biological Examples}
\label{sec:ex}
Below we describe applications using the ACME method in computing the
time-evolving and the steady state probability landscapes of several
biological reaction networks. We study the genetic toggle switch, the phage lambda
lysogenic-lytic epigenetic switch, and the MAPK cascade reaction
network.  We first show how minimal buffer capacities required for
specific error tolerance can be determined \textit{a priori}.  The
time-evolving and the steady state probability landscapes of these
networks are then computed.  We further generate the distributions of
first passage times to study the probabilities of rare transition
events.  Although these three networks are well known, results
reported here are significant, as the full stochasticity and the
time-evolving probability landscapes have not been computed by solving 
the underlying dCME for the latter two networks. 
Furthermore, estimating rare event probabilities such
as short first passage time of transition between different states has
been a very challenging problem, even for the relatively simple
one dimensional Schl\"{o}gl model~\cite{Gillespie2009,Donovan2013}.

\subsection{Genetic Toggle Switch and Its $6$-Dimensional Probability Landscapes}
The genetic toggle switch consists of two genes repressing
each other through binding of their protein dimeric products on the
promoter sites of the other genes.  This genetic network has been
studied extensively~\cite{Gardner2000,Kepler2001,Kim2007,Schultz_JCP07}.  
We follow~\cite{Schultz_JCP07,CaoBMCSB08} and study a detailed
model of the genetic toggle switch with a more realistic
control mechanism of gene regulations.  Different from simpler toggle
switch models~\cite{Munsky2008,Deuflhard2008,Sjoberg2009,Kazeev2014},
in which gene binding and unbinding reactions are approximated by Hill
functions, here detailed negative feedback regulation of gene expressions 
are modeled explicitly through gene binding and unbinding reactions. 
Although Hill functions are useful to curve-fit gene regulation models with 
experimental observations~\cite{Santillan2008}, it may be inaccurate 
to model stochastic networks~\cite{Santillan2008,Kim2012}. 
It is also difficult to 
obtain the cooperativity parameters in Hill functions and relate them 
to the detailed rate constants~\cite{Santillan2008,Kim2012}. 
Furthermore, Hill function-based model may not capture 
important multistability characteristics of the reaction network. 
The genetic toggle switch studied in this example and other previous 
studies~\cite{Schultz_JCP07,CaoBMCSB08} using detailed reaction network 
with explicit gene binding and unbinding exhibits $4$ distinct stable 
states (on/off, off/on, on/on, and off/off) for the $GeneX$ and $GeneY$. 
However, similar genetic toggle switch modeled using Hill function 
exhibits only two stable states (on/off and off/on)~\cite{Kim2007,Kazeev2014}.

The molecular species, reactions, and their
rate constants for the genetic toggle switch are listed in
Table~\ref{eqn:tgrxns}. 
Specifically, two genes $GeneX$ and $GeneY$ express protein products
$X$ and $Y$, respectively.  Two $X$/$Y$ protein monomers can bind on
the promoter site of $GeneY$/$GeneX$ to form protein-DNA complexes
$BGeneY$/$BGeneX$, and turn off the expression of $GeneY$/$GeneX$, 
respectively.

\begin{table}[ht]
	\centering
		\begin{tabular}{| c | r | r | r |}
		  \hline
			Sizes of buffer queues & mb-dCME & Hypercube method & Reduction factor \\
			\hline
	\multicolumn{4}{|l|}{Bistable genetic toggle switch} \\
			\hline
			  	10, 10 & 400 & 1,936 & 4.84 \\
				20, 20 & 1,600 & 7,056 & 4.41 \\
				30, 30 & 3,600 & 15,376 & 4.27 \\
				40, 40 & 6,400 & 26,896 & 4.20 \\
			\hline
	\multicolumn{4}{|l|}{Phage lambda epigenetic switch network} \\
            		\hline
            		    10, 10 & 2,151 & 61,952 & 28.80  \\
            		    20, 20 & 9,711 & 225,792 & 23.25 \\
            		    30, 30 & 22,671 & 492,032 & 21.70 \\
            		    40, 40 & 41,031 & 860,672 & 20.98 \\
            		\hline
	\multicolumn{4}{|l|}{MAPK signaling network} \\
			\hline
			        $3$, $3$ & $2,176$ & $4.3 \times 10^{9}$ & $2.0 \times 10^6$ \\
				$6$, $6$ & $209,304$ & $3.3 \times 10^{13}$ & $1.6 \times 10^8$ \\
				$9$, $9$ & $6,210,644$ & $1.0 \times 10^{16}$ & $1.6 \times 10^9$ \\
				$14$, $6$ & $2,706,935$ & $1.1 \times 10^{11}$ & $4.1 \times 10^4$ \\
			\hline
		\end{tabular}
  \caption{Size comparison of enumerated state spaces for the genetic
  toggle switch, the epigenetic switch network of phage lambda, and
  the MAPK network.  Column 1 lists sizes of buffer queues used in the
    mb-dCME method, columns 2 and 3 sizes of the space enumerated by
    the dCME and the traditional hypercube methods, respectively.
    Column 4 lists the reduction factors using the mb-dCME method over
    the hypercube method. }
  \label{tab:ssc3}
\end{table}

\begin{figure}[ht]
{\centering\includegraphics[scale=0.8]{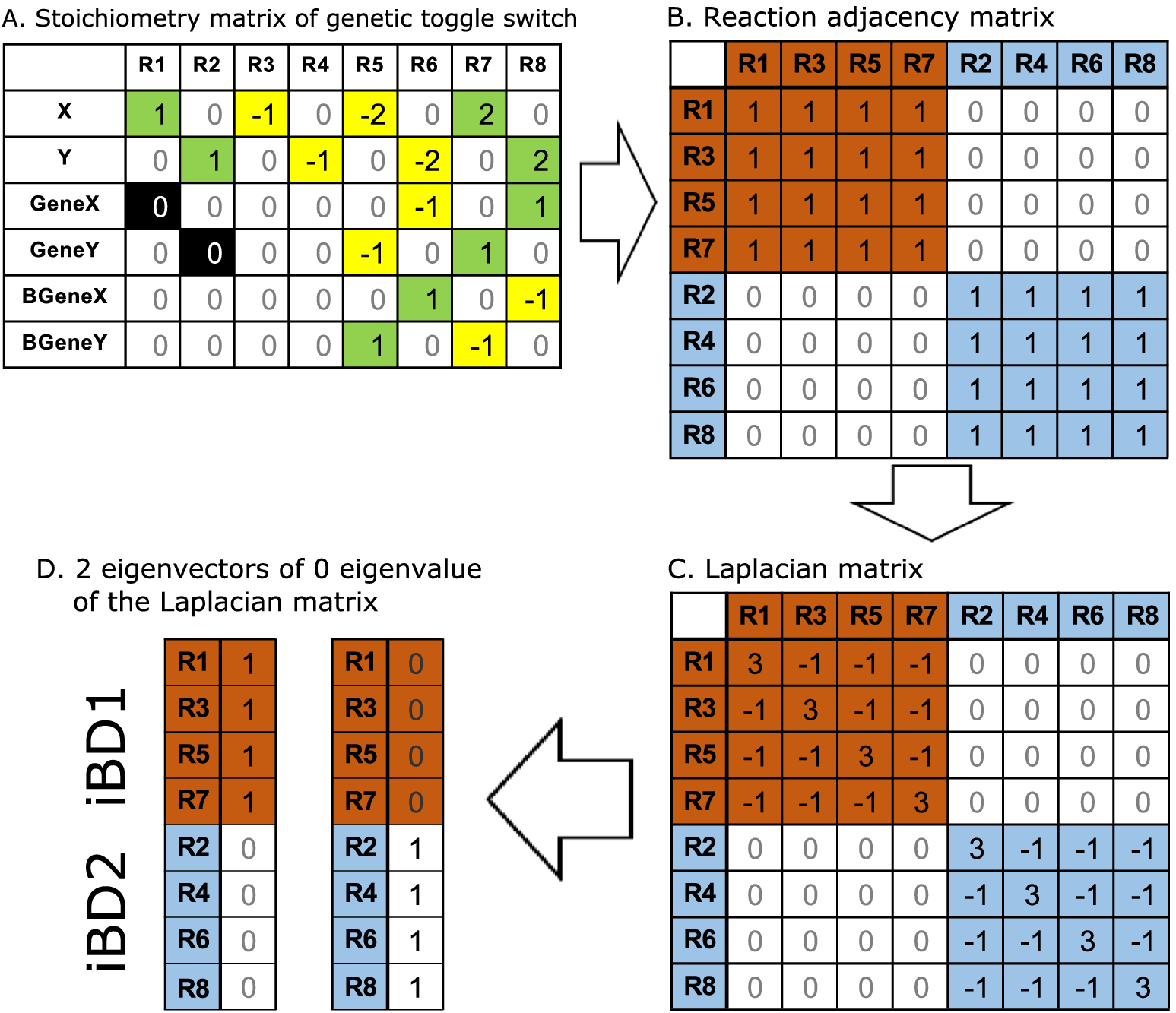}}
\caption{Partitioning the bistable genetic toggle switch network into multiple
independent Birth-Death (iBD) components using Algorithm~\ref{alg:IBD}. (A) Stoichiometry
matrix of the genetic toggle switch constructed from the reaction network in
Eqn.~(\ref{eqn:tgrxns}) in the Appendix. (B) The reaction adjacency matrix constructed from
the stoichiometry matrix according to Eqn.~(\ref{eqn:adjC}). (C) The Laplacian matrix of
the reaction network constructed using Eqn.~(\ref{eqn:lapL}).  There are two 0 eigenvalues
for the Laplacian matrix in (C). (D) The 2 eigenvectors corresponding to the
two 0 eigenvalues give the partition of the reaction network.  }
\label{fig:alg1}
\end{figure}

\paragraph{Number of buffer queues and comparison of state space sizes} 
According to Algorithm~\ref{alg:IBD}, there are two iBDs in this network, namely, iBD$_1$
with reactions $R1$, $R3$, $R5$, and $R7$, and iBD$_2$ with $R2$, $R4$, $R6$, and $R8$. Each is
assigned a separate buffer queue. Detailed steps of iBD partition for the
genetic toggle switch network using the Algorithm~\ref{alg:IBD} are illustrated in Fig.~\ref{fig:alg1}.
In this network, reaction $R1$ generates a new molecule $X$ and does not alter the
copy number of all other species.  Therefore, row $X$ of column $R1$ is 1 (Fig.~\ref{fig:alg1}A),
and 0 for all other rows of column $R1$).  Reaction $R8$ converts one copy of
bound gene $X$ ($BGeneX$) into an unbound gene $X$ ($GeneX$) and generates two copies
of $Y$ molecules. Therefore, row $BGeneX$ of column $R8$ in the stoichiometry matrix
is -1, row $GeneX$ is 1, and row $Y$ is 2.  All other rows of column $R8$ are 0s
(Fig.~\ref{fig:alg1}A).  The remaining column vectors of the stoichiometry matrix for other
reactions can be obtained similarly.  Each row in the resulting stoichiometry
matrix records the stoichiometry of a molecular species participating in all of
the reactions. The reaction graph can then be constructed by examining which
pairs of reactions share reactant(s) and/or product(s).  Molecular species $X$
changes copy numbers in both reaction $R1$ and $R3$, therefore we have the edge
$e_{{R1},\, {R3}}=1$ in the reaction graph $\bG_R$.  We use an adjacency matrix
to encode the graph, and the entry for row $R1$ and column $R3$ is therefore 1
(Fig.~\ref{fig:alg1}B).  Similarly, $R1$ and $R5$ both involve copy number changes in $X$, hence
$e_{{R1},\, {R5}}=1$.  As $R1$ and $R7$ also involve copy number changes in $X$, we
have $e_{{R1},\, {R7}} =1$.  In contrast, as $X$ is the only species that
changes copy number in reaction $R1$, and $X$ does not participate either as a
reactant or a product with altered copy number in $R2$, $R4$, $R6$, and $R8$, the
corresponding entries in the adjacency matrix of $\bG_R$ therefore have 0s as
entries.  More generally, if the dot product of the stoichiometry vectors of
two reactions $R_i$ and $R_j$ is nonzero, $e_{{Ri},\, {Rj}} =1$,  otherwise the
entry is zero. Once the full adjacency matrix for the reaction graph is
complete (Fig.~\ref{fig:alg1}B), the Laplacian matrix (Fig.~\ref{fig:alg1}C) can be obtained following
Eqn.~(\ref{eqn:lapL}) in the Appendix.  The number of the eigenvectors of the Laplacian
matrix corresponding to the eigenvalue of 0 gives the number of iBDs in the
reaction network, and the non-zero entries of each eigenvector gives the
membership of the corresponding iBD (Fig.~\ref{fig:alg1}D). In this example of genetic
toggle switch, 0 is an eigenvalue of multiplicity of 2 of the Laplacian matrix.
The two eigenvectors associated with the eigenvalue of 0 give the two iBDs
(Fig.~\ref{fig:alg1}D). Specifically, the reactions with nonzero entries in each eigenvector
form the corresponding iBD: iBD$_1$ consists of reactions $R1$, $R3$, $R5$, and $R7$, and
iBD$_2$ consists of $R2$, $R4$, $R6$, and $R8$ (Fig.~\ref{fig:alg1}D).

The genetic toggle switch is 
sufficiently complex to exhibit reduced sizes of the enumerated
state spaces using the multi-finite buffer algorithm, when compared with the
traditional hypercube method.  Table~\ref{tab:ssc3} lists the sizes
of the state spaces using these two methods.  The size of enumerated
state space for the hypercube method is the product of the maximum
number of possible states of each individual species.  For example,
when both buffer queues have a buffer capacity of $40$, the state
space size is $41^2 \times 2^4 = 26,896$, in which $40+1=41$ is the
total number of all possible different copy numbers of protein $X$ and
protein $Y$, and $2^4$ is the total different binding and unbinding
configurations for each of $GeneX$, $GeneY$, $BGeneX$ and $BGeneY$.
The traditional approach generates a state space that is about $4$
times larger than that generated by the mb-dCME method in this case.

\begin{figure}[ht]
{\centering\includegraphics[scale=1.0]{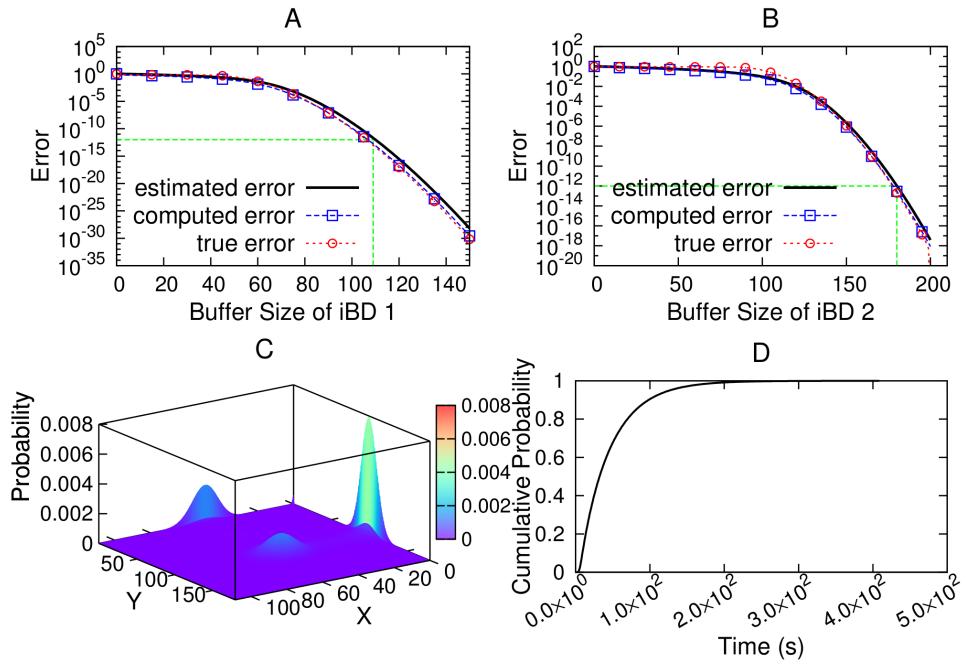}}
\caption{ Error estimation and computing the steady state probability landscape and the
  first passage time of the genetic toggle switch network. (A) and (B):
  The {\it a priori}\/ estimated error (black solid curve), the
  computed error (blue dashed line and squares), and the true error
  (red dotted line and circles) of the steady state probability
  landscape for iBD$_1$ and iBD$_2$, respectively.  The {\it a
    priori}\/ estimated error is always larger than the computed
  error.  The green dashed lines indicate the estimated minimal buffer
  size required so the error is within the predefined tolerance
  of $1 \times 10^{-12}$.  (C): Steady state probability landscape.
  (D): The cumulative distribution of the first passage time from the
  initial state ($\bx_s = \{X = 49, Y=0, GeneX=1, GeneY=1, BGeneX=0,
  BGeneY=0\}$) to the end state ($\bx_e = \{ X = 0, Y = 99\}$).}
\label{fig:TG_Fig1}
\end{figure}

\paragraph{Errors and buffer size determinations}
The sizes combination of buffer queues $\bB = (200, 400)$ is found to
be sufficient to obtain the exact steady state probability landscape
(estimated error $<10^{-30}$) according to calculations using
Eqn.~(\ref{eqn:upb}).  With the exact steady state probability
landscape known, true errors calculated using Eqn.~(\ref{eqn:trueerr})
for different sizes of the two buffer queues are shown in Fig.~\ref{fig:TG_Fig1}A and
Fig.~\ref{fig:TG_Fig1}B (red dotted lines and circles), both of which
decrease monotonically with increasing buffer sizes.

The computed error estimates by solving the boundary probability 
from the underlying dCME (Fig.~\ref{fig:TG_Fig1}A and B, blue
dashed lines and squares) also decrease monotonically with increasing
buffer size.  The computed error estimates for the 1-st and 2-nd iBD
are larger than the true error when the buffer size is larger than $89$ and
$163$, respectively, as would be expected from Fact~\ref{thm:1}.

To estimate {\it a priori}\/ the required minimum buffer sizes for
both buffer queues for a predefined error tolerance of $\epsilon = 1.0
\times 10^{-12}$ so that the total error does not exceed $2.0 \times
10^{-12}$, we use Eqn.~(\ref{eqn:upb}) to estimate errors at different
buffer sizes (black solid lines in Fig.~\ref{fig:TG_Fig1}A and B). We
follow Eqn.~(\ref{eqn:alu}) and (\ref{eqn:blu}) to compute
$\overline{\alpha}_{i}=k_1$ and
$\underline{\beta}_{(i+1)}=[(i+1)-2]\cdot k_3$ for the first iBD,
where the subscript $(i+1)$ is the total copy number of species $X$ in
the system, and the subtraction of $2$ is necessary because upto $2$
copies of $X$ can be protected from degradation by binding to $GeneY$.
This corresponds to the extreme case when $GeneX$ is constantly turned
on and $GeneY$ is constantly turned off. The {\it a priori}\/ error
estimates at different buffer size are shown in
Fig.~\ref{fig:TG_Fig1}A (black solid lines).  Similarly, we have
$\overline{\alpha}_{i}=k_2$ and $\underline{\beta}_{i+1}=[(i+1)-2]
\cdot k_4$ following Eqn.~(\ref{eqn:alu}) and (\ref{eqn:blu}) for the
second iBD. This corresponds to the other extreme case when the
$GeneY$ is constantly turned on, and $GeneX$ is constantly turned off
(Fig.~\ref{fig:TG_Fig1}B, black solid lines).  
As discussed earlier, the \textit{a priori} estimated error bounds 
can be easily computed by examining the maximum and minimum 
reaction rates. There is not need for the transition rate matrix. 
For both buffer queues,
the {\it a priori}\/ estimated errors are conservative and are larger
than computed errors at all buffer sizes.  They are also larger than the true
errors when the buffer sizes are sufficiently large.  We can therefore
determine that the minimal buffer size to satisfy the predefined error
tolerance of $\epsilon = 2.0 \times 10^{-12}$ is $109$ for the first
iBD (green dashed lines in Fig.~\ref{fig:TG_Fig1}A) and $180$ for the
second iBD (green dashed lines in Fig.~\ref{fig:TG_Fig1}B).  This
combination of buffer sizes $\bB = (109, 180)$ is used for all
subsequent calculations.  The enumerated state space has a total of
$78,480$ states. The $78,480\times78,480$ transition rate matrix is
sparse and contains a total of $468,564$ non-zero elements.

\begin{figure}[ht]
{\centering
\includegraphics[scale=0.6]{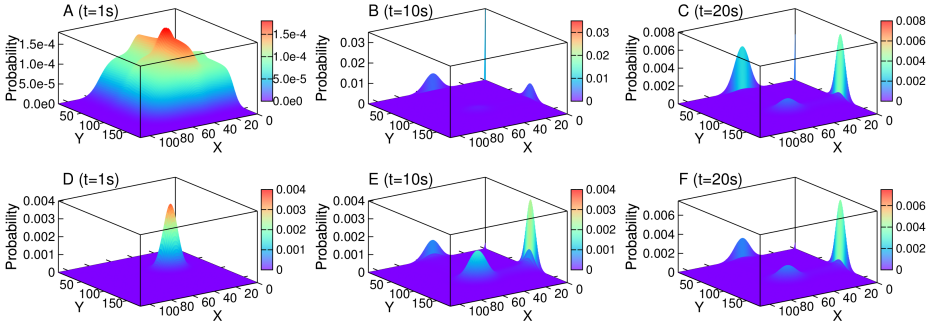}
}
\caption{ The time evolving probability landscapes of the genetic
  toggle switch network.  (A), (B), and (C): Probability landscapes at
  $t=1s$, $t=10s$, and $t=20s$, starting from the uniform
  distribution, respectively.  (D), (E), and (F): Probability
  landscapes at $t=1s$, $t=10s$, and $t=20s$ starting from the initial
  distribution with $p(X=0, Y=0, GeneX=1, GeneY=1, BGeneX=0, BGeneY=0;
  t=0) = 1$, respectively.  }
\label{fig:Toggle_time}
\end{figure}

\paragraph{Steady state and time-evolving probability landscapes}
The time-evolving probability landscape from two different initial
conditions are shown in Fig.~\ref{fig:Toggle_time}.  We use a time
step $\Delta t = 0.5 s$ and a total simulation time of $t = 50 s$.
The probability landscape in Fig.~\ref{fig:Toggle_time}A-C starts
from the uniform initial distribution, in which each state takes the
same initial probability of $1/78,480$.  The probability landscape in
Fig.~\ref{fig:Toggle_time}D-F starts from
an initial probability distribution, in which the state $(X=0,\, Y=0,
\, GeneX=1, \, GeneY=1, \, BGeneX=0, \, BGeneY=0)$ has probability $1$
and all other states have probability $0$.  

The time-evolving probability landscapes for both initial conditions 
converges to the same steady state
(Fig.~\ref{fig:TG_Fig1}C) at time $t = 40 s$, with the computed error
for buffer queues 1 and 2 being $1.741\times 10^{-13}$ and
$2.881\times 10^{-13}$ for results in Fig.~\ref{fig:Toggle_time}A-C,
and $1.716\times 10^{-13}$ and $2.898\times 10^{-13}$ for results in
Fig.~\ref{fig:Toggle_time}D-F, respectively.  Note that the Z-scale 
is different for the time-evolving probability landscapes.  The calculation 
is completed within $2$ minutes using one single core of a 1GHz Quad-Core
AMD CPU.

The steady state probability landscape is also computed separately
(Fig.~\ref{fig:TG_Fig1}C for species $X$ and $Y$).  It has
four peaks that centered at $(X = 0, Y=99)$ with a probability of
$7.910 \times 10^{-3}$; at $(X = 49, Y=0)$ with a probability of
$2.473 \times 10^{-3}$, at $(X = 49, Y=99)$ with a probability of
$1.269 \times 10^{-3}$, and at $(X = 0, Y=0)$ with a probability of
$5.909\times 10^{-4}$, respectively.  The computed error estimates of
$1.715 \times 10^{-13}$ for the first iBD and $2.899 \times 10^{-13}$
for the second iBD are both smaller than the predefined error
tolerance of $\epsilon = 1.0 \times 10^{-12}$.  The computing time is
within $1$ minute.

\paragraph{First passage time distribution and rare event
  probabilities}  We study the problem of the first passage time when
the system travels from the initial starting state $\bx_s = \{X = 49,
Y=0, GeneX=1, GeneY=1, BGeneX=0, BGeneY=0\}$ to the end state $\bx_e =
\{ X = 0, Y = 99 \}$.  We modified the transition rate matrix by
making the end state an absorbing
state~\cite{Gross1984,Cao2013JCP}. The time evolving probability
landscape using the absorbing transition rate matrix $\bA_{abs}$ is
then calculated using a time step $\Delta t = 0.5$ for a total of $500
s$ simulation time.

When the duration is short, the transition from the initial starting
state to the end state is of very low probability.  When the first
passage time is set to $t \leq 3 s$, the probability is calculated to
be $1.993 \times 10^{-5}$, with a computation time of about $10$
seconds.  Our method enables accurate and rapid calculations of
probabilities of such rare events.  As the sampling space of the
toggle switch is two-dimension ($X, Y$), the rare event probability
estimations in this network is far more challenging than the Schl\"ogl
model, which was already beyond the original SSA algorithm~\cite{Gillespie_JPC77} and 
a number of biased stochastic simulation 
algorithms~\cite{Kuwahara2008,Gillespie2009,Daigle2011,Roh2011}. To
our knowledge, no other methods have succeeded in calculating
accurately the rare event probabilities in this model of genetic
switch.

The computed full cumulative probability distribution of the first
passage time is plotted in Fig.~\ref{fig:TG_Fig1}D. It increases
monotonically with time, and approaching probability $1$.  The full
calculation is completed within $10$ minutes.

\begin{figure}[ht]
    \centering
    \includegraphics[scale=0.5]{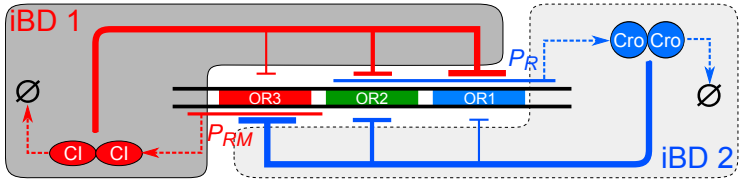}
    \caption{ The network model of the lysogeny-lysis decision circuit
      of phage lambda. CI and Cro proteins can repress the expression
      of each other by differentially binding to three operator sites
      ({\sc OR1, OR2}, and {\sc OR3}). The network can be partitioned
      into two iBDs using Algorithm~\ref{alg:IBD}, as shown in two
      shaded areas of different color. There are a total of $11$
      molecular species and $50$ reactions in this network (see
      Table~\ref{eqn:plrxns} in the Appendix).  }
    \label{fig:phage}
\end{figure}

\subsection{Phage Lambda Epigenetic Switch and Its $11$-Dimensional Probability Landscapes}

The epigenetic switch for lysogenic maintenance and lytic induction in
phage lambda is a classic problem in systems
biology~\cite{Ptashne2004}.  The efficiency and stability of the
decision circuit of the lysogeny-lysis switch have been studied
extensively~\cite{Arkin1998,Aurell2002PRE,Aurell2002PRL,ZhuJBCB2004,Zhu2004}.
Here we use a more realistic model of the reaction network adapted
from reference~\cite{Cao2010}. It consists of $11$ molecular species and
$50$ reactions. The network diagram is shown in Fig.~\ref{fig:phage}
and detailed reaction schemes and rate constants are based on previous 
studies~\cite{Hawley_PNASUSA80,Hawley_JMB82,Shea1985,Li_PNASUSA97,Arkin1998,Kuttler2006,Cao2010}
and are listed in Table~\ref{eqn:plrxns} in the Appendix. 
Molecular species enclosed in parenthesis are 
required for the specific reactions to occur, but with no changes in 
stoichiometry. Here $COR(i)$ denotes operator sites OR$_i$ bounded by Cro$_2$ dimer, 
$ROR(i)$ for OR$_i$ bounded by CI$_2$ dimer, $i = 1, 2, 3$.

\paragraph{Number of buffer queues and comparison of state space sizes} 
There are two iBDs in this  network according to
Algorithm~\ref{alg:IBD}.  The first iBD contains all reactions
involving $CI$ (dark gray shaded area in Fig.~\ref{fig:phage}), and
the second iBD contains all reactions involving $Cro$ (light gray
shaded area in Fig.~\ref{fig:phage}).  Each iBD is therefore assigned
a separate buffer queue.

Table~\ref{tab:ssc3} lists the sizes of the state spaces using the
mb-dCME method and the traditional hypercube method.  As before, the latter is
the product of the maximum number of possible states of each
individual species.  The size of the state space by the traditional
approach is about 21--29 times larger than that by the mb-dCME
method.

\begin{figure}[ht]
{\centering
\includegraphics[scale=1.0]{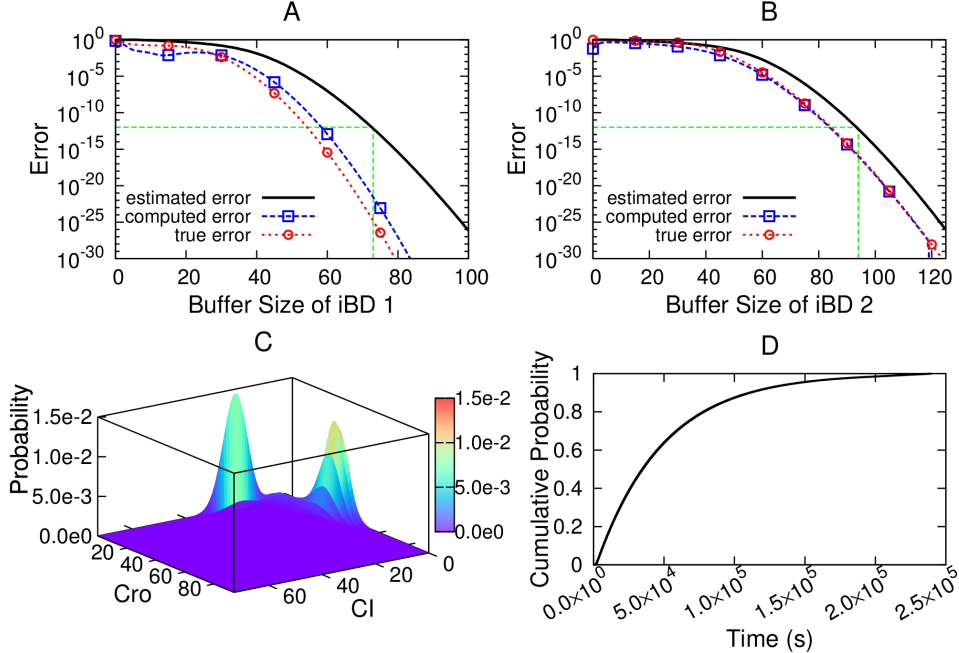}
}
\caption{ Computing the 11-dimension steady state probability
  landscape and the first passage time of the network of epigenetic
  switch of phage lambda.  (A) and (B): The {\it a priori}\/ estimated
  error (black solid curve), the computed error (blue dashed line and
  squares), and the true error (red dotted line and circles) of the
  steady state probability landscape for iBD$_1$ and iBD$_2$,
  respectively.  The computed errors and the true errors are always
  smaller than the {\it a priori}\/ estimated errors.  The green
  dashed lines indicate the estimated minimal buffer sizes required so
  the error is within the predefined tolerance of $10^{-12}$.  (C):
  The 11-dimensional steady state probability landscape projected onto
  the $CI$--$Pro$ plane.  (D): The cumulative distribution of first
  passage time from the initial state ($\bx_s = \{CI = 21, Cro=0,
  OR1=OR2=OR3=0, ROR1=ROR2=ROR3=0, COR1=COR2=COR3=0\}$) to the end
  state ($\bx_e = \{ CI = 2, Cro = 33\}$). }
\label{fig:phage_Fig1}
\end{figure}

\paragraph{Errors and buffer size determinations}
The size combination of buffer queues of $\bB = (150, 150)$ is
sufficient to obtain the exact steady state probability landscape
according to calculations using Eqn.~(\ref{eqn:upb}) (estimated error
$<10^{-30}$).  The true errors calculated using
Eqn.~(\ref{eqn:trueerr}) for different sizes of two buffer queues are shown in
Fig.~\ref{fig:phage_Fig1}A and B (red dotted lines and circles),
both of which decrease monotonically with increasing buffer size. 

The computed error estimates by solving the boundary probability 
from the underlying dCME (Fig.~\ref{fig:phage_Fig1}A and B,
blue dashed lines and squares) also decrease monotonically with
increasing buffer size, when buffer sizes are larger than $23$ and $6$
for the 1-st and 2-nd iBD, respectively.  The computed error estimates
for the 1-st and 2-nd iBD are larger than the true error when the
buffer size is larger than $28$ and $69$, respectively, as would be
expected from Fact~\ref{thm:1}.

To estimate {\it a priori}\/ the required minimum buffer sizes for a
predefined error tolerance of $\epsilon = 1.0 \times 10^{-12}$, we use
Eqn.~(\ref{eqn:upb}) to estimate {\it a priori}\/ errors at different
buffer sizes (black solid lines in Fig.~\ref{fig:phage_Fig1}A and
B). We follow Eqn.~(\ref{eqn:alu}) and (\ref{eqn:blu}) to compute
$\overline{\alpha}_{i}=s_{CI}^1$ and $\underline{\beta}_{(i+1)}=[(i+1)-6]
\cdot d_{CI} $ for the first iBD, where subscript $(i+1)$ is the total
copy number of species $CI$ in the system, and the subtraction of $6$
is because there can be maximally $6$ copies of $CI$ molecules
protected from degradation by binding on the three operator sites
$OR1$, $OR2$, and $OR3$.  This corresponds to the extreme case when
$CI$ is constantly synthesized at the maximum rate, and degraded at
the minimum rate.  Similarly, we assign values of
$\overline{\alpha}_{i}=s_{Cro}$ and $\underline{\beta}_{i+1}= [(i+1)-6]
\cdot d_{Cro}$ in Eqn.~(\ref{eqn:alu}) and (\ref{eqn:blu}) to calculate
the estimated error for the 2nd iBD, which corresponds to the other
extreme case when the $Cro$ is constantly synthesized at its maximum
rate, and degraded at the minimum rate.  In both cases, {\it a
  priori}\/ estimated errors are larger than computed errors at all
buffer sizes.  We can therefore determine 
conservatively {\it a priori}\/ that the minimal buffer size necessary to satisfy the
predefined error tolerance of $1.0 \times 10^{-12}$ is $73$ for the
first iBD (green straight dashed lines in
Fig.~\ref{fig:phage_Fig1}A) and $94$ for the second iBD (green
straight dashed lines in Fig.~\ref{fig:phage_Fig1}B).  This
combination of buffer sizes $\bB = (73, 94)$ is used for all
subsequent calculations.  The enumerated state space has a total of
$180,756$ states. The $180,756 \times 180,756$ transition rate matrix
is sparse and contains a total of $1,330,838$ non-zero elements.

\begin{figure}[ht]
{\centering
\includegraphics[scale=0.6]{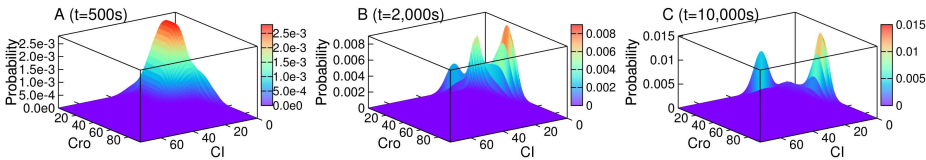}
}
\caption{ Projection of the 11-dimensional time evolving probability landscape of the
  epigenetic switch of phage lambda projected to the $CI$--$Cro$ plane
  starting from the uniform distribution, with the probability
  landscape (A) at $t=500$s; (B) at $t=2,000$s; and (C) at
  $t=10,000$s.  }
\label{fig:Phage_time}
\end{figure}

\paragraph{Steady state and time-evolving probability landscapes}
A projection of the time-evolving 11-dimension probability landscape starting from the
uniform initial distribution is shown in Fig.~\ref{fig:Phage_time}, in
which each state takes the same initial probability of $1/180,756$.
We use a time step $\Delta t = 5 s$ and a total simulation time of $t
= 300,000 s$.  The time-evolving probability landscape converges to
the steady state (shown separately on Fig.~\ref{fig:phage_Fig1}C) at around $t = 250,000
s$, with the computed error of $1.496 \times 10^{-21}$ for buffer
queue 1 and $2.722 \times 10^{-16}$ for buffer queue 2.  The
calculation took $18$ hours using one single core of a 1GHz Quad-Core
AMD CPU.

The steady state probability landscape is also computed separately.
Its projection to the $CI$--$Cro$ plane is plotted in
Fig.~\ref{fig:phage_Fig1}C, which has two peaks centered at $(X =
21, Y=0)$, with a probability of $1.447 \times 10^{-2}$, and at $(X =
2, Y=33)$, with a probability of $1.211 \times 10^{-2}$, respectively.
The computed error of $1.503 \times 10^{-21}$ for the first iBD and
$2.711 \times 10^{-16}$ for the second iBD are both significantly
smaller than the predefined error tolerance of $\epsilon = 1.0 \times
10^{-12}$.  The computation of the steady state probability landscape
is completed within $50$ minutes.

\paragraph{First passage time distribution and rare event probabilities} 
We study the problem of the first passage time when the system travels from
the initial state $\bx_s = \{CI = 21, Cro=0, OR1=OR2=OR3=0,
ROR1=ROR2=ROR3=0, COR1=COR2=COR3=0\}$ in the peak of $CI$ on the
$CI-Cro$ plane, to the end state of $\bx_e = \{ CI = 2, Cro = 33 \}$,
which contains $27$ different microstates at the peak of $Cro$.  We
modified the transition rate matrix by making these end microstates
absorbing~\cite{Gross1984,Cao2013JCP}. The time evolving probability landscape using the absorbing
transition rate matrix $\bA_{abs}$ is then calculated using a time
step $\Delta t = 5$ for a total of $250,000 s$ simulation time.

When the duration is short, the transition from the initial starting
state to the end state is of very low probability.  When the first
passage time is set to $t \leq 500 s$, the probability is calculated
to be $7.184 \times 10^{-9}$, with a computation time of $9$ minutes.
Similar results would require billions of trajectories when using the
alternative method of the stochastic simulation algorithm.  Similar to
the toggle switch example, this rare event problem is two-dimensional
($CI$ and $Cro$), and no current methods we are aware of can
accurately calculate such rare event probabilities.

The computed full cumulative probability distribution of the first
passage time is plotted in Fig.~\ref{fig:phage_Fig1}D. It increases
monotonically with time, and approaching probability $1$.  That is,
given enough time, the system will reach the end state $\bx_e = \{ X =
2, Y = 33 \}$ with certainty $1$.  The full calculation is completed
within $25$ hours.


\subsection{Bistable MAPK Signaling Cascade and Its $16$-Dimensional Probability Landscapes}

The mitogen-activated protein kinase (MAPK) cascades play critical
roles in controlling cell responses to external signals and in
regulating cell behavior, including proliferation, migration,
differentiation, and polarization~\cite{Johnson2002}.  
There are multiple levels of signal
transduction in a MAPK cascade, where activated kinase at each level
phosphorylates the kinase at the next level.  The MAP kinase 
is activated by dual phosphorylations at two conserved threonine
(T) and tyrosine (Y) residues.  Phosphorylated MAPKs can also be
dephosphorylated by specific MAP kinase phosphatases (MKPs). Numerous
mathematical models have been developed to study the complex behavior
of the MAPK cascade in signal
transduction~\cite{Huang1996,Wang2006,Qiao2007,Markevich_JCB04,Smolen2008}.

We examine in details both the time-evolving and the steady state
probability landscapes of a MAPK cascade model consisting of two levels of
kinases, namely, the extracellular signal-regulated kinase (ERK) and
its kinase MEK.  This network model of $16$ molecular species is an open network, in which the
phosphorylation processes for ERK (reactions $R_5$ to $R_{21}$ in
Table~\ref{eqn:mkrxns})~\cite{Markevich_JCB04}, as well as
the synthesis and degradation of both ERK and MEK (reactions $R_1$ to
$R_4$ in Table~\ref{eqn:mkrxns}) are modeled in details.  A
feedback loop in the network enhances the synthesis of MEK by
activating ERKs (Fig.~\ref{fig:mk0}), 
leading to bistability~\cite{Smolen2008}.  The full
network is shown in Fig.~\ref{fig:mk1}.  It includes a total of $16$
molecular species and $35$ individual reactions.  Details of the
molecular species are listed in Table~\ref{tab:mkspecies}, and 
reaction schemes and rate constants are specified in
Table~\ref{eqn:mkrxns} in Appendix.  We set the copy number of 
MKP3 to $1$ and assume that phosphorylations do not protect the ERK 
from degradation.  To our knowledge, this is the largest network 
where full stochastic probability landscapes are computed by solving 
the underlying dCME.

\begin{figure}[ht]
{\centering
\includegraphics[scale=0.35]{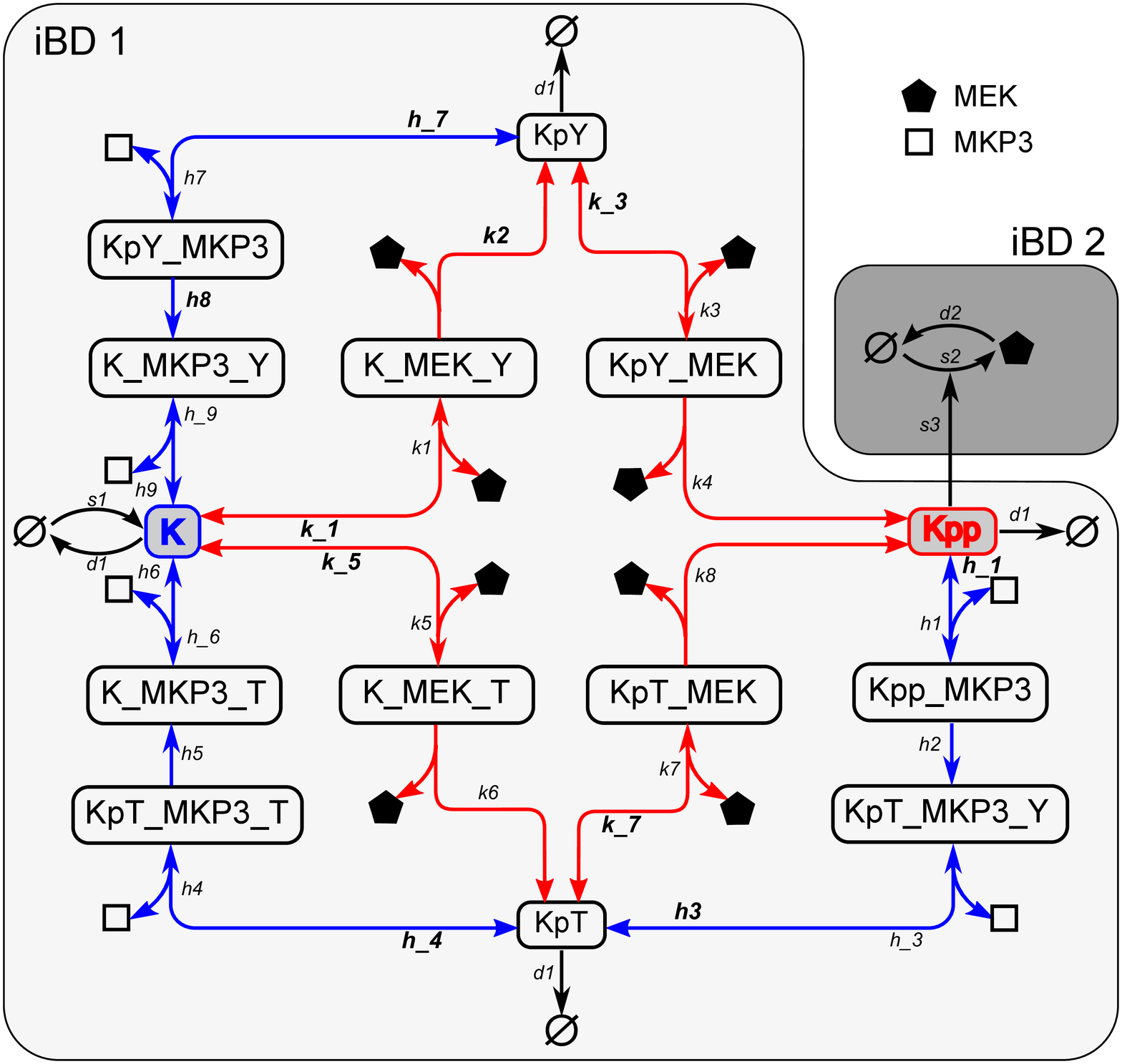}
}
\caption{ A detailed network model of the MAPK cascade.  The ERK(K)
  phosphorylation is catalyzed by the kinase MEK, whereas MEK
  synthesis is up-regulated by dual phosphorylated ERK(Kpp). Detailed
  reactions during the dual phosphorylation process of the ERK(K), the
  synthesis and degradation of MEK are explicitly modeled. Red and
  blue arrows represent phosphorylation and dephosphorylation
  reactions, respectively.  Bidirectional arrows represent reversible
  reactions. The network can be partitioned into two iBDs using
  Algorithm~\ref{alg:IBD}, as shown in two shaded areas of different
  color. There are a total of $16$ molecular species and $35$
  individual reactions in the network (see Tables~\ref{tab:mkspecies}
  in Appendix and \ref{eqn:mkrxns} for more details).
}
\label{fig:mk1}
\end{figure}

\paragraph{Number of buffer queues and comparison of state space sizes} 
According to Algorithm~\ref{alg:IBD}, there are two iBDs in the
network.  The first iBD contains all reactions related to the ERK, labeled as $K$, 
(reactions 3--21 in Table~\ref{eqn:mkrxns} and species in the
lightly shaded area in Fig.~\ref{fig:mk1}). The second iBD contains
reactions of synthesis and degradation of MEK (reactions 1--2 in
Table~\ref{eqn:mkrxns} and species in the darkly shaded box in
Fig.~\ref{fig:mk1}). Each iBD is assigned a separate buffer queue.

To demonstrate the advantage of the mb-dCME state space enumeration
method over the traditional hypercube method, Table~\ref{tab:ssc3}
lists the sizes of the state space with three different choices of the
buffer queues.  The state spaces generated using the traditional
hypercube approach is about $10^4$ to $10^9$ times larger than
that generated by the mb-dCME method.  For example, when both buffer
queues have a capacity of $9$, the size of the enumerated state space
using the traditional hypercube method is $(9+1)^{16}$, in which 16 is
the number of molecular species.  Compared to the size of $6,210,644$
using the mb-dCME method, the reduction factor is approximately $1.6
\times 10^9$. Without this dramatic reduction, it would not be
feasible to compute the exact probability landscape of this model of
MAPK cascade network.

\begin{figure}[ht]
{\centering
\includegraphics[scale=1.0]{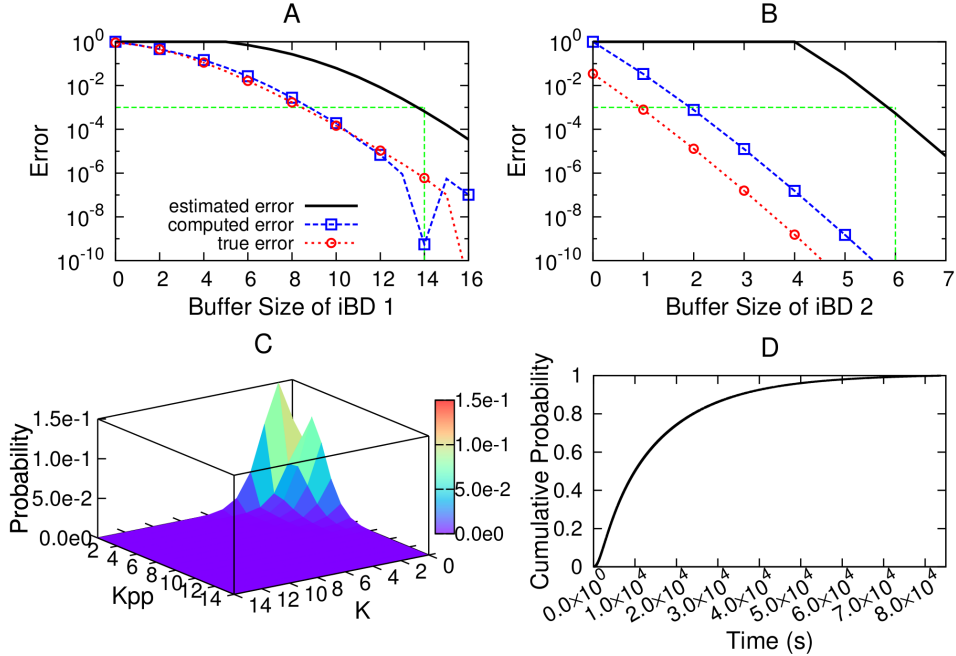}
}
\caption{ Computing the 16-dimension steady state probability
  landscape and the first passage time of the MAPK cascade network
  model.  (A) and (B): The {\it a priori}\/ estimated error (black
  solid curve), the computed error (blue dash line and squares), and
  the true error (red dotted line and circles) of the steady state
  probability landscape for iBD$_1$ and iBD$_2$, respectively.  The
  computed error is significantly smaller than the {\it a priori}\/
  estimated error.  The green straight dashed lines indicate the
  estimated minimal buffer size required so the error is within the
  predefined tolerance of $10^{-3}$.  (C): The steady state
  probability landscape projected to the $K$--$K_{pp}$ plane.  (D):
  The cumulative distribution of first passage time from the initial
  state ($K=3, MKP3=1$) to the end state ($Kpp=2, MKP3=1$). }
\label{fig:mapk_Fig2}
\end{figure}

\paragraph{Errors and buffer size determinations}
The size combination of buffer queues $\bB = (16, 7)$ is used to
approximate the exact solution to the steady state probability
landscape (estimated error $\epsilon <10^{-4}$) according to
calculations using Eqn.~(\ref{eqn:upb}).  Although this estimated $\epsilon$ is
larger than what is used in other models, it is still quite small, as
it is the summation of differences in probabilities of the whole state
space.  This is due to the complexity of this MAPK model and the
limitation of the 3GB CUDA memory of the GPU processor we used.
Access to more capable computing facility would allow a different
choice of sizes of buffer queues such that a smaller \textit{a priori} 
$\epsilon$ can be used.  Note that the computed errors for the steady state 
are considerably smaller ($10^{-8} - 10^{-11}$) as described below. 
With the landscape computed using $\bB = (16, 7)$ 
regarded as approximately the true steady state probability landscape, the
approximated true errors calculated using Eqn.~(\ref{eqn:trueerr}) for
different sizes of two buffer queues are shown in Fig.~\ref{fig:mapk_Fig2}A 
and \ref{fig:mapk_Fig2}B (red dotted lines and circles), both of which
decrease monotonically with increasing buffer sizes.

To estimate {\it a priori}\/ the required minimum buffer sizes for
both buffer queues for a predefined error tolerance of $\epsilon =
10^{-3}$, we use Eqn.~(\ref{eqn:upb}) to estimate errors {\it a
  priori}\/ at different buffer size (black solid lines in
Fig.~\ref{fig:mapk_Fig2}A and B). We follow Eqn.~(\ref{eqn:alu}) and
(\ref{eqn:blu}) to compute $\overline{\alpha}_{i}=s_1$ and
$\underline{\beta}_{(i+1)}= [(i+1)-5] \cdot d_1 $ for the first iBD.
Here the subscript $(i+1)$ is the total copy number of ERK.  As an
ERK molecule can be protected from degradation by forming as many as
$5$ copies of ERK-MKP3 and ERK-MEK complexes in our model (one copy
for each of the four species involving ``$\_ MEK \_$'', and one copy
for all species involving ``$\_ MKP3$'', Table~\ref{tab:ssc3}), the actual minimum
degradation rates are conservatively calculated to be $ [(i+1)-5]
\cdot d_1 $, where $d_1 = 0.0001$ is the degradation rate of $ERK$
(Table~\ref{eqn:mkrxns}). This corresponds to the extreme case when the $ERK$ is
constantly synthesized at its maximum rate, and degraded at the
minimum rate.  Similarly, we have $\overline{\alpha}_{i}=s_3$ and
$\underline{\beta}_{i+1}= [(i+1)-4] \cdot d_2$ for
Eqn.~(\ref{eqn:alu}) and (\ref{eqn:blu}) for the 2-nd iBD.  As $MEK$
can be protected from degradation by forming as many as of $4$ copies
of complexes with $ERK$, the actual minimum degradation rates $\underline{\beta}_{i+1}$ are then
conservatively calculated as $ [(i+1)-4] \cdot d_2$, where $d_2 =
0.15$ is the degradation rate of $MEK$ (Table~\ref{eqn:mkrxns}). This corresponds to
the other extreme case when the $MEK$ is constantly synthesized at its
maximum rate, and degraded at the minimum rate.  For both buffer
queues, estimated errors are larger than computed errors and true
errors at all buffer sizes.  We can therefore determine from {\it a
  priori}\/ estimated errors that the minimal buffer size to satisfy
the predefined error tolerance $10^{-3}$ is $14$ for the first iBD
(green straight dashed lines in Fig.~\ref{fig:mapk_Fig2}A), and $6$
for the second iBD (green straight dashed lines in
Fig.~\ref{fig:mapk_Fig2}B).  This combination of buffer sizes $\bB =
(14, 6)$ is used for all subsequent calculations.  The enumerated
state space has a total of $2,706,935$ states. The $2,706,935 \times
2,706,935$ transition rate matrix is sparse and contains a total of
$36,869,845$ non-zero elements.

\begin{figure}[ht]
{\centering
\includegraphics[scale=0.6]{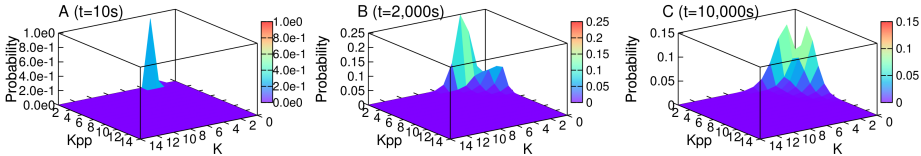}
}
\caption{The projected time evolving 16-dimension probability landscape of the
  MAPK cascade reaction network starting from the initial probability
  distribution with $p (K = 3, MKP3 = 1, All Other = 0) = 1$.  (A) The
  probability landscape projected to the $K$--$K_{pp}$ plane at $t =
  10 s$. (B) Projected probability landscape at $t = 2,000 s$.  (C)
  Projected probability landscape at $t = 10,000 s$. }
\label{fig:mapk_time}
\end{figure}

\paragraph{Steady state and time-evolving probability landscapes}
The 16-dimension time-evolving probability landscapes starting from the initial
probability distribution with $p (K = 3, MKP3 = 1, \rm{others} = 0) = 1$ 
are shown in Fig.~\ref{fig:mapk_time}.  We use a time step
$\Delta t = 10 s$ and a total simulation time of $t = 30,000 s$.  The
time-evolving probability landscape converges to the steady state
(Fig.~\ref{fig:mapk_Fig2}C) at about $t = 80,000 s$.  The calculation took $160$
minutes using a GPU workstation with an nVidia GeForce GTX 580 card
(3GB CUDA memory)~\cite{Maggioni2013GPU}.

The steady state probability landscape is also solved separately
(Fig.~\ref{fig:mapk_Fig2}C, projected onto the $K$-$Kpp$ plane).  It
has two peaks centered at $(K = 1, Kpp = 0)$ with the probability of
$0.1495$, and $(K = 0, Kpp = 2)$ with probability $0.1133$,
respectively.  The computed errors of $3.447 \times 10^{-8}$ for the
1st iBD and $1.335 \times 10^{-11}$ for the 2nd iBD are both
significantly smaller than the predefined error tolerance of $\epsilon
= 10^{-3}$.  The computation is completed within $50$ minute using the
same GPU workstation.

\paragraph{First passage time distribution and rare event probabilities} 
We study the problem of first passage time when the system travels from
an initial start state of $\bx_s = \{K = 3, MPK3 = 1 \}$, with all
other species $0$ copies, to an end state of $\bx_e = \{ Kpp = 2, MPK3
= 1 \}$, with all other species $0$ copies.  We modified the
transition rate matrix by making the end state an absorbing
state~\cite{Gross1984,Cao2013JCP}. The time evolving probability
landscape using the absorbing transition rate matrix $\bA_{abs}$ is
then calculated using a time step $\Delta t = 1 s$ for a total of
$85,000 s$ simulation time.

When the duration is short, the transition from the initial starting
state to the end state is of very low probability.  When the first
passage time is set to $t \leq 10 s$, the probability is calculated to
be $6.047 \times 10^{-9}$, with a computation time of about $22$
seconds.  Similar results would require billions of trajectories when
using the alternative method of the stochastic simulation algorithm.
As the toggle switch model, this rare event problem is
two-dimensional ($K, Kpp$) and no current methods we are aware of can
accurately calculate such rare event probabilities.

The computed full cumulative probability distribution of the first
passage time is plotted in Fig.~\ref{fig:mapk_Fig2}D. It increases
monotonically with time, and approaching probability $1$.  That is,
given enough time, the system will reach the end state $\bx_e = \{ Kpp
= 2, MPK3 = 1 \}$ with certainty $1$.  The full calculation is
completed within $41$ hours.

\section{Discussions and Conclusions}

Direct solution to the discrete chemical master equation (dCME) is of
fundamental importance.  Because the dCME plays the role in system
biology analogous to that of the Schr\"{o}dinger equation in quantum
mechanics~\cite{Beard2008}, developing methods for solving
the dCME  has important implications, just as
developing techniques for solving the Schr\"{o}dinger equation for
systems with many atoms does.

Without the truncation of higher order expansions of the discrete jump
operator and without assumptions of lower order noise as in the chemical
Langevin and the Fokker-Planck equations, accurate direct computation
of the time-evolving as well as the steady state probability landscapes
allows the stochastic properties of a biological network to be fully
characterized.  The overall stochastic behavior of a network,
including the presence or absence of multi-stabilities, the often
small probabilities of transitions between states, as well as the
overall dynamic behavior of the network can all be fully assessed.

A key challenge to obtain direct solution to the dCME is the
obstacle of the enormous discrete state space.  Conventional hypercube method
for state enumeration is easy to implement, but rapidly becomes
intractable when the network architecture is nontrivial.
In this study, we develop the
ACME algorithm using multi-buffers for directly 
solving the discrete chemical master equation.  By decomposing the
reaction network into independent components of birth-death processes,
multiple buffer queues for these components are employed for more
effective state enumeration.  With orders of magnitude reduction in
the size of the enumerated state space, our algorithm enables accurate
solution of the dCME for a large class of problems, whose solutions
were previously unobtainable.  As the network inside each birth-death
component becomes more complex, significant reduction can be
achieved.  For example, computational studies of the MAPK network
shows that a reduction factor of 6--9 orders ({\it e.g.}\/, from $1.0
\times 10^{16}$ to $6.2\times 10^{6}$ ) can be achieved, allowing a
stochastic problem otherwise unsolvable to be computed on a desktop
computer.

As truncation of the state space will eventually occur for systems of
a given fixed finite buffer capacity with fast synthesis reactions, it
is essential to quantify the truncation error and to establish a
conservative upper bound of the error, so one can assess whether the
computed results are within a predefined error tolerance and are
therefore trustworthy.  This critically important task is made possible
through theoretical analysis of the boundary states and their
associated steady state probability, via the construction of an
aggregated continuous-time Markov process based on factoring of the
state space by the buffer queue usage.  With explicit formulae for
calculating conservative error bounds for the steady state, one can
easily calculate error bounds {\it a priori}\/ for a finite 
state space associated with a given buffer capacity.  
One can also determine the minimal buffer capacity required
if a predefined error tolerance is to be satisfied.  This eliminates
the need of multiple iterations of costly trial computations to solve
the dCME for determining the appropriate buffer capacity necessary to
ensure small truncation errors.  Furthermore, for a given fixed
memory, we can also strategically allocate the memory to
different buffer queues so the overall error is minimized, 
or computing efficiency optimized.

The analysis of the truncation error also enables accurate computation
of the steady state probability landscape of a stochastic network.
This differs significantly from the finite state projection (FSP)
method, which was developed to compute the transient time evolving probability
landscape~\cite{Munsky2006,Munsky2007}.  The FSP method treats all
boundary states effectively as one absorbing state, which will
eventually trap all probability mass, resulting in a truncation error
that can increase to $1$ as time proceeds.  The error certificate in
the FSP method is used for bounding this leaked probability mass, and
requires trial solutions to the dCME, which can be costly.  This error certificate 
therefore may be unsuitable for studying long-time behavior or the steady state of
the probability landscape, as time proceeds it approaches to 1.0,
and becomes uninformative~\cite{Munsky2006,Munsky2007}.  In contrast,
no absorbing states are introduced in the mb-dCME method, the error
bound is based on analysis of the probability mass on the boundary
states.  To our knowledge, the ACME method is 
among the first general methods 
that can directly compute the steady state probabilistic landscape of
stochastic networks.

We have also provided computational results of three well-known
stochastic networks, namely, the
toggle switch, the phage lambda epigenetic circuit, and the MAPK
cascade. They are bi- or multi-stable networks. Both the time-evolving and the steady state probability
landscapes are computed, all with error less than a predefined
threshold.  Many biologically critical but rare events, such as the
spontaneous induction of latent lysogeny of phage lambda provirus into
lysis~\cite{Cao2010,Aurell2002PRE,Little1999}, or the cancerogenesis
of a normal cell~\cite{Hanahan2000}, can in principle be formulated as
a problem of estimating the distribution of the first-passage time.
The ACME method can be used to directly compute the exact
probability of rare events in a stochastic model occurring in an arbitrary time interval.
This has been demonstrated in all three examples. Our method can
provide solutions to this challenging problem 
that various forms of specifically designed stochastic
simulation algorithms have difficulties to
resolve~\cite{Allen2009,Daigle2011,Roh2011,Cao2013JCP}.

In this study, we use the {\sc Expokit}, a Krylov subspace projection
method~\cite{Sidje1998}, to compute all time-evolving probability
landscapes.  Exploiting the special structure of the state space,
recent development in methods of tensor train decomposition offers
another attractive approach to compute the probability landscape by
decomposing the dCME transition rate matrix into multiplication of
smaller tensors~\cite{Kazeev2014}.  It would be interesting to explore
how this technique can be applied to a state space enumerated by the
mb-dCME method. Although the ACME method dramatically reduces the state space
and can quantify the truncation error asymptotically, it can still fail when
the biological network in question is so large that a reduction factor of $O(n!)$
is insufficient. In addition, the \textit{a priori} error estimate may not be tight for
some complex networks. Further improvements and developments will be the focus of future studies.

Since we have a quantitative estimation of the truncation error, we 
can be sure that all major probability peaks are contained in the 
computed solution of probability landscapes when the estimated errors 
are sufficiently small, as are the cases for the three examples given here. 
Overall, the goal of this study is to provide a methodology for high
precision solutions to the dCME that can be applied to a
large class of problems. Important unknown features such as basins,
attractors, and transitions for many biological networks can be
uncovered, analyzed, and their biological significance assessed.  It
is now possible to analyze details of the topological, topographical,
as well as dynamic properties of the probability landscape for a large
number of biologically important stochastic networks that are
previously not amenable to computational investigations.

\section{ACKNOWLEDGMENTS}

This work was supported by NIH grant GM079804, NSF grant MCB1415589,
and the Chicago Biomedical Consortium with support from the Searle
Funds at The Chicago Community Trust. 
We thank Dr. Ao Ma for helpful comments, 
and Alan Perez-Rathke for helpful discussions. 
YC acknowledges the support of the 
U.S. Department of Energy through the LANL/LDRD Program for this work. 
YC thanks Dr. Alan Perelson for helpful comments.

\bibliography{dcme}

\begin{thebibliography}{10}

\bibitem{Ao2005}
Ping Ao.
\newblock Laws in darwinian evolutionary theory.
\newblock {\em Physics of life Reviews}, 2(2):117--156, 2005.

\bibitem{Stewart2012}
Jacob Stewart-Ornstein and Hana El-Samad.
\newblock Stochastic modeling of cellular networks.
\newblock {\em Computational Methods in Cell Biology}, 110:111, 2012.

\bibitem{Qian2012}
Hong Qian.
\newblock Cooperativity in cellular biochemical processes: noise-enhanced
  sensitivity, fluctuating enzyme, bistability with nonlinear feedback, and
  other mechanisms for sigmoidal responses.
\newblock {\em Annual Review of Biophysics}, 41:179--204, 2012.

\bibitem{Arkin1998}
Adam Arkin, John Ross, and Harley~H. McAdams.
\newblock {Stochastic kinetic analysis of developmental pathway bifurcation in
  phage $\lambda$-infected {E}scherichia coli cells}.
\newblock {\em Genetics}, 149(4):1633--1648, 1998.

\bibitem{Swain2002}
P.S. Swain, M.B. Elowitz, and E.D. Siggia.
\newblock {Intrinsic and extrinsic contributions to stochasticity in gene
  expression}.
\newblock {\em Proceedings of the National Academy of Sciences of the United
  States of America}, 99(20):12795, 2002.

\bibitem{Elowitz2002}
M.B. Elowitz, A.J. Levine, E.D. Siggia, and P.S. Swain.
\newblock {Stochastic gene expression in a single cell}.
\newblock {\em Science}, 297(5584):1183, 2002.

\bibitem{Cao2010}
Youfang Cao, Hsiao-Mei Lu, and Jie Liang.
\newblock {Probability landscape of heritable and robust epigenetic state of
  lysogeny in phage lambda}.
\newblock {\em Proceedings of the National Academy of Sciences of the United
  States of America}, 107(43):18445--18450, 2010.

\bibitem{McAdams1999}
H.H. McAdams and A.~Arkin.
\newblock {It's a noisy business! Genetic regulation at the nanomolar scale}.
\newblock {\em Trends in Genetics}, 15(2):65--69, 1999.

\bibitem{Wilkinson2009}
Darren~J Wilkinson.
\newblock Stochastic modelling for quantitative description of heterogeneous
  biological systems.
\newblock {\em Nature Reviews Genetics}, 10(2):122--133, 2009.

\bibitem{Gillespie_JPC77}
D.~T. Gillespie.
\newblock Exact stochastic simulation of coupled chemical reactions.
\newblock {\em Journal of Physical Chemistry}, 81:2340--2361, 1977.

\bibitem{Gillespie-PhysicaA-1992}
Daniel~T. Gillespie.
\newblock A rigorous derivation of the chemical master equation.
\newblock {\em Physica A}, 188:404--425, 1992.

\bibitem{vanKampen2007}
N.G. Van~Kampen.
\newblock {\em Stochastic processes in physics and chemistry, 3rd Edition}.
\newblock Elsevier Science and Technology books, 2007.

\bibitem{Beard2008}
D.A. Beard and H.~Qian.
\newblock {\em {Chemical biophysics: quantitative analysis of cellular
  systems}}.
\newblock Cambridge Univ Pr, 2008.

\bibitem{Gillespie2009-jcp}
Daniel~T. Gillespie.
\newblock A diffusional bimolecular propensity function.
\newblock {\em The Journal of Chemical Physics}, 131(16):--, 2009.

\bibitem{Darvey-1966-jcp}
I.G. Darvey, B.W. Ninham, and P.J. Staff.
\newblock Stochastic models for second order chemical reaction kinetics. the
  equilibrium state.
\newblock {\em The Journal of Chemical Physics}, 45:2145--2155, 1966.

\bibitem{McQuarrie-1967-JAppProb}
D.A. McQuarrie.
\newblock Stochastic approach to chemical kinetics.
\newblock {\em Journal of Applied Probability}, 4:413--478, 1967.

\bibitem{Laurenzi-2000-jcp}
I.J. Laurenzi.
\newblock An analytical solution of the stochastic master equation for
  reversible bimolecular reaction kinetics.
\newblock {\em The Journal of Chemical Physics}, 113:3315--3322, 2000.

\bibitem{Vellela2007}
Melissa Vellela and Hong Qian.
\newblock A quasistationary analysis of a stochastic chemical reaction:
  {K}eizer’s paradox.
\newblock {\em Bulletin of Mathematical Biology}, 69(5):1727--1746, 2007.

\bibitem{VanKampen-1961}
N.~G. Van~Kampen.
\newblock A power series expansion of the master equation.
\newblock {\em Canadian Journal of Physics}, 39(4):551--567, 1961.

\bibitem{Gillespie_JCP2000}
Daniel.~T. Gillespie.
\newblock The chemical langevin equation.
\newblock {\em The Journal of Chemical Physics}, 113:297--306, 2000.

\bibitem{Gillespie2002}
Daniel~T. Gillespie.
\newblock The chemical {L}angevin and {F}okker−{P}lanck equations for the
  reversible isomerization reaction†.
\newblock {\em The Journal of Physical Chemistry A}, 106(20):5063--5071, 2002.

\bibitem{Haseltine2002}
Eric~L. Haseltine and James~B. Rawlings.
\newblock Approximate simulation of coupled fast and slow reactions for
  stochastic chemical kinetics.
\newblock {\em The Journal of Chemical Physics}, 117(15):6959--6969, 2002.

\bibitem{Gardiner2004-book}
C.~W. Gardiner.
\newblock {\em Handbook of Stochastic Methods for Physics, Chemistry and the
  Natural Sciences}.
\newblock Springer, New York, 2004.

\bibitem{Ao2010}
P~Ao, D~Galas, L~Hood, L~Yin, and XM~Zhu.
\newblock Towards predictive stochastic dynamical modeling of cancer genesis
  and progression.
\newblock {\em Interdisciplinary Sciences: Computational Life Sciences},
  2(2):140--144, 2010.

\bibitem{Shi2012}
Jianghong Shi, Tianqi Chen, Ruoshi Yuan, Bo~Yuan, and Ping Ao.
\newblock Relation of a new interpretation of stochastic differential equations
  to ito process.
\newblock {\em Journal of Statistical Physics}, 148(3):579--590, 2012.

\bibitem{Grima-2011-jcp}
R.~Grima, P.~Thomas, and A.~V. Straube.
\newblock {{H}ow accurate are the nonlinear chemical {F}okker-{P}lanck and
  chemical {L}angevin equations?}
\newblock {\em The Journal of Chemical Physics}, 135(8):084103, Aug 2011.

\bibitem{Grima-2013-bcmgenomics}
P.~Thomas, H.~Matuschek, and R.~Grima.
\newblock {{H}ow reliable is the linear noise approximation of gene regulatory
  networks?}
\newblock {\em BMC Genomics}, 14 Suppl 4:S5, 2013.

\bibitem{Allen2005Sampling}
R.J. Allen, P.B. Warren, and P.R. Ten~Wolde.
\newblock Sampling rare switching events in biochemical networks.
\newblock {\em Physical Review Letters}, 94(1):18104, 2005.

\bibitem{Kuwahara2008}
H.~Kuwahara and I.~Mura.
\newblock An efficient and exact stochastic simulation method to analyze rare
  events in biochemical systems.
\newblock {\em The Journal of Chemical Physics}, 129:165101, 2008.

\bibitem{Daigle2011}
B.J. Daigle, M.K. Roh, D.T. Gillespie, and L.R. Petzold.
\newblock Automated estimation of rare event probabilities in biochemical
  systems.
\newblock {\em The Journal of Chemical Physics}, 134:044110, 2011.

\bibitem{Jiao2011}
Shuyun Jiao, Yanbo Wang, Bo~Yuan, and Ping Ao.
\newblock Kinetics of muller's ratchet from adaptive landscape viewpoint.
\newblock In {\em Systems Biology (ISB), 2011 IEEE International Conference
  on}, pages 27--32. IEEE, 2011.

\bibitem{Cao2013JCP}
Youfang Cao and Jie Liang.
\newblock Adaptively biased sequential importance sampling for rare events in
  reaction networks with comparison to exact solutions from finite buffer
  d{CME} method.
\newblock {\em The Journal of Chemical Physics}, 139(2):025101, 2013.

\bibitem{Wang2014}
Gaowei Wang, Xiaomei Zhu, Jianren Gu, and Ping Ao.
\newblock Quantitative implementation of the endogenous molecular--cellular
  network hypothesis in hepatocellular carcinoma.
\newblock {\em Interface focus}, 4(3):20130064, 2014.

\bibitem{Munsky2006}
Brian Munsky and Mustafa Khammash.
\newblock The finite state projection algorithm for the solution of the
  chemical master equation.
\newblock {\em The Journal of Chemical Physics}, 124(4):044104, 2006.

\bibitem{CaoBMCSB08}
Youfang Cao and Jie Liang.
\newblock {Optimal enumeration of state space of finitely buffered stochastic
  molecular networks and exact computation of steady state landscape
  probability}.
\newblock {\em BMC Systems Biology}, 2(1):30, 2008.

\bibitem{Wolf2010}
Verena Wolf, Rushil Goel, Maria Mateescu, and Thomas Henzinger.
\newblock Solving the chemical master equation using sliding windows.
\newblock {\em BMC Systems Biology}, 4(1):42, 2010.

\bibitem{Jahnke2011}
Tobias Jahnke.
\newblock On reduced models for the chemical master equation.
\newblock {\em Multiscale Modeling \& Simulation}, 9(4):1646--1676, 2011.

\bibitem{Sidje1998}
Roger~B Sidje.
\newblock Expokit: a software package for computing matrix exponentials.
\newblock {\em ACM Transactions on Mathematical Software (TOMS)},
  24(1):130--156, 1998.

\bibitem{Munsky2007}
Brian Munsky and Mustafa Khammash.
\newblock A multiple time interval finite state projection algorithm for the
  solution to the chemical master equation.
\newblock {\em Journal of Computational Physics}, 226(1):818 -- 835, 2007.

\bibitem{Cormen2001}
Thomas~H Cormen, Charles~E Leiserson, Ronald~L Rivest, Clifford Stein, et~al.
\newblock {\em Introduction to algorithms}, volume~2.
\newblock MIT press Cambridge, 2001.

\bibitem{Chung1997}
Fan~RK Chung.
\newblock {\em Spectral graph theory}, volume~92.
\newblock American Mathematical Soc., 1997.

\bibitem{Cao2016a}
Youfang Cao, Anna Terebus, and Jie Liang.
\newblock State space truncation with quantified errors for accurate solutions
  to discrete chemical master equation.
\newblock {\em Bulletin of Mathematical Biology}, page in press, 2016.

\bibitem{Taylor1998}
H.M. Taylor and S.~Karlin.
\newblock {\em {An Introduction to Stochastic Modeling, 3rd Ed.}}
\newblock Academic Press, 1998.

\bibitem{Fox1988}
Bennett~L Fox and Peter~W Glynn.
\newblock Computing poisson probabilities.
\newblock {\em Communications of the ACM}, 31(4):440--445, 1988.

\bibitem{Truffet1997}
Laurent Truffet.
\newblock Near complete decomposability: bounding the error by a stochastic
  comparison method.
\newblock {\em Advances in Applied Probability}, pages 830--855, 1997.

\bibitem{Irle2003}
A~Irle.
\newblock Stochastic ordering for continuous-time processes.
\newblock {\em Journal of Applied Probability}, pages 361--375, 2003.

\bibitem{Kazeev2014}
Vladimir Kazeev, Mustafa Khammash, Michael Nip, and Christoph Schwab.
\newblock Direct solution of the chemical master equation using quantized
  tensor trains.
\newblock {\em PLoS Computational Biology}, 10(3):e1003359, 03 2014.

\bibitem{Lehoucq_Arpack}
R.~Lehoucq, D.~Sorensen, and C.~Yang.
\newblock {\em Arpack users' guide: Solution of large scale eigenvalue problems
  with implicitly restarted {A}rnoldi methods}.
\newblock SIAM, Philadelphia, 1998.

\bibitem{Stewart1994}
W.J. Stewart.
\newblock {\em {Introduction to the numerical solution of Markov chains}}.
\newblock Princeton University Press NJ, 1994.

\bibitem{Saad1986gmres}
Youcef Saad and Martin~H Schultz.
\newblock {GMRES}: A generalized minimal residual algorithm for solving
  nonsymmetric linear systems.
\newblock {\em SIAM Journal on Scientific and Statistical Computing},
  7(3):856--869, 1986.

\bibitem{Gross1984}
Donald Gross and Douglas~R. Miller.
\newblock The randomization technique as a modeling tool and solution procedure
  for transient {M}arkov processes.
\newblock {\em Operations Research}, 32(2):343--361, 1984.

\bibitem{Gillespie2009}
Dan~T Gillespie, Min Roh, and Linda~R Petzold.
\newblock Refining the weighted stochastic simulation algorithm.
\newblock {\em The Journal of Chemical Physics}, 130(17):174103, 2009.

\bibitem{Donovan2013}
Rory~M Donovan, Andrew~J Sedgewick, James~R Faeder, and Daniel~M Zuckerman.
\newblock Efficient stochastic simulation of chemical kinetics networks using a
  weighted ensemble of trajectories.
\newblock {\em The Journal of Chemical Physics}, 139(11):115105, 2013.

\bibitem{Gardner2000}
Timothy~S Gardner, Charles~R Cantor, and James~J Collins.
\newblock Construction of a genetic toggle switch in {E}scherichia coli.
\newblock {\em Nature}, 403(6767):339--342, 2000.

\bibitem{Kepler2001}
Thomas~B Kepler and Timothy~C Elston.
\newblock Stochasticity in transcriptional regulation: origins, consequences,
  and mathematical representations.
\newblock {\em Biophysical Journal}, 81(6):3116--3136, 2001.

\bibitem{Kim2007}
Keun-Young Kim and Jin Wang.
\newblock Potential energy landscape and robustness of a gene regulatory
  network: toggle switch.
\newblock {\em PLoS Computational Biology}, 3(3):e60, 2007.

\bibitem{Schultz_JCP07}
D.~Schultz, J.~N. Onuchic, and P.~G. Wolynes.
\newblock Understanding stochastic simulations of the smallest genetic
  networks.
\newblock {\em The Journal of Chemical Physics}, 126(24):245102, 2007.

\bibitem{Munsky2008}
Brian Munsky and Mustafa Khammash.
\newblock The finite state projection approach for the analysis of stochastic
  noise in gene networks.
\newblock {\em Automatic Control, IEEE Transactions on}, 53(Special
  Issue):201--214, 2008.

\bibitem{Deuflhard2008}
Peter Deuflhard, Wilhelm Huisinga, T~Jahnke, and Michael Wulkow.
\newblock Adaptive discrete {G}alerkin methods applied to the chemical master
  equation.
\newblock {\em SIAM Journal on Scientific Computing}, 30(6):2990--3011, 2008.

\bibitem{Sjoberg2009}
Paul Sj{\"o}berg, Per L{\"o}tstedt, and Johan Elf.
\newblock {F}okker--{P}lanck approximation of the master equation in molecular
  biology.
\newblock {\em Computing and Visualization in Science}, 12(1):37--50, 2009.

\bibitem{Santillan2008}
Moises Santill{\'a}n.
\newblock On the use of the hill functions in mathematical models of gene
  regulatory networks.
\newblock {\em Mathematical Modelling of Natural Phenomena}, 3(02):85--97,
  2008.

\bibitem{Kim2012}
Haseong Kim and Erol Gelenbe.
\newblock Stochastic gene expression modeling with hill function for
  switch-like gene responses.
\newblock {\em IEEE/ACM Transactions on Computational Biology and
  Bioinformatics}, 9(4):973--979, 2012.

\bibitem{Roh2011}
M.K. Roh, B.J. Daigle, D.T. Gillespie, and L.R. Petzold.
\newblock State-dependent doubly weighted stochastic simulation algorithm for
  automatic characterization of stochastic biochemical rare events.
\newblock {\em Journal of Chemical Physics}, 135(23):234108, 2011.

\bibitem{Ptashne2004}
Mark Ptashne.
\newblock {\em A Genetic Switch: Phage Lambda Revisited.}
\newblock Cold Spring Harbor Laboratory Press; 3 edition, 2004.

\bibitem{Aurell2002PRE}
Erik Aurell, Stanley Brown, Johan Johanson, and Kim Sneppen.
\newblock {Stability puzzles in phage $\lambda$}.
\newblock {\em Physical Review E}, 65(5):051914, 2002.

\bibitem{Aurell2002PRL}
Erik Aurell and Kim Sneppen.
\newblock {Epigenetics as a first exit problem}.
\newblock {\em Physical Review Letters}, 88(4):048101, 2002.

\bibitem{ZhuJBCB2004}
X.-M. Zhu, L.~Yin, L.~Hood, and P.~Ao.
\newblock Robustness, stability and efficiency of phage lambda genetic switch:
  dynamical structure analysis.
\newblock {\em Journal of Bioinformatics and Computational Biology},
  2:785--817, 2004.

\bibitem{Zhu2004}
X.-M. Zhu, L.~Yin, L.~Hood, and P.~Ao.
\newblock {Calculating biological behaviors of epigenetic states in the phage
  $\lambda$ life cycle}.
\newblock {\em Functional \& Integrative Genomics}, 4(3):188--195, 2004.

\bibitem{Hawley_PNASUSA80}
D.~Hawley and W.~McClure.
\newblock In vitro comparison of initiation properties of bacteriophage lambda
  wild-type {PR} and x3 mutant promoters.
\newblock {\em Proceedings of the National Academy of Sciences of the United
  States of America}, 77(11):6381--6385, 1980.

\bibitem{Hawley_JMB82}
D.~Hawley and W.~McClure.
\newblock Mechanism of activation of transcription initiation from the lambda
  {PRM} promoter.
\newblock {\em Journal of Molecular Biology}, 157(3):493--525, 1982.

\bibitem{Shea1985}
Madeline~A. Shea and Gary~K. Ackers.
\newblock {The $OR$ control system of bacteriophage lambda a physical-chemical
  model for gene regulation}.
\newblock {\em Journal of Molecular Biology}, 181(2):211--230, 1985.

\bibitem{Li_PNASUSA97}
M.~Li, W.~McClure, and M.~Susskind.
\newblock Changing the mechanism of transcriptional activation by phage lambda
  repressor.
\newblock {\em Proceedings of the National Academy of Sciences of the United
  States of America}, 94(8):3691--3696, 1997.

\bibitem{Kuttler2006}
C\'{e}line Kuttler and Joachim Niehren.
\newblock {Gene Regulation in the Pi Calculus: Simulating Cooperativity at the
  Lambda Switch}.
\newblock {\em Transactions on Computational Systems Biology VII}, 4230:24--55,
  2006.

\bibitem{Johnson2002}
Gary~L Johnson and Razvan Lapadat.
\newblock Mitogen-activated protein kinase pathways mediated by erk, jnk, and
  p38 protein kinases.
\newblock {\em Science}, 298(5600):1911--1912, 2002.

\bibitem{Huang1996}
Chi-Ying Huang and James~E Ferrell.
\newblock Ultrasensitivity in the mitogen-activated protein kinase cascade.
\newblock {\em Proceedings of the National Academy of Sciences of the United
  States of America}, 93(19):10078--10083, 1996.

\bibitem{Wang2006}
Xiao Wang, Nan Hao, Henrik~G Dohlman, and Timothy~C Elston.
\newblock Bistability, stochasticity, and oscillations in the mitogen-activated
  protein kinase cascade.
\newblock {\em Biophysical Journal}, 90(6):1961--1978, 2006.

\bibitem{Qiao2007}
Liang Qiao, Robert~B Nachbar, Ioannis~G Kevrekidis, and Stanislav~Y Shvartsman.
\newblock Bistability and oscillations in the {H}uang-{F}errell model of {MAPK}
  signaling.
\newblock {\em PLoS Computational Biology}, 3(9):e184, 2007.

\bibitem{Markevich_JCB04}
N.~I. Markevich, J.~B. Hoek, and B.~N. Kholodenko.
\newblock Signaling switches and bistability arising from multisite
  phosphorylation in protein kinase cascades.
\newblock {\em The Journal of Cell Biology}, 164(3):353--359, 2004.

\bibitem{Smolen2008}
Paul Smolen, Douglas~A. Baxter, and John~H. Byrne.
\newblock Bistable {MAP} kinase activity: a plausible mechanism contributing to
  maintenance of late long-term potentiation.
\newblock {\em American Journal of Physiology - Cell Physiology},
  294(2):C503--C515, 2008.

\bibitem{Maggioni2013GPU}
Marco Maggioni, Tanya Berger-Wolf, and Jie Liang.
\newblock Gpu-based steady-state solution of the chemical master equation.
\newblock In {\em Parallel and Distributed Processing Symposium Workshops \&
  PhD Forum (IPDPSW), 2013 IEEE 27th International}, pages 579--588. IEEE,
  2013.

\bibitem{Little1999}
John~W. Little, Donald~P. Shepley, and David~W. Wert.
\newblock {Robustness of a gene regulatory circuit}.
\newblock {\em The EMBO Journal}, 18(15):4299--4307, 1999.

\bibitem{Hanahan2000}
Douglas Hanahan and Robert~A Weinberg.
\newblock The hallmarks of cancer.
\newblock {\em Cell}, 100(1):57--70, 2000.

\bibitem{Allen2009}
Rosalind~J Allen, Chantal Valeriani, and Pieter~Rein ten Wolde.
\newblock Forward flux sampling for rare event simulations.
\newblock {\em Journal of Physics: Condensed Matter}, 21(46):463102, 2009.

\bibitem{Merris1994Laplacian}
Russell Merris.
\newblock Laplacian matrices of graphs: a survey.
\newblock {\em Linear Algebra and Its Applications}, 197:143--176, 1994.

\end{thebibliography}

\clearpage

\section*{Appendix}

\subsection*{Graph of Reaction Network, its Adjacency and Laplacian Matrices}

$\bG_R$ can be represented by an $m \times m$ adjacency
matrix $\bC$, where:
\begin{equation}
\bC^{m \times m} = || \bC_{i,\,j} || = \left\{ 
  \begin{array}{l l}
    1, & \quad \text{if $e_{ij}$ exists,} \\
		0, & \quad \text{otherwise}
 \\
  \end{array} \right.
\label{eqn:adjC}
\end{equation}
The diagonal degree matrix $\bD$ of the graph $\bG_R$ is: 
\begin{equation}
\bD^{m \times m} = || \bD_{i,\,j} || = \left\{ 
  \begin{array}{l l}
    \sum_{k=1}^m {\bC_{i,k}}, & \quad \text{if $i = j$}, \\
		0, & \quad \text{if $i \neq j$}, \\
  \end{array} \right.
\label{eqn:degD}
\end{equation}
where each diagonal element $\bD_{i,i}$ is the vertex degree of 
the corresponding reaction $R_i$.
The Laplacian matrix $\bL$ of the graph 
$\bG_R$ can be then written as~\cite{Merris1994Laplacian}: 
\begin{equation}
\bL = \bD - \bC.
 \label{eqn:lapL}
\end{equation}


\setlength{\tabcolsep}{1pt}
\begin{table}[!h]
\caption{Detailed reactions and rate constants of genetic toggle switch.}
\label{eqn:tgrxns}
\begin{center}
\begin{scriptsize}
\begin{tabular}{l l}
  \hline
$R_1: GeneX \stackrel{k_1}{\rightarrow} GeneX + X, \quad k_1 = 50 \, s^{-1} $ & 
$R_3: X \stackrel{k_3}{\rightarrow} \emptyset, \quad k_3 = 1 \, s^{-1} $ \\
$R_2: GeneY \stackrel{k_2}{\rightarrow} GeneY + Y, \quad k_2 = 100 \, s^{-1} $ & 
$R_4: Y \stackrel{k_4}{\rightarrow} \emptyset, \quad k_4 = 1 \, s^{-1} $ \\
$R_5: 2X + GeneY \stackrel{k_5}{\rightarrow} BGeneY, \quad k_5 = 1 \times 10^{-5} \, nM^{-2} \cdot s^{-1} $ &
$R_7: BGeneY \stackrel{k_7}{\rightarrow} 2X + GeneY, \quad k_7 = 0.1 \, s^{-1} $ \\
$R_6: 2Y + GeneX \stackrel{k_6}{\rightarrow} BGeneX, \quad k_6 = 1 \times 10^{-5} \, nM^{-2} \cdot s^{-1} $ &
$R_8: BGeneX \stackrel{k_8}{\rightarrow} 2Y + GeneX, \quad k_8 = 0.1 \, s^{-1} $ \\
  \hline
\end{tabular}
\end{scriptsize}
\end{center}
\end{table}


\setlength{\tabcolsep}{1pt}
\begin{table}[!h]
\caption{Detailed reactions and rate constants of phage lambda epigenetic switch.}
\label{eqn:plrxns}
\begin{center}
\begin{tiny}
\begin{tabular}{l l}
  \hline
$R_1: \emptyset + (OR3 + OR2) \stackrel{s_{CI}^0}{\rightarrow} CI + (OR3 + OR2), s_{CI}^0 = 0.0069 /s,  $ & 
$R_7: 2CI + OR1 \stackrel{b_{CI}}{\rightarrow} ROR1, b_{CI} = 0.0021 /nM^2 \cdot s, $ \\
$R_2: \emptyset + (OR3 + COR2) \stackrel{s_{CI}^0}{\rightarrow} CI + (OR3 + COR2), s_{CI}^0 = 0.0069 /s, $ & 
$R_8: 2CI + OR2 \stackrel{b_{CI}}{\rightarrow} ROR2, b_{CI} = 0.0021 /nM^2 \cdot s, $ \\
$R_3: \emptyset + (OR3 + ROR2) \stackrel{s_{CI}^1}{\rightarrow} CI + (OR3 + ROR2), s_{CI}^1 = 0.066 /s, $ & 
$R_9: 2CI + OR3 \stackrel{b_{CI}}{\rightarrow} ROR3, b_{CI} = 0.0021 /nM^2 \cdot s, $ \\
$R_4: \emptyset + (OR1 + OR2) \stackrel{s_{Cro}}{\rightarrow} Cro + (OR1 + OR2), s_{Cro} = 0.0929 /s, $ & 
$R_{10}: 2Cro + OR1 \stackrel{b_{Cro}}{\rightarrow} COR1, b_{Cro} = 0.01289 /nM^2 \cdot s, $ \\
$R_5: CI \stackrel{d_{CI}}{\rightarrow} \emptyset, d_{CI} = 0.0027 /s, $ & 
$R_{11}: 2Cro + OR2 \stackrel{b_{Cro}}{\rightarrow} COR2, b_{Cro} = 0.01289 /nM^2 \cdot s, $ \\
$R_6: Cro \stackrel{d_{Cro}}{\rightarrow} \emptyset, d_{Cro} = 0.0025 /s, $ & 
$R_{12}: 2Cro + OR3 \stackrel{b_{Cro}}{\rightarrow} COR3, b_{Cro} = 0.01289 /nM^2 \cdot s, $ \\
\multicolumn{2}{l}{$R_{13}:  ROR1 + (OR2) \stackrel{u_{ROR1}}{\rightarrow} 2CI + OR1 + (OR2), u_{ROR1} = 0.03998/s, $} \\
\multicolumn{2}{l}{$R_{14}:  ROR1 + (ROR2 + OR3) \stackrel{u_{ROR1}^{12}}{\rightarrow} 2CI + OR1 + (ROR2 + OR3), u_{ROR1}^{12} = 0.0005/s, $} \\
\multicolumn{2}{l}{$R_{15}:  ROR1 + (ROR2 + ROR3) \stackrel{u_{ROR1}^{123}}{\rightarrow} 2CI + OR1 + (ROR2 + ROR3), u_{ROR1}^{123} = 0.05531/s, $} \\
\multicolumn{2}{l}{$R_{16}:  ROR1 + (ROR2 + COR3) \stackrel{u_{ROR1}^{12}}{\rightarrow} 2CI + OR1 + (ROR2 + COR3), u_{ROR1}^{12} = 0.0005/s, $} \\
\multicolumn{2}{l}{$R_{17}:  ROR1 + (COR2) \stackrel{u_{ROR1}}{\rightarrow} 2CI + OR1 + (COR2), u_{ROR1} = 0.03998/s, $} \\
\multicolumn{2}{l}{$R_{18}:  ROR2 + (OR1 + OR3) \stackrel{u_{ROR2}}{\rightarrow} 2CI + OR2 + (OR1 + OR3), u_{ROR2} = 1.026/s, $} \\
\multicolumn{2}{l}{$R_{19}:  ROR2 + (ROR1 + OR3) \stackrel{u_{ROR2}^{12}}{\rightarrow} 2CI + OR2 + (ROR1 + OR3), u_{ROR2}^{12} = 0.01284/s, $} \\
\multicolumn{2}{l}{$R_{20}:  ROR2 + (OR1 + ROR3) \stackrel{u_{ROR2}^{23}}{\rightarrow} 2CI + OR2 + (OR1 + ROR3), u_{ROR2}^{23} = 0.00928/s, $} \\
\multicolumn{2}{l}{$R_{21}:  ROR2 + (ROR1 + ROR3) \stackrel{u_{ROR2}^{123}}{\rightarrow} 2CI + OR2 + (ROR1 + ROR3), u_{ROR2}^{123} = 0.01284/s, $} \\
\multicolumn{2}{l}{$R_{22}:  ROR2 + (COR1 + OR3) \stackrel{u_{ROR2}}{\rightarrow} 2CI + OR2 + (COR1 + OR3), u_{ROR2} = 1.026/s, $} \\
\multicolumn{2}{l}{$R_{23}:  ROR2 + (OR1 + COR3) \stackrel{u_{ROR2}}{\rightarrow} 2CI + OR2 + (OR1 + COR3), u_{ROR2} = 1.026/s, $} \\
\multicolumn{2}{l}{$R_{24}:  ROR2 + (COR1 + COR3) \stackrel{u_{ROR2}}{\rightarrow} 2CI + OR2 + (COR1 + COR3), u_{ROR2} = 1.026/s, $} \\
\multicolumn{2}{l}{$R_{25}:  ROR2 + (ROR1 + COR3) \stackrel{u_{ROR2}^{12}}{\rightarrow} 2CI + OR2 + (ROR1 + COR3), u_{ROR2}^{12} = 0.01284/s, $} \\
\multicolumn{2}{l}{$R_{26}:  ROR2 + (COR1 + ROR3) \stackrel{u_{ROR2}^{23}}{\rightarrow} 2CI + OR2 + (COR1 + ROR3), u_{ROR2}^{23} = 0.00928/s, $} \\
\multicolumn{2}{l}{$R_{27}:  ROR3 + (OR2) \stackrel{u_{ROR3}}{\rightarrow} 2CI + OR3 + (OR2), u_{ROR3} = 5.19753/s, $} \\
\multicolumn{2}{l}{$R_{28}:  ROR3 + (ROR2 + OR1) \stackrel{u_{ROR3}^{23}}{\rightarrow} 2CI + OR3 + (ROR2 + OR1), u_{ROR3}^{23} = 0.04702/s, $} \\
\multicolumn{2}{l}{$R_{29}:  ROR3 + (ROR2 + ROR1) \stackrel{u_{ROR3}^{123}}{\rightarrow} 2CI + OR3 + (ROR2 + ROR1), u_{ROR3}^{123} = 5.19753/s, $} \\
\multicolumn{2}{l}{$R_{30}:  ROR3 + (ROR2 + COR1) \stackrel{u_{ROR3}^{23}}{\rightarrow} 2CI + OR3 + (ROR2 + COR1), u_{ROR3}^{23} = 0.04702/s, $} \\
\multicolumn{2}{l}{$R_{31}:  ROR3 + (COR2) \stackrel{u_{ROR3}}{\rightarrow} 2CI + OR3 + (COR2), u_{ROR3} = 5.19753/s, $} \\
\multicolumn{2}{l}{$R_{32}:  COR1 + (OR2) \stackrel{u_{COR1}}{\rightarrow} 2Cro + OR1 + (OR2), u_{COR1} = 0.08999/s, $} \\
\multicolumn{2}{l}{$R_{33}:  COR1 + (ROR2) \stackrel{u_{COR1}}{\rightarrow} 2Cro + OR1 + (ROR2), u_{COR1} = 0.08999/s, $} \\
\multicolumn{2}{l}{$R_{34}:  COR1 + (COR2 + OR3) \stackrel{u_{COR1}^{12}}{\rightarrow} 2Cro + OR1 + (COR2 + OR3), u_{COR1}^{12} = 0.01776/s, $} \\
\multicolumn{2}{l}{$R_{35}:  COR1 + (COR2 + ROR3) \stackrel{u_{COR1}^{12}}{\rightarrow} 2Cro + OR1 + (COR2 + ROR3), u_{COR1}^{12} = 0.01776/s, $} \\
\multicolumn{2}{l}{$R_{36}:  COR1 + (COR2 + COR3) \stackrel{u_{COR1}^{123}}{\rightarrow} 2Cro + OR1 + (COR2 + COR3), u_{COR1}^{123} = 0.05531/s, $} \\
\multicolumn{2}{l}{$R_{37}:  COR2 + (OR1 + OR3) \stackrel{u_{COR2}}{\rightarrow} 2Cro + OR2 + (OR1 + OR3), u_{COR2} = 0.6306/s, $} \\
\multicolumn{2}{l}{$R_{38}:  COR2 + (ROR1 + OR3) \stackrel{u_{COR2}}{\rightarrow} 2Cro + OR2 + (ROR1 + OR3), u_{COR2} = 0.6306/s, $} \\
\multicolumn{2}{l}{$R_{39}:  COR2 + (OR1 + ROR3) \stackrel{u_{COR2}}{\rightarrow} 2Cro + OR2 + (OR1 + ROR3), u_{COR2} = 0.6306/s, $} \\
\multicolumn{2}{l}{$R_{40}:  COR2 + (ROR1 + ROR3) \stackrel{u_{COR2}}{\rightarrow} 2Cro + OR2 + (ROR1 + ROR3), u_{COR2} = 0.6306/s, $} \\
\multicolumn{2}{l}{$R_{41}:  COR2 + (COR1 + OR3) \stackrel{u_{COR2}^{12}}{\rightarrow} 2Cro + OR2 + (COR1 + OR3), u_{COR2}^{12} = 0.12448/s, $} \\
\multicolumn{2}{l}{$R_{42}:  COR2 + (OR1 + COR3) \stackrel{u_{COR2}^{23}}{\rightarrow} 2Cro + OR2 + (OR1 + COR3), u_{COR2}^{23} = 0.23822/s, $} \\
\multicolumn{2}{l}{$R_{43}:  COR2 + (COR1 + COR3) \stackrel{u_{COR2}^{123}}{\rightarrow} 2Cro + OR2 + (COR1 + COR3), u_{COR2}^{123} = 0.14641/s, $} \\
\multicolumn{2}{l}{$R_{44}:  COR2 + (ROR1 + COR3) \stackrel{u_{COR2}^{23}}{\rightarrow} 2Cro + OR2 + (ROR1 + COR3), u_{COR2}^{23} = 0.23822/s, $} \\
\multicolumn{2}{l}{$R_{45}:  COR2 + (COR1 + ROR3) \stackrel{u_{COR2}^{12}}{\rightarrow} 2Cro + OR2 + (COR1 + ROR3), u_{COR2}^{12} = 0.12448/s, $} \\
\multicolumn{2}{l}{$R_{46}:  COR3 + (OR2) \stackrel{u_{COR3}}{\rightarrow} 2Cro + OR3 + (OR2), u_{COR3} = 0.00928/s, $} \\
\multicolumn{2}{l}{$R_{47}:  COR3 + (ROR2) \stackrel{u_{COR3}}{\rightarrow} 2Cro + OR3 + (ROR2), u_{COR3} = 0.00928/s, $} \\
\multicolumn{2}{l}{$R_{48}:  COR3 + (COR2 + OR1) \stackrel{u_{COR3}^{23}}{\rightarrow} 2Cro + OR3 + (COR2 + OR1), u_{COR3}^{23} = 0.00351/s, $} \\ 
\multicolumn{2}{l}{$R_{49}:  COR3 + (COR2 + ROR1) \stackrel{u_{COR3}^{23}}{\rightarrow} 2Cro + OR3 + (COR2 + ROR1), u_{COR3}^{23} = 0.00351/s, $} \\
\multicolumn{2}{l}{$R_{50}:  COR3 + (COR2 + COR1) \stackrel{u_{COR3}^{123}}{\rightarrow} 2Cro + OR3 + (COR2 + COR1), u_{COR3}^{123} = 0.01092/s $} \\
  \hline
\end{tabular}
\end{tiny}
\end{center}
\end{table}

\begin{table}[!h]
\caption{Molecular species in the network of bistable MAPK signaling cascade.}
\label{tab:mkspecies}
\begin{center}
\begin{footnotesize}
\begin{tabular}{|l|l|}
  \hline
  Molecular species & Descriptions \\
  \hline
  MEK & ERK kinase \\
  MKP3 & ERK phosphatase \\
	K & ERK, extracellular signal-regulated kinase \\
  KpY & Single phosphorylated ERK on Y residue \\
  KpT & Single phosphorylated ERK on T residue \\
  Kpp & Dual phosphorylated ERK on both Y and T residue \\
  K\_MEK\_Y & K bound by MEK at residue Y \\
  K\_MEK\_T & K bound by MEK at residue T \\
  KpY\_MEK & KpY bound by MEK \\
  KpT\_MEK & KpT bound by MEK \\
  Kpp\_MKP3 & Kpp associated with MKP3 \\
  KpY\_MKP3 &  KpY associated with MKP3 \\
  KpT\_MKP3\_Y & KpT associated with MKP3 at residue Y \\
  KpT\_MKP3\_T & KpT associated with MKP3 at residue T \\
  K\_MKP3\_T & K associated with MKP3 at residue T \\
  K\_MKP3\_Y & K associated with MKP3 at residue Y \\
  \hline
\end{tabular}
\end{footnotesize}
\end{center}
\end{table}


\begin{figure}[!h]
{\centering
\includegraphics[scale=0.4]{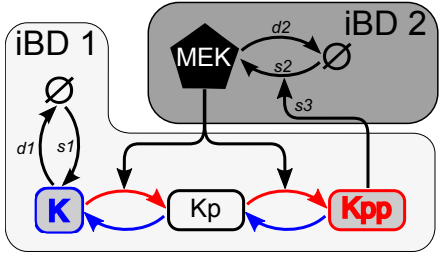}
}
\caption{A simplified conceptual model of the MAPK network. 
The MEK and ERK (K) form a positive feedback loop. 
}
\label{fig:mk0}
\end{figure}


\setlength{\tabcolsep}{1pt}

\begin{table}[!h]
\caption{Detailed reactions and rate constants in MAPK signaling network.}
\label{eqn:mkrxns}
\begin{center}
\begin{tiny}
\begin{tabular}{l l}
  \hline
$R_1: \emptyset \overset{\, s_{2} \,}{\underset{\, d_{2} \,}{\rightleftharpoons}} \text{MEK}, s_{2} = 0.001 /s, d_{2} = 0.15 /s, $ & 
$R_2: \emptyset + \text{(Kpp)} \stackrel{s_{3}}{\rightarrow} \text{MEK} + \text{(Kpp)}, s_{3} = 0.005 /s, $ \\
$R_5: \text{K} + \text{MEK} \overset{\, k_1 \,}{\underset{\, k_{-1} \,}{\rightleftharpoons}} \text{K\_MEK\_Y}, k_1 = 0.375 /nM \cdot s, k_{-1} = 1.0 /s, $ & 
$R_3: \emptyset \overset{\, s_{1} \,}{\underset{\, d_{1} \,}{\rightleftharpoons}} \text{K}, s_1 = 0.00024 /s, d_1 = 0.0001 /s, $ \\
$R_{7}: \text{KpY} + \text{MEK} \overset{\, k_3 \,}{\underset{\, k_{-3} \,}{\rightleftharpoons}} \text{KpY\_MEK}, k_3 = 0.375 /nM \cdot s, k_{-3} = 1.0 /s, $ & 
$R_4: \text{KpY} \stackrel{d_1}{\rightarrow} \emptyset, \text{KpT} \stackrel{d_1}{\rightarrow} \emptyset, \text{Kpp} \stackrel{d_1}{\rightarrow} \emptyset, d_1 = 0.0001 /s, $ \\
$R_{9}: \text{K} + \text{MEK} \overset{\, k_5 \,}{\underset{\, k_{-5} \,}{\rightleftharpoons}} \text{K\_MEK\_T}, k_5 = 0.375 /nM \cdot s, k_{-5} = 1.0 /s, $ & 
$R_{6}: \text{K\_MEK\_Y} \stackrel{k_2}{\rightarrow} \text{KpY} + \text{MEK}, k_2 = 0.06 /s, $ \\
$R_{11}: \text{KpT} + \text{MEK} \overset{\, k_7 \,}{\underset{\, k_{-7} \,}{\rightleftharpoons}} \text{KpT\_MEK}, k_7 = 0.375 /nM \cdot s, k_{-7} = 1.0 /s, $ & 
$R_{8}: \text{KpY\_MEK} \stackrel{k_4}{\rightarrow} \text{Kpp} + \text{MEK}, k_4 = 4.5 /s, $ \\
$R_{13}: \text{Kpp} + \text{MKP3} \overset{\, h_1 \,}{\underset{\, h_{-1} \,}{\rightleftharpoons}} \text{Kpp\_MKP3}, h_1 = 0.015 /nM \cdot s, h_{-1} = 1.0 /s, $ &
$R_{10}: \text{K\_MEK\_T} \stackrel{k_6}{\rightarrow} \text{KpT} + \text{MEK}, k_6 = 0.06 /s, $ \\
$R_{15}: \text{KpT\_MKP3\_Y} \overset{\, h_3 \,}{\underset{\, h_{-3} \,}{\rightleftharpoons}} \text{KpT} + \text{MKP3}, h_3 = 0.31 /s, h_{-3} = 0.01 /nM \cdot s, $ & 
$R_{12}: \text{KpT\_MEK} \stackrel{k_8}{\rightarrow} \text{Kpp} + \text{MEK}, k_8 = 4.5 /s, $ \\
$R_{16}: \text{KpT} + \text{MKP3} \overset{\, h_4 \,}{\underset{\, h_{-4} \,}{\rightleftharpoons}} \text{KpT\_MKP3\_T}, h_4 = 0.01 /nM \cdot s, h_{-4} = 1.0 /s, $ & 
$R_{14}: \text{Kpp\_MKP3} \stackrel{h_2}{\rightarrow} \text{KpT\_MKP3\_Y}, h_2 = 0.032 /s, $ \\
$R_{18}: \text{K\_MKP3\_T} \overset{\, h_6 \,}{\underset{\, h_{-6} \,}{\rightleftharpoons}} \text{K} + \text{MKP3}, h_6 = 0.086 /s, h_{-6} = 0.0011 /nM \cdot s, $ & 
$R_{17}: \text{KpT\_MKP3\_T} \stackrel{h_5}{\rightarrow} \text{K\_MKP3\_T}, h_5 = 0.5 /s, $ \\
$R_{19}: \text{KpY} + \text{MKP3} \overset{\, h_7 \,}{\underset{\, h_{-7} \,}{\rightleftharpoons}} \text{KpY\_MKP3}, h_7 = 0.01 /nM \cdot s, h_{-7} = 1.0 /s, $ & 
$R_{20}: \text{KpY\_MKP3} \stackrel{h_8}{\rightarrow} \text{K\_MKP3\_Y}, h_8 = 0.47 /s, $ \\
$R_{21}: \text{K\_MKP3\_Y} \overset{\, h_9 \,}{\underset{\, h_{-9} \,}{\rightleftharpoons}} \text{K} + \text{MKP3}, h_9 = 0.14 /s, h_{-9} = 0.0018 /nM \cdot s. $ & \\
  \hline
\end{tabular}
\end{tiny}
\end{center}
\end{table}

\end{document}